\documentclass[prb,longbibliography,twocolumn,include graphics]{revtex4-2}

\usepackage[utf8]{inputenc}

\usepackage{amssymb}

\usepackage{amsmath}

\usepackage{amsfonts}

\usepackage[usenames,dvipsnames]{xcolor}

\usepackage[pdftex]{graphicx}

\usepackage{bm}

\usepackage{hyperref}

\usepackage{nccmath} 

\hypersetup{plainpages=false,colorlinks=true,linkcolor=Red, citecolor=blue, urlcolor=blue}

\usepackage{array}

\newcolumntype{L}[1]{>{\raggedright\let\newline\\\arraybackslash\hspace{0pt}}m{#1}}

\newcolumntype{C}[1]{>{\centering\let\newline\\\arraybackslash\hspace{0pt}}m{#1}}

\newcolumntype{R}[1]{>{\raggedleft\let\newline\\\arraybackslash\hspace{0pt}}m{#1}}

\newcommand{\dd}{\text{d}}

\newcommand{\fref}[1]{Fig.~\ref{#1}}

\newcommand{\eref}[1]{Eq.(\ref{#1})}

\usepackage{pifont}

\begin{document}


\title{Mott transition, Widom line and pseudogap in \\the half-filled triangular lattice Hubbard model}

\author{P.-O.~\surname{Downey}$^1$}

\author{O.~\surname{Gingras}$^{1,2,3}$}

\author{J.~\surname{Fournier}$^1$}

\author{C.-D.~\surname{Hébert}$^1$}

\author{M.~\surname{Charlebois}$^4$}

\author{A.-M.~S.~\surname{Tremblay}$^1$}

\affiliation{$^1$D\'epartement de physique and Institut quantique, Universit\'e de Sherbrooke, Qu\'ebec, Canada J1K 2R1}
\affiliation{$^2$D\'epartement de physique, Universit\'e de Montr\'eal, Qu\'ebec, Canada H2V 2S9}
\affiliation{$^3$Center for Computational Quantum Physics, Flatiron Institute, 162 Fifth Avenue, New York, New York 10010, USA}
\affiliation{$^4$D\'epartement de Chimie, Biochimie et Physique, Institut de Recherche sur l’Hydrog\`ene, Universit\'e du Qu\'ebec \`a Trois-Rivi\`eres, Trois-Rivi\`eres, Qu\'ebec G9A 5H7, Canada}

\date{\today}



\begin{abstract}

The Mott transition is observed experimentally in materials that are magnetically frustrated so that long-range order does not hide the Mott transition at finite temperature. The Hubbard model on the triangular lattice at half filling is a paradigmatic model to study the interplay of interactions and frustration on the normal-state phase diagram. We use the dynamical cluster approximation with continuous-time auxiliary-field quantum Monte Carlo to solve this model for 1, 4, 6, 12, and 16-site clusters with detailed analysis performed for the 6-site cluster. We show that (a) for every cluster there is an inflection point in the double occupancy as a function of interaction, defining a Widom line that extends above the critical point of the first-order Mott transition; (b) the presence of this line and the cluster size dependence argue for the observability of the Mott transition at finite temperature in the thermodynamic limit; (c) the loss of spectral weight in the metal-to-Mott-insulator transition as a function of temperature and for strong interactions is momentum dependent, the hallmark of a pseudogap. That pseudogap spans a large region of the phase diagram near the Mott transition. 

\end{abstract}

\maketitle



\section{Introduction}

\label{sec:Introduction}
The existence of a sharp Fermi surface (FS) and of quasiparticles at zero temperature in the presence of interactions is one of the remarkable emergent phenomena in strongly correlated electron systems. Correlation-induced insulating behavior is also an important emergent phenomenon.
The transition from metal to Mott insulator (MI), called the Mott transition (MT), is considered a hallmark of strong electronic correlations in condensed matter physics.
The simplest model that captures the MT is the Hubbard model~\cite{hubbard1963electron, hubbard1964electron2, hubbard1964electron3, Gutzwiller:1963,kanamori1963electron}. Recent reviews can be found in Refs.~\cite{qinHubbardModelComputational2022a,Arovas_Berg_Kivelson_Raghu_2022}.
Several numerical studies based on dynamical mean-field theory (DMFT) and its extensions~\cite{Georges:1992,Jarrell:1992,Georges:1996,maier_quantum_2005,KotliarRMP:2006,LTP:2006} have found that when interactions and kinetic energy become comparable, there is a MT, namely, a first-order transition with hysteresis between a metallic and an insulting state~\cite{Metzner:1989,Georges:1992,Jarrell:1992,Georges:1996,maier_quantum_2005,KotliarRMP:2006,LTP:2006,kyung:2006,park:2008,Balzer:2009,Dang:2015,WalshSordiEntanglement:2019}. This was observed experimentally in materials~\cite{mcwhan_metal-insulator_1973,granados1993,Lefebvre:2000,dumm:2009, kanoda_mott_2011,Pustogow_Bories_2018}. 

Other emergent phenomena are associated with this first-order transition. Increasing the temperature $T$, the first-order transition~\cite{mcwhan_metal-insulator_1973, Metzner:1989, Georges:1992, Jarrell:1992, granados1993, Georges:1996, Lefebvre:2000, maier_quantum_2005, KotliarRMP:2006, LTP:2006, kyung:2006, park:2008, Balzer:2009, dumm:2009, kanoda_mott_2011, Dang:2015, Pustogow_Bories_2018, WalshSordiEntanglement:2019} ends at a critical point~\cite{Lefebvre:2000, park:2008, Sordi:2011, Semon:2012, vucicevic:2013, Dang:2015, hebert_superconducting_2015,WalshSordiEntanglement:2019}, followed by a Widom line~\cite{Lefebvre:2000, park:2008, Sordi:2011, WalshSordiEntanglement:2019} in close analogy with phenomena in first-order liquid-gas phase transitions~\cite{XuStanleyWidom:2005,McMillan_Stanley_2010}.
The Widom line acts as a high-temperature indicator of the underlying presence of a MT, sometimes hidden by long-range-ordered states~\cite{fratino2016organizing,Fratino_Semon_Charlebois_Sordi_Tremblay_2017}.

The pseudogap (PG) is another important phenomenon associated with the Mott transition. Pseudogaps have been found in the $U$-$T$ phase diagram on the square lattice. Limiting ourselves to half filling, for values of $U$ smaller than what is needed for a MT, the PG is a precursor of the long-range-ordered antiferromagnetic (AFM) state~\cite{Vilk:1995,Vilk:1997,Huscroft:2001,rost:2012, merinoPseudogapSingletFormation2014a,Ye_Chubukov_2019}. It appears when the AFM correlation length exceeds the thermal de Broglie wavelength, the so-called Vilk criterion~\cite{Vilk:1995,Vilk:1997}. 
If a PG also appears at half filling when interactions are stronger than those needed for the MT, one should find out whether long AFM correlation lengths are also needed. In any case, in analogy with the finite-doping case, it would be useful to differentiate between the PG mechanisms for weak and strong interactions~\cite{Hankevych:2006,Senechal:2004}. 

In this paper, we study the normal-state phase diagram of the Hubbard model on the triangular lattice at half filling. This model is appropriate for several experimental systems including organic conductors with $\kappa$-ET structure~\cite{powell_spin_2009, PowellMcKenzieReview:2011, kanoda_mott_2011,Maegawa_Itou_Oyamada_Kato_2011,Lefebvre:2000,Limelette:2003,Shimizu:2003,Kurosaki:2005,Itou_Oyamada_Maegawa_Tamura_Kato_2007,Kandpal:2009,Maegawa_Itou_Oyamada_Kato_2011,Isono:2014}, some transition metal oxides~\cite{Lee_LiNbO:2007,Soma_Yoshimatsu_Ohtomo_2020,Rawl_BaCoNbO:2017} or sulfides~\cite{Guratinder_FeGaS}, cobaltates~\cite{Cobaltates:2004}
and artificial platforms like one-third monolayer of Sn atoms on a Si(111) surface~\cite{Ming_Johnston_Sn_Si:2017}, optical lattices~\cite{Yang_Liu_Mongkolkiattichai_Schauss_2021} and moiré materials like twisted-bilayer graphene and transition-metal dichalcogenides~\cite{cao_unconventional_2018, cao_correlated_2018, yankowitz_tuning_2019, wu2018hubbard, wu2019topological,Zang_Wang_Cano_Georges_Millis_2022}.

From a theoretical perspective, the triangular lattice offers the possibility to study the pristine finite-temperature normal-state phase diagram, including the Widom line, 
the PG and the MT, unencumbered by phase transitions to long-range-ordered states. 
In particular, close to the value of $U$ where one finds the MT, a number of studies~\cite{morita_nonmagnetic_2002, kyung:2006, sahebsara_hubbard_2008, laubachPhaseDiagramHubbard2015a, Misumi_Mott_triangular:2017, Tocchio_backflow_Mott:2008, Yoshioka_triangular:2009, yang_effective_2010, szasz_chiral_2020, chenQuantumSpinLiquid2022}, including a recent authoritative one~\cite{wietek_mott_2021}, find that frustration is sufficiently strong to lead to a nonmagnetic ground state, barring disagreement of variational calculations~\cite{Tocchio_Montorsi_Becca_2020} and ladder dual fermion approximation at finite temperature\cite{Yu_Li_Iskakov_Gull_2023}. While superconductivity may arise~\cite{kyung:2006,hebert_superconducting_2015}, it occurs at very low temperature. 
The absence of ordered states even at very low temperature also offers us the possibility to find out whether there is a PG at half filling and, if there is one, to verify whether the AFM condition is crucial for PG formation in the large $U$ limit.




We apply the dynamical cluster approximation (DCA)~\cite{hettler_dynamical_2000,maier_quantum_2005} to the Hubbard model on a triangular lattice at half filling. The DCA is a cluster generalization of DMFT, detailed in Sec.~\ref{sec: metho}. 
In Sec.~\ref{sec: phases}, we unveil the $U$-$T$ phase diagram close to the MT, shown in Fig.~\ref{fig: nc6_half_filled_phase_diag}. We prove at high temperature the existence of three phases: a correlated Fermi liquid (cFL), a MI, and another one characterized by a momentum-dependent loss of spectral weight, a strong indication of a PG phase.
In Sec.~\ref{sec: diagram_features}, we present the first-order MT and, stemming from its critical point, a Widom line that extends at higher temperatures. We show how PG, cFL, and MI phases compete around this critical point. 
We also discuss the validity of our results with respect to the cluster choice, arguing that the results for the Widom line are valid in the thermodynamic limit.
Since the Widom line emerges from the critical point that ends the first-order MT, we argue that this strongly suggests the observability of the first-order MT at finite temperature and in the thermodynamic limit, a result that has been recently questioned~\cite{wietek_mott_2021}.


\begin{figure}[b]
    \centering
    \includegraphics[width=\linewidth]{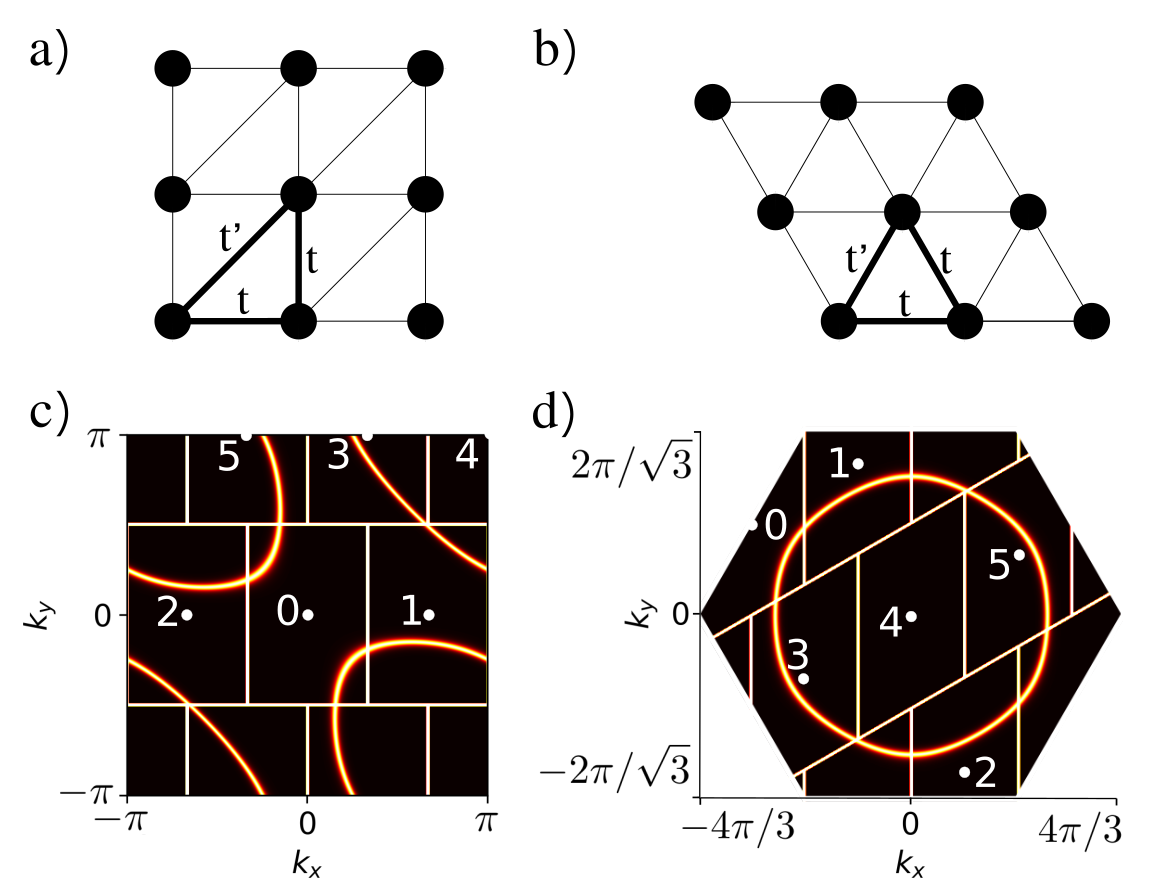}
    \caption{(Color online.) (a) Part of the triangular lattice in the computational basis. Note that only half of the bonds usually labeled $t'$ on the square lattice are considered, making this lattice topologically triangular. When $t'=|t|$, it is also physically triangular for all observables that do not require real-space representations. 
    (b) Part of the real-space triangular lattice. These two lattices can easily be mapped from one to the other by applying an affine transformation.
    (c) Fermi surface of the triangular lattice in the computational basis for the noninteracting case at half filling with $t=1$ and $t'=-1$. This figure shows how the Fermi surface is distributed among the patches of the $N_c=6$ cluster. The white dots with an index $i$ in the middle of the patches are the allowed wave-vectors $\textbf{K}_i$ of the $N_c=6$ cluster.
    (d) Fermi surface and patches transformed to the real-space basis vectors in (b) for $t=-1$ and $t'=-1$. Note that patch 4 is at the center of a hole Fermi surface. }
    \label{fig: lattice}
\end{figure}

\section{\label{sec: metho}Methodology}

We first present the model, then the DCA solution. 

\subsection{The model.}
One of the simplest and most studied models that incorporates both kinetic interactions on a lattice and strong local electronic correlations is the Hubbard model~\cite{Arovas_Berg_Kivelson_Raghu_2022,qinHubbardModelComputational2022a,LeBlanc_2015,Schaefer_Wentzell_2021}, defined in second quantization as
\begin{align}
  \label{eq:H}
    H= & 
      -\sum_{i, j,\sigma} t_{ij} \hat{c}^\dagger_{i\sigma} \hat{c}_{j\sigma} 
      + U \sum_i \hat{n}_{i\uparrow} \hat{n}_{i\downarrow} 
      - \mu \sum_{i\sigma} \hat{n}_{i\sigma},
\end{align} 
where $\hat{c}^\dagger_{i\sigma} $ and $\hat{c}_{i\sigma}$ are the operators that respectively create and annihilate an electron on site $i$ with spin $\sigma$, $\hat{n}_{i\sigma}= \hat{c}^\dagger_{i\sigma} \hat{c}_{i\sigma}$ is the corresponding number operator, $U$ is the energy cost of double occupancy, $\mu$ is the chemical potential, and $t_{ij}$ is the kinetic energy of an electron hopping from site $j$ to $i$.

We allow hopping $t$ ($t'$) between $i$ and $j$ as nearest (next-nearest) neighbors only, as depicted in Fig.~\ref{fig: lattice}(a), which presents the computational basis that we use for a triangular lattice ($|t'|=|t|$).
Every vertex represents a site that is connected to six neighbors by bonds that represent hopping.
When $|t'|$ is equal to $|t|$, then the topology is exactly that of a triangular lattice shown in Fig.~\ref{fig: lattice}(b). 
On a square lattice, there would be another $t'$ hopping in the other diagonal direction. 

Transformation between the computational basis vectors and the real-space basis vectors in Fig.~\ref{fig: lattice}(b) can be done with the combination of a linear transformation and a scale change, a special case of an affine transformation. 
The set of basis vectors ${\bf a}_j$ and corresponding reciprocal lattice vectors ${\bf G}_i$ in either the computational basis or the real-space basis satisfy ${\bf G}_i\cdot {\bf a}_j=2\pi\delta_{ij}$. 
It is that property and the fact that the location of every site in the computational basis is obtained by exactly the same linear combination of basis vectors as in the real-space basis that make the computational basis and the real-space basis generate identical dispersion relations. 
It is only at the end of the calculation, when there is a need to make predictions for scattering probes with actual wave vectors, that we need to transform back to the real-space basis, as in going from Fig.~\ref{fig: lattice}(c) to Fig.~\ref{fig: lattice}(d). 
Phase diagrams, that we focus on, are basis independent.

We work in units where $t$, $\hbar$, $k_B$, and lattice spacing are unity. Thus $U$, $t'$, $\mu$, and temperature $T$ are given in units of $t$.

\subsection{Solving the model.}

Although there is no known exact solution to the two-dimensional Hubbard model, DMFT captures the local quantum fluctuations generated by the large on-site Coulomb repulsion, which allows to solve it exactly in infinite dimensions~\cite{Georges:1992, Georges:1996, Jarrell:1992}.
In DMFT, the interacting atom of a lattice is mapped to an Anderson impurity model (AIM) embedded in an infinite bath of noninteracting electrons that describes the surrounding lattice.
The dynamical hopping from the bath to the impurity and vice versa is described by a hybridization function.
The parameters of both the bath and the hybridization function are self-consistently determined by requiring that the self-energy of the impurity coincide identically with that of the infinite lattice.
However, since DMFT is exact only in infinite dimensions, it does not capture the spatial fluctuations that are quite important in the two-dimensional case we want to study.
To correct this discrepancy, we use one of DMFT's cluster generalizations, DCA~\cite{hettler_dynamical_2000, maier_quantum_2005,Pavarini_Koch_Coleman:2015}.

In DCA, the Brillouin zone is divided into disconnected patches whose center is positioned at the coarse-grained wave vectors $\textbf{K}_i$. 
The wave vectors $\textbf{K}_i$ are the reciprocal lattice vectors of a small cluster with periodic boundary conditions. The wave vectors $\Tilde{\textbf{k}}$ label every wave vector within a patch.
We define $N$ as the number of wave vectors $\textbf{k}=\textbf{K}+\Tilde{\textbf{k}}$ in the Brillouin zone and $N_c$ as the number of coarse-grained patches. With these definitions, the noninteracting dispersion relation in the cluster is defined using the average of the dispersion relation $\epsilon_{\textbf{k}}$ over each patch. The coarse-grained Green's function in each patch is
\begin{align}
    \bar G(\textbf{K}, i\omega_n) &=\frac{N_c}{N}\sum_{\Tilde{\textbf{k}}} G(\textbf{K}+\Tilde{\textbf{k}},i\omega_n).
\end{align}
The cluster self-energy $\Sigma_c(\textbf{K}, i\omega_n)$ for each patch is calculated as a functional of $G_0^{-1}(\textbf{K}, i\omega_n)$ defined by
\begin{align}
    G_0^{-1}(\textbf{K}, i\omega_n) \equiv \bar G^{-1}(\textbf{K}, i\omega_n) + \Sigma_c(\textbf{K}, i\omega_n)
\end{align}
that contains, here implicitly, the hybridization function. Self-consistency is achieved by modifying the hybridization function until the equality 
\begin{align}
    \bar G(\textbf{K}, i\omega_n) =\frac{N_c}{N}\sum_{\Tilde{\textbf{k}}} \frac{1}{i\omega_n-\epsilon_{\textbf{K}+\Tilde{\textbf{k}}}-\Sigma_c(\textbf{K},i\omega_n)}.
\end{align} 
is satisfied for each patch. The solution to the cluster in a bath was obtained with continuous-time auxiliary-field quantum Monte Carlo~\cite{Gull_continuous_2008, Gull:2011}.


%

DCA is exact in the limit of an infinitely large cluster. 
When there is no sign problem, quantum Monte Carlo solves the cluster in a bath problem in polynomial time. Because of the sign problem, the computational resources required scale exponentially as system size increases and temperature decreases~\cite{troyer2005computational}.
Moreover, construction of the cluster is arbitrary and can be optimized to converge faster when there are few sites~\cite{GullFerrero:2010,SakaiSize:2012}.
A study of cluster size and geometry is presented in Sec.~\ref{sec: diagram_features-critical_values}.
In Appendix~\ref{sec: biparticity}, we discuss how biparticity can affect the results.
The results in the following section are obtained using the cluster labeled as $N_c=6$. They are accurate at high temperature according to a cluster-size-dependent study (see Fig.~\ref{fig: cluster_conv}) and are a good compromise in the temperature range of the MT. Details on the estimation of errors can be found in Appendix~\ref{sec: data_compile_algo}.





\section{\label{sec: phases}Phase diagram}

\begin{figure}
    \centering
    \includegraphics[width=\linewidth]{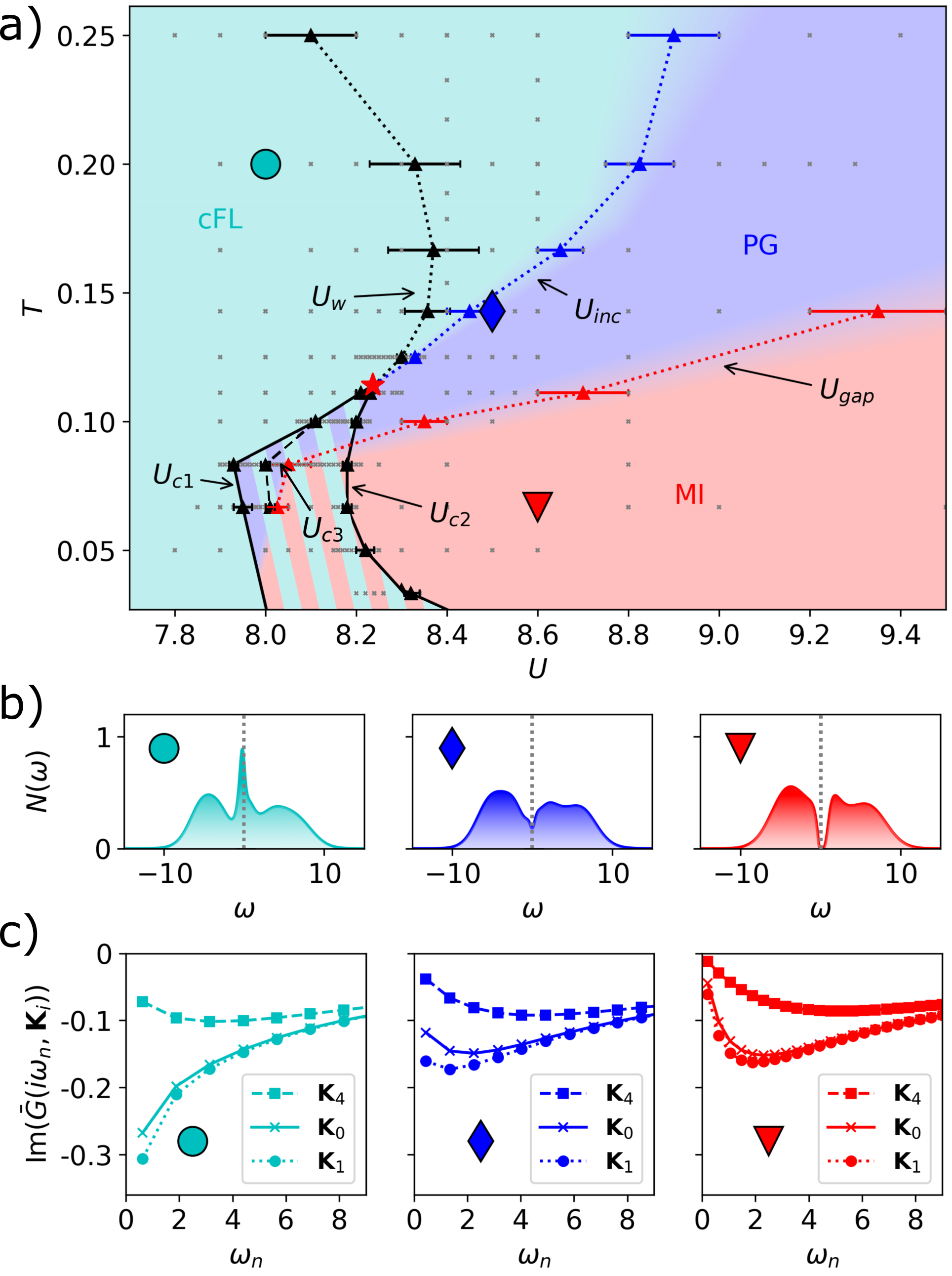}
    \caption{ (Color online) 
    (a)~$U$-$T$ phase diagram for the $N_c=6$ cluster size. The cFL, PG and MI phases are assigned as described in Sec.~\ref{sec: phases}. These regions are displayed in cyan, blue and red respectively, hatched when coexisting. The faint grey $x$'s indicate all parameters for which calculations were performed.
    The blurring between the phases indicates the crossover nature of the transitions. Further details on the transitions can be found in Appendix~\ref{sec: Ugap and Uinc}. Each phase has a representative point selected as a circle, diamond and triangle, respectively.
    The critical values $U_{c1}$, $U_{c2}$, $U_{c3}$, the crossover line $U_W$, and the critical point ($U_c$, $T_c$) are discussed in Sec.~\ref{sec: diagram_features}. The critical point ($U_c$, $T_c$), identified by a red star, is found using a linear extrapolation of $U_{c1}$ and $U_{c2}$. The blue dotted line $U_{inc}$ represents the disappearance of the quasiparticle peak. The red dotted line $U_{gap}$ represents the approximate line where the gap fully opens.
    Error bars depend on the step between two $U$s and on Monte Carlo statistical noise.
    (b)~For each representative point of (a), the DOS $N(\omega)$ calculated using maximum entropy~\cite{MaxEntBergeron} is shown. They demonstrate the disappearance of the quasiparticle peak at the Fermi level, indicated by a dotted line.
    (c)~For each representative point of (a), the imaginary part of the Green's function in Matsubara frequencies is plotted for three different $\textbf{K}$ points. The plain line with crosses, the dotted line with circles, and the dashed line with squares correspond to $\textbf{K}_0=(0,0)$, $\textbf{K}_1=(2/3\pi,0)$, and $\textbf{K}_4=(\pi,\pi)$, respectively. Every calculation of this figure was performed with $t=1$ and $t'=-1$. }
    \label{fig: nc6_half_filled_phase_diag}
\end{figure}

The main results of this work are summarized by the phase diagram in Fig.~\ref{fig: nc6_half_filled_phase_diag}(a).
We highlight three different regions: the correlated Fermi liquid, cFL, where the conduction charge carriers retain their metallic behavior even in the presence of strong electronic correlations, the MI, where the system becomes gapped at the Fermi level, and finally the PG, an intermediate state where the FS is partially gapped depending on the momentum \textbf{k} of the quasiparticles. Note that we use ``correlated Fermi liquid'' instead of just ``Fermi liquid'' because we have verified only the metallicity of the phase, not all properties pertaining to Fermi liquids, that are also correlated phases.

As discussed in the next sections, various criteria are used to classify each region. They are based on the behavior of the density of states (DOS) along with the $\textbf{k}$-dependence of the spectral function at the Fermi level.
The Widom line $U_W$, the incoherence line $U_{inc}$, the gap opening line $U_{gap}$, and the lines of critical interaction strengths $U_{c1}$, $U_{c2}$ and $U_{c3}$, along with the critical point depicted by a red star in Fig.~\ref{fig: nc6_half_filled_phase_diag}(a), are discussed in Sec.~\ref{sec: diagram_features}.

\subsection{\label{sec: cFL}Correlated Fermi liquid.}

At half filling, the noninteracting system described only by kinetic energy is a trivial Fermi liquid with lattice and FS shown in the computational basis in Fig.~\ref{fig: lattice}.
Once interactions are introduced through the Hubbard interaction term $U$, the bandwidth near the Fermi level narrows and upper and lower Hubbard bands begin to appear, as illustrated in the left-hand panel of Fig.~\ref{fig: nc6_half_filled_phase_diag}(b). The lack of particle-hole symmetry on the triangular lattice is clear.
Here, the electrons near the Fermi level are still itinerant~\cite{dengHowBadMetals2013}.

In order to systematically identify the cFL phase, we assume that the modulations of the one-particle propagator are not strong enough to convert the behavior of conduction electrons from itinerant to localized.
Also, according to Luttinger's theorem, one should expect a quasicircular FS not to be very distorted by weak interactions.
In other words, every patch that contains states at the Fermi level in the noninteracting case $U=0$ should have their metallic-like behavior preserved, even in the presence of interactions.
The DOS at the Fermi level in a given DCA patch \textbf{K}$_i$ is related to the Green's function by
\begin{equation}
    \label{eq: dos_imG}
    A(\omega=0, \mathbf{K}_i) = -2\text{Im} \bar{G}(\omega \rightarrow 0, \mathbf{K}_i).
\end{equation}

Although we are working in Matsubara frequencies $\omega_n$, we define a cFL through the presence of a quasiparticle peak close to $\omega=0$. This signature of the Fermi liquid is not always sharp. Nevertheless, a clear change in the shape of the spectral weight at $\omega=0$ indicates the end of the cFL phase, as captured in Fig.~\ref{fig: nc6_half_filled_phase_diag}(a) by the incoherence line $U_{inc}$. This line is defined by the absence of a maximum of the the spectral weight at $\omega=0$. Appendix~\ref{sec: Ugap and Uinc} gives more details on how we define $U_{inc}$.
This condition gives results similar to the criterion $\text{Im}(\bar G(i\omega_0,\textbf{K}_i)) < \text{Im}(\bar G(i\omega_1,\textbf{K}_i))$ that, when it is violated, also identifies a clear non-Fermi liquid behavior. 
The latter condition works with any patch apart from $\textbf{K}_4$ since that patch does not have significant FS weight at $U=0$ [see Figs.~\ref{fig: lattice}(c) and \ref{fig: lattice}(d)] and is thus not expected to be metallic at finite $U$.
An example of this behavior in the cFL phase is again shown in the left-hand panel of Fig.~\ref{fig: nc6_half_filled_phase_diag}(c).
As expected, this phase is found in the high-temperature and low-interaction-strength regions of the phase diagram, where kinetic energy is dominant over potential energy.
Lowering the temperature, scattering decreases, resulting in an increase of the spectral weight in the quasiparticle peak close to the Fermi level already present at high temperature [left-hand panel of Fig.~\ref{fig: nc6_half_filled_phase_diag}(b)]. This can be seen in Figs~\ref{fig: nc6_K_dependant_loses_and_hotspots}(a) and \ref{fig: nc6_K_dependant_loses_and_hotspots}(c), which show the spectral weight at the Fermi level in the metallic state for $U=8.0$ as a function of temperature. Recall that in a Fermi liquid, the lifetime becomes infinite at $\omega=0$ and $T=0$.

\subsection{\label{sec: MI}Mott insulator.}
A key characteristic of a MI state is that it is completely gapped, leaving no spectral weight at $\omega=0$.
In Matsubara frequencies, we do not directly have access to this quantity. One could extrapolate the $\omega=0$ result from the first few Matsubara frequencies. This method can be useful to grasp the general behavior at the Fermi level, but as shown of Fig.~\ref{fig: nc6_K_dependant_loses_and_hotspots}(a), it is surely not reliable since this method gives spectral weight at the Fermi level in the MI. For this reason, to find the MI, we instead perform analytic continuation with the maximum entropy method~\cite{MaxEntBergeron, Jarrell:1996} and directly verify if there is a gap. Appendix~\ref{sec: Ugap and Uinc} gives more details on how we define $U_{gap}$.
An example of the DOS of a MI is given for the representative point with an inverted red triangle in Fig.~\ref{fig: nc6_half_filled_phase_diag}(a) and in the right-hand panel of Fig.~\ref{fig: nc6_half_filled_phase_diag}(b). Those results highly suggest the presence of a MI.

Additionally, the loss of spectral weight characteristic of the MI should be visible for every wave vector and thus every DCA patch.
This is observed for the representative point of the MI region of the phase diagram on the right-hand panel of Fig.~\ref{fig: nc6_half_filled_phase_diag}(c) and at the lowest temperature in Fig.~\ref{fig: nc6_K_dependant_loses_and_hotspots}(d).
As expected during the MT, when we lower the temperature with $U>U_c$, we lose spectral weight on every patch until it becomes negligible.
The shape of the MT and the Clausius-Clapeyron equation have implications for the understanding of the MI~\cite{Sordi:2011, WalshSordiEntanglement:2019}. This is discussed further in Sec.~\ref{sec: first_order_trans}.


\subsection{\label{sec: PG}Pseudogap}
We defined the cFL state in Sec.~\ref{sec: cFL} as having a quasiparticle peak at the Fermi level and metallic-like behavior for each DCA patch, while the MI phase in Sec.~\ref{sec: MI} was defined as fully gapped at the Fermi level, with each DCA patch having insulating character. 
This leaves us with an intermediate regime, that we will also call a phase, in Fig.~\ref{fig: nc6_half_filled_phase_diag}(a) that we identify as a PG. 

The middle panel of Fig.~\ref{fig: nc6_half_filled_phase_diag}(b) shows the DOS for a point in parameter space, marked by a diamond in Fig.~\ref{fig: nc6_half_filled_phase_diag}(a), with PG features.
There is no gap, but a dip indicating the absence of quasiparticles, a clear sign of a non-Fermi liquid behavior.

\begin{figure}[t]
    \centering
    \includegraphics[width=\linewidth]{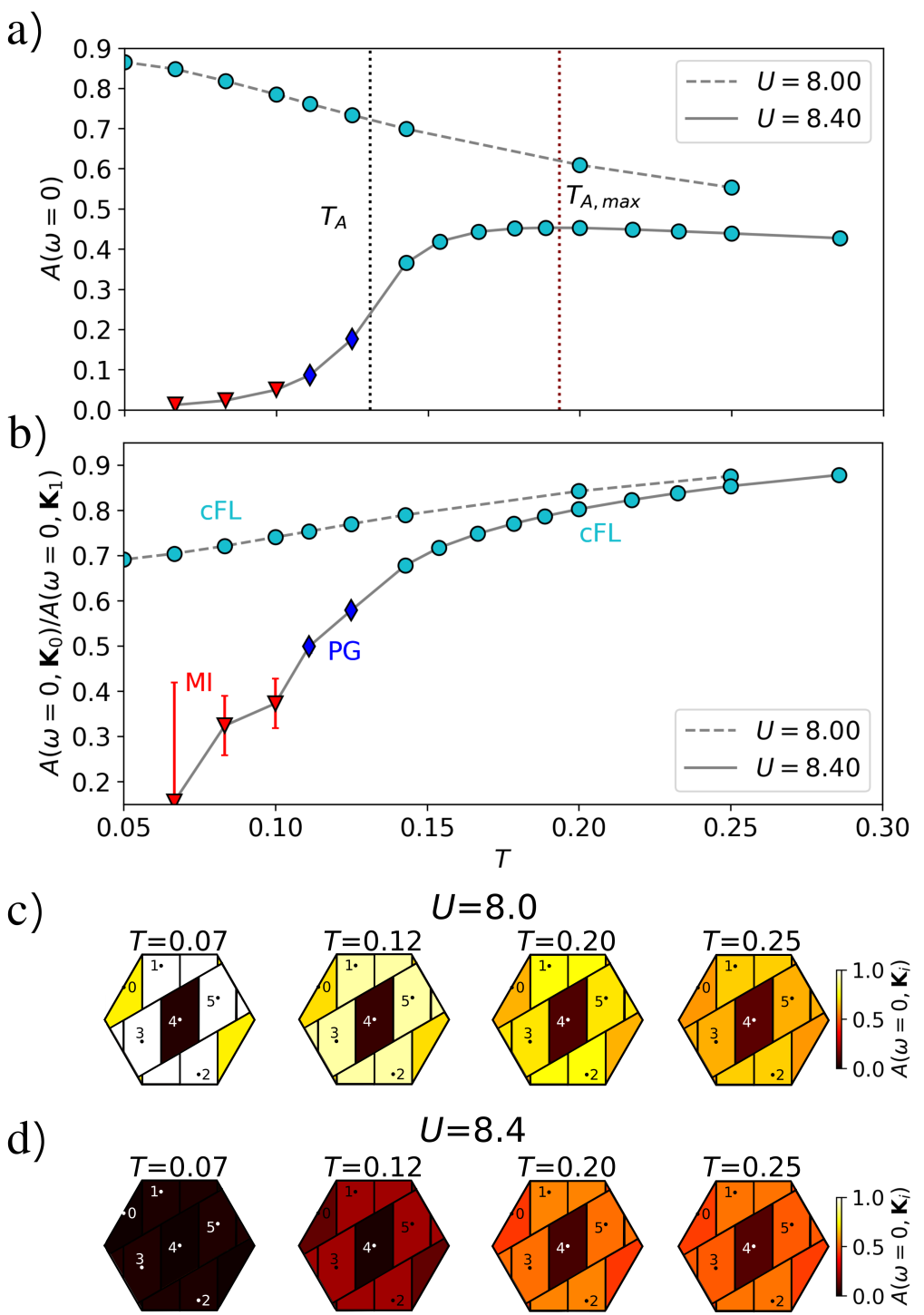}
    \caption{ (Color online) 
    (a) Spectral weight at the Fermi level at $U=8.0<U_c$ and $U=8.4>U_c$ obtained by fitting a second-degree polynomial to the mean of the Green's function in Matsubara frequencies and extrapolating to $\omega=0$. The crossover of the spectral weight $T_A$ and its associated maximum $T_{A,max}$ are indicated. The uncertainties, smaller than the points, are the ones of the first Matsubara frequency that are calculated using the technique described in Appendix~\ref{sec: data_compile_algo}.
    (b) Evolution in temperature of the ratio between the spectral weight at $\omega=0$ of the $\textbf{K}_0$ and the $\textbf{K}_1$ patches, for two different values of $U$. The phases are defined using the same criteria as in Fig.~\ref{fig: nc6_half_filled_phase_diag}(a). Uncertainties are calculated using the error on the first Matsubara frequency and Ref.~\cite{lebigotUncertaintiesPythonPackagea}. Momentum-dependent spectral weight extrapolated to $\omega=0$ for (c) $U=8.0$ and (d) $U=8.4$ using a second-degree polynomial fit on the first three Matsubara frequencies.}
    \label{fig: nc6_K_dependant_loses_and_hotspots}
\end{figure}

According to this criterion, this phase, intermediate between the cFL and the MI, could simply be classified as a bad metal as defined in Ref.~\cite{dengHowBadMetals2013}.
In fact, two different phases are plausible: either a very small uniform gap or a PG.
In order to differentiate the two, we inspect if characteristics of a pseudogap are present: 
\begin{enumerate}
    \item There is a crossover at a temperature $T_A$ where the quasiparticle peak in the spectral weight disappears.
    \item There is a temperature $T_{A,max}$ where the spectral weight is maximal.
    \item The loss of spectral weight is momentum-dependent.
\end{enumerate}

There are many other valid criteria to verify that the state is a pseudogap, for example measuring a maximum in the susceptibility, $\chi_{max}$. However, such calculations are much harder and the program we are using does not measure two-particle observables. Nevertheless, $T_{A,max}$ is very correlated to $\chi_{max}$ as was shown in Ref.~\citenum{SordiResistivity:2013}. 
In Fig.~\ref{fig: nc6_K_dependant_loses_and_hotspots}(a), we indicated both $T_A$ and $T_{A,max}$. Although it argues favorably for the presence of a PG, it is not a definitive proof.

Unfortunately, we are numerically restricted by the coarse grid and our current choice of patches is not strategically selected to emphasize possible magnetic hot spots.
Furthermore, as seen in the middle panel of Fig.~\ref{fig: nc6_half_filled_phase_diag}(c), both the $\textbf{K}_0$ and the \textbf{K}$_1$ patches give a clear indication of loss in spectral weight at $\omega=0$ characteristic of an insulating like behavior. 
Also, every patch loses spectral weight when going from the cFL to the MI, as illustrated in Fig.~\ref{fig: nc6_K_dependant_loses_and_hotspots}(d). This indicates that the loss of spectral weight is either generalized on all the FSs or that the PG hot spots are distributed on many patches, a phenomenon that would be caused by our coarse grid. 

To find another signature that a PG is present, we study the distribution of hot spots over patches of the FS by computing, as a function of temperature, the ratio of the spectral weights at the Fermi level on different patches.
More specifically, we plot in Fig.~\ref{fig: nc6_K_dependant_loses_and_hotspots}(b) the spectral weight ratio between the \textbf{K}$_0$ and \textbf{K}$_1$ patches for two values of interaction: $U=8.0$ in the cFL and $U=8.4$ that undergoes a MT.
In the $U=8.0$ case, which is clearly in the cFL regime as supported by Figs.~\ref{fig: nc6_K_dependant_loses_and_hotspots}(a) and (c), the ratio is quite stable and evolves smoothly with temperature.
In the second case where $U=8.4$, two different regimes are identified.
At high temperatures, the ratio evolves similarly to the $U=8.0$ case, interpreted as the cFL regime. 
At lower temperature, although Fig.~\ref{fig: nc6_K_dependant_loses_and_hotspots}(d) suggests that the spectral weight is lost uniformly, the ratio between the patches change dramatically, indicating that the loss of spectral weight is momentum-dependent when changing temperature.
Momentum-dependent decrease of the spectral weight at the FS has also been seen with cluster dynamical mean-field theory (CDMFT) on a $2\times2$ plaquette~\cite{Parcollet:2004} and with 12-site cluster perturbation theory~\cite{Kang_Yu_Xiang_Li_2011}.
One should note that in Figs.~\ref{fig: nc6_K_dependant_loses_and_hotspots}(a)(b) the low temperature results in the MI depending sensitively on polynomial extrapolation of the first Matsubara frequencies and hence they are not meaningful. 

Additionally, one might have noticed that the patch losing spectral weight the fastest is $\textbf{K}_0$, the only one with significant spectral weight that is not crossed by the AFM $(\pi,\pi)$ Brillouin zone. This is relevant because for weak interaction strength, long-wavelength AFM fluctuations that are precursor to long-range AFM order lead to a PG at the crossing with the AFM Brillouin zone~\cite{VilkShadow:1997,Kyung:2004,Motoyama:2007,merinoPseudogapSingletFormation2014a}. For sure, the many different orders possible at half filling on the triangular lattice~\cite{laubachPhaseDiagramHubbard2015a,kyung:2006,Sahebsara:2006,yang_effective_2010,laubachPhaseDiagramHubbard2015a} could lead to an analogous phenomenon, but it is unlikely. Although one could argue that the loss of spectral weight is correlated to spiral order, in the same way as discussed in Appendix~\ref{sec: Nc=12} for the 12 site cluster, here we do not want to speculate further on the origin of the PG since this is not the main goal of this paper.

\section{\label{sec: diagram_features}Widom line, phase transitions and cluster size study}


In addition to the different phases identified in Fig.~\ref{fig: nc6_half_filled_phase_diag}, the phase diagram contains a region bounded by lines labeled by $U_{c1}$ and $U_{c2}$.
These spinodal lines correspond to values where observables such as the double occupancy, or equivalently the potential energy, show a hysteretic behavior between the cFL and MI phases, with a discontinuity characteristic of first-order transitions. For $U$ smaller than $U_{c1}$ the MI is unstable whereas for $U$ larger than $U_{c2}$ the cFL is unstable. 
Going up in temperature, $U_{c1}$ and $U_{c2}$ evolve and eventually connect at a critical point $(U_c, T_c)$ in parameter space, depicted by a red star on the phase diagram.
Although there is no longer a first-order transition above this temperature, the discontinuities that were present at the first-order transition along the spinodal lines now appear as an inflection point, forming a so-called Widom line~\cite{XuStanleyWidom:2005,Sordi:2012}.


In Sec.~\ref{sec: diagram_features-critical_values}, we formally define the spinodal lines, the critical point, and the Widom line and discuss their appearance in the phase diagram of Sec.~\ref{sec: phases} obtained using the $N_c=6$ cluster.
We argue that the Widom line is a strong indicator of an underlying MT.
In Sec.~\ref{sec: diagram_cluster}, we provide a convergence study on the size and shape of the cluster, summarized in Fig.~\ref{fig: cluster_conv}. It demonstrates unequivocally the existence of the Widom line in the thermodynamic limit, which in turn provides very strong evidence for the MT on the triangular lattice. This has been observed experimentally, for example, in layered organic conductors~\cite{Lefebvre:2000,Shimizu:2003,Kurosaki:2005, Pustogow_Bories_2018}.

\subsection{\label{sec: diagram_features-critical_values}Phase transitions and crossovers.}

The MT is in the universality class of the liquid-gas phase transition~\cite{KotliarLange:2000, imadaQuantumMottTransition2005}.
Both possess a Widom line and a first-order transition that ends at a critical point~\cite{park:2008, WalshSordiEntanglement:2019, Sordi:2011, Semon:2012, Dang:2015, vucicevic:2013}.
In this section, we first discuss the first-order transition and then define and discuss the Widom line.

\begin{figure}[b]
    \centering
    \includegraphics[width=\linewidth]{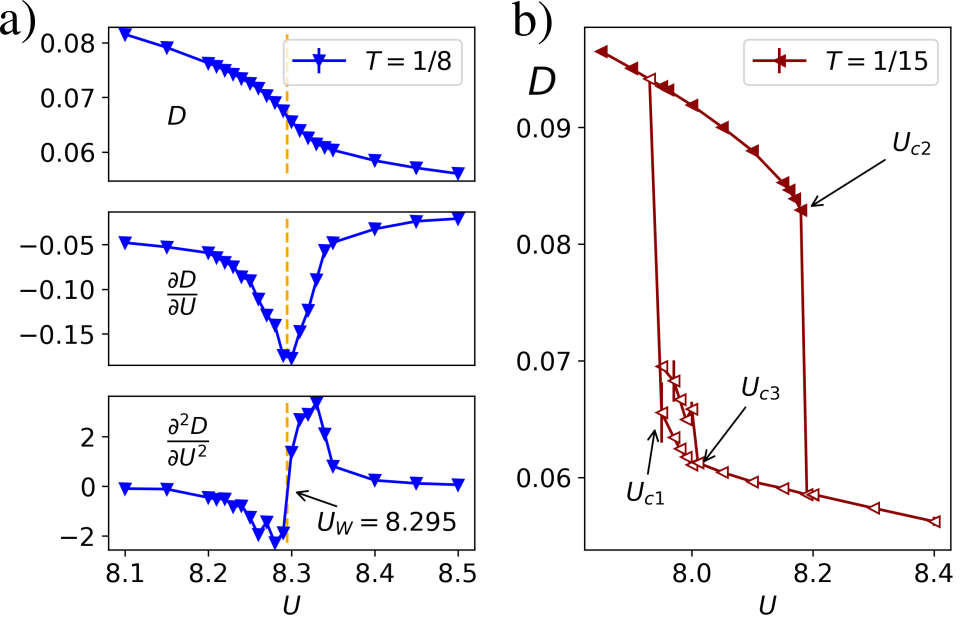}
    \caption{ (Color online) Widom line and first-order transition from the double occupancy point of view for the $N_c=6$ cluster. At a given temperature, we calculate the double occupancy and its first and second derivatives with respect to interaction strength, allowing identification of the Widom line $U_W$ defined by the inflection criterion [Eq.~(\ref{eq:widom})] for a second-order transition. The critical values $U_{c1}$ and $U_{c2}$ of the interaction strength, in the right-hand panel, are defined by the points where the double occupancy is discontinuous, signaling a first-order metal-insulator transition. The values of $U_{c1}$ and $U_{c2}$ as a function of temperature define the spinodal lines. $U_{c3}$ defines the bistable limit in the MI, discussed further in Appendix~\ref{sec: bistability}. The uncertainties for all calculated points are estimated using the technique described in Appendix~\ref{sec: data_compile_algo}. }
    \label{fig: docc_U}
\end{figure}
\subsubsection{\label{sec: first_order_trans}First-order transition.}


We adopt the usual procedure to monitor double occupancy $D$ to find the first-order MT. 
Figure~\ref{fig: docc_U}(a) shows that at high temperature, double occupancy has an inflection point that turns, at low enough temperature, into a discontinuity with hysteresis, characteristic of a first-order transition.
This situation is depicted in Fig.~\ref{fig: docc_U}(b) for $T=1/15$.
The lower and upper boundaries in interaction strength of the resulting hysteresis are the spinodal lines labeled $U_{c1}$ and $U_{c2}$ in Fig.~\ref{fig: nc6_half_filled_phase_diag}(a).
As temperature decreases, $U_{c1}$ exhibits a sudden change in direction.
There we find a small region of bistability between $U_{c1}$ and $U_{c3}$, namely two solutions with different double occupancy but similar DOS.
This small region is further discussed in Appendix~\ref{sec: bistability}.
To stay in the insulating like phase in this bistable region is extremely computationally demanding, which could explain why the behavior of $U_{c1}$ changes abruptly around $T\approx0.083$. If a new physical phenomenon is involved, we were not able to identify it. 
Let us now compare the $U_{c2}$ line in Fig.~\ref{fig: nc6_half_filled_phase_diag}(a) with that reported in previous works.
Near the critical point where the first-order transition ends, the slope of this line is similar to the one observed on square lattices~\cite{park:2008, WalshSordiEntanglement:2019} where it was argued that this behavior reflects the fact that the entropy is smaller in the MI and PG phases compared to the cFL.

Following Ref.~\citenum{Sordi:2011} to explain the slope of the first-order transition in our case, recall that the change in the free-energy density at constant volume extracted from the partition function in the canonical ensemble is given by 
\begin{equation}
    \dd F (T, N, U) = -s \dd T + \mu\dd n + D\dd U
\end{equation}
where $s$ is the entropy per number of sites.
At half filling ($n=1$), along the line of thermodynamic phase transition $U_{pt}$ (found for the square lattice in Ref.~\citenum{WalshSordiEntanglement:2019})  the free energies in the cFL and the MI and PG states are equal, leading to $\dd F_{ins}\lvert_{U_{pt}} = \dd F_{met}\lvert_{U_{pt}}$.
At half filling ($dn=0$), this implies the Clausius-Clapeyron equation
\begin{align}
    \frac{\dd T}{\dd U} \Big \lvert_{U_{pt}} =\frac{D_{ins}-D_{met}}{s_{ins}-s_{met}}.
    \label{Clausius}
\end{align}

\begin{figure*}
    \centering
    \includegraphics[width=\linewidth]{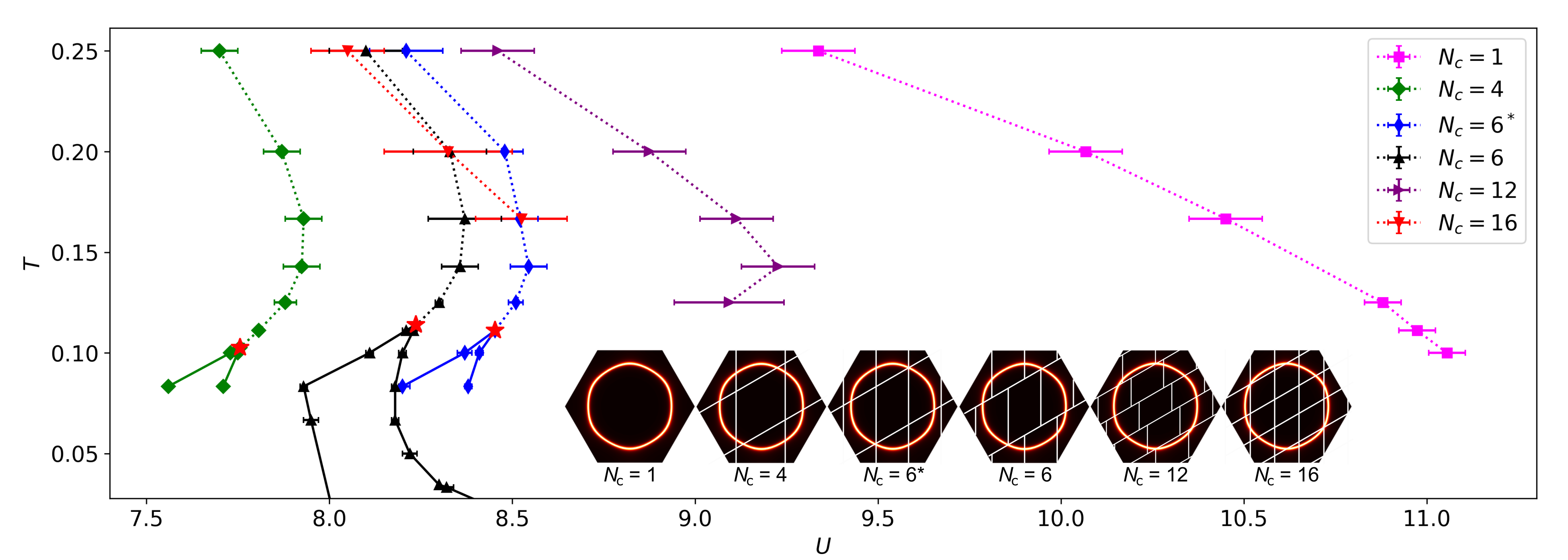}
    \caption{ 
        (Color online) Cluster-size and -shape dependence of the Widom line and of some features of the Mott transition. 
        Six different clusters each corresponding to different divisions of the Brillouin zone are compared and shown in the inset.
        They correspond to five lattice sizes: $N_c=1, 4, 6, 12$ and $16$, with two different definitions for the 6-patch cluster: a nonbipartite one and a bipartite one, denoted by $N_c=6^*$ and $N_c=6$ respectively. What we mean by bipartite is discussed in Appendix~\ref{sec: biparticity}.
        The single site cluster corresponds to the DMFT case.
        The dotted lines correspond to the Widom line $U_W$, the plain lines to the spinodals $U_{c1}$ and $U_{c2}$, and the dashed line to $U_{c3}$.
        We only looked for $U_{c3}$ in the $N_c=6$ cluster.
        The critical points, indicated by the red stars, are found in the same way as for Fig.~\ref{fig: nc6_half_filled_phase_diag}.
        The coordinates of the critical points are $(U_c,\ T_c)_{N_c=4}\sim(7.8,\ 0.10)$, $(U_c,\, T_c)_{N_c=6}\sim(8.2,\, 0.11)$ and $(U_c,\, T_c)_{N_c=6^*}\sim(8.5,\, 0.11)$.
        Error bars depend on the step in $U$ and on Monte Carlo statistical noise. 
    }
    \label{fig: cluster_conv}
\end{figure*}

Focusing on the lowest temperatures far from the critical point, assuming that the slope of $U_{c2}$ follows that of $U_{pt}$ and extracting the inequality $\frac{\dd T_{c2}}{\dd U_{c2}} < 0$ from the $N_c=6$ cluster in black in Fig.~\ref{fig: cluster_conv}, we conclude that since the inequality $D_{ins}<D_{met}$ is satisfied, the thermodynamic entropy satisfies $s_{ins}>s_{met}$ as in single-site DMFT.
Increasing temperature, the slope of the spinodal $U_{c2}$ becomes positive. Assuming again that this is the same trend as $U_{pt}$, either the numerator or the denominator of the Clausius-Clapeyron equation must change sign. 
Physically, we expect that the change in sign of the slope must come from a decrease in entropy at higher temperature, leading to the inequality $s_{ins}<s_{met}$. 
This is consistent with the result of Ref.~\cite{wietek_mott_2021} that increasing temperature increases 120$^\circ$ spin correlations. 
It must manifest as a decrease of entropy of the paramagnetic insulating state. 

There are plenty of different orders that could appear on a triangular lattice at low temperature~\cite{laubachPhaseDiagramHubbard2015a, szaszPhaseDiagramAnisotropic2021, Yu_Li_Iskakov_Gull_2023}, a question we do not address here. But as mentioned in the Introduction, for the range of $U$ where the first-order MT occurs, a number of studies~\cite{morita_nonmagnetic_2002, kyung:2006, sahebsara_hubbard_2008, laubachPhaseDiagramHubbard2015a, Misumi_Mott_triangular:2017, Tocchio_backflow_Mott:2008, Yoshioka_triangular:2009, yang_effective_2010, szasz_chiral_2020, chenQuantumSpinLiquid2022,wietek_mott_2021} find a nonmagnetic ground state.
Our calculations are in the paramagnetic state, so they are not accurate once true long-range order is established. 

\subsubsection{Widom line and critical point.}

At high temperature above the critical point of the liquid-gas transition, thermodynamic quantities do not exhibit singularities. However, there are precursors of the first-order transition. Thermodynamic quantities, such as the specific heat, for example, exhibit an extremum along a line that ends at the critical point. Different thermodynamic quantities yield different lines, but they all become asymptotically identical near the critical point, a consequence of the diverging correlation length~\cite{XuStanleyWidom:2005,McMillan_Stanley_2010,reymbautMottTransition2020a}. Strictly speaking, the Widom line is this asymptotic line close to the critical point. Here we call the Widom line just the line associated with double occupancy, as we explain below. 


Here we focus on the behavior of the double occupancy $D=\langle n_{i\uparrow}n_{i\downarrow}\rangle$, a first derivative of the free-energy with respect to interaction $U$, and on its derivative with respect to $U$, a second-order derivative of the free energy analogous to more usual thermodynamic quantities such as the compressibility~\cite{Sordi:2012,WalshSordiEntanglement:2019}.
The corresponding Widom line is defined by
\begin{align}
    \label{eq:widom}
    U_W(T,n)=\min_U\Big(\frac{\partial D}{\partial U}\Big\lvert_{T,n} \Big).
\end{align}
In other words, $U_W$ is the value of $U$ at fixed $T$ and occupation $n=1$ where the change in double occupancy has a minimal value.
Figure~\ref{fig: docc_U}(a) shows $D$ and its first two derivatives for an example case at $T = 1/8$ in the vicinity of the intersection with the Widom line.
Because the change in curvature at the inflection point decreases with increasing temperature, the Widom line becomes difficult to determine. This line eventually becomes undetectable at high temperature so we did not investigate it for $T>1/4$.
With more precision and computing resources, the Widom line might be found at higher temperatures. 


Using this method, we find the Widom line in the $T$-$U$ phase diagram at half filling for the triangular lattice. It is identified in Fig.~\ref{fig: nc6_half_filled_phase_diag}(a).
We first note that it closely resembles what was found for various crossover lines on the square lattice using cluster DMFT (CDMFT)~\cite{park:2008,Sordi:2012, WalshSordiEntanglement:2019, reymbautMottTransition2020a}.
Indeed, for $T>0.2$ we find a negative slope, a rather constant slope for $0.14<T<0.20$ and a positive one for $0.11<T<0.14$.
In Ref.~\citenum{park:2008}, they argue that at temperatures above $T>0.15$, the increased entropy of lightly correlated spins in the more localized phase favors that phase, leading to a negative slope. 
This is the sign of the slope found in single-site DMFT. Continuing this line of arguments, even though we are outside the first-order transition, we argue that around $0.11<T<0.14$, the formation of singlets decreases the entropy of the more insulating phase compared with the metallic state, which explains the positive slope.
Note that the MI-PG and the PG-cFL crossovers in the phase diagram of Fig.~\ref{fig: nc6_half_filled_phase_diag}(a) all have positive slope. The more positive slope of the PG-cFL crossover line, compared with the MI-PG crossover line, suggests that the singlets in the PG region are more effective at decreasing the entropy compared with the cFL region than the singlets in the MI region are at decreasing the entropy compared with the PG region. In any case, this argument suggests that there are still many singlets in the PG region.


At the Widom line's lowest point in temperature, we find the critical point, namely where the slope $\partial D/\partial U$ becomes minus infinity. At the critical point, the correlations are largest and all extrema in thermodynamic quantities converge. 
The critical point's coordinates obtained in this work are found to be a factor of two larger than the one of Ref.~\onlinecite{WalshSordiEntanglement:2019} on the square lattice with nearest-neighbor hopping only.
Increasing $|t'/t|$ on the anisotropic triangular lattice increases frustration and suppresses the Néel antiferromagnetism~\cite{laubachPhaseDiagramHubbard2015a, Yu_Li_Iskakov_Gull_2023}, concomitant with an increase in both $U_c$~\cite{kyung:2006}
and $T_c$.

It is interesting to note, again in Fig.~\ref{fig: nc6_half_filled_phase_diag}(a), that the critical point lies directly on the boundary between the PG and the cFL states.
 This PG-cFL crossover line is linked to the inflection point of a dynamical quantity, such as that of the total density of states, whose concavity at the Fermi level changes from a maximum to a minimum along a line~\cite{reymbautMottTransition2020a} in the $U$-$T$ plane.

\subsection{\label{sec: diagram_cluster} Mott transition and thermodynamic limit for the Widom line.}




The choice of the cluster labeled $N_c=6$ shown in Fig.~\ref{fig: cluster_conv}(a) for the results presented up to now is a compromise between the necessity to have a large cluster and the numerical resources required to perform the calculations.
This cluster does not support three-sublattice order, is relatively small, although larger than many of the previously used clusters for the MT~\cite{park:2008, Sordi:2010,Sordi:2011,Balzer:2009,Dang:2015,reymbautMottTransition2020a}, and following the definition of Ref.~\citenum{betts:1999}, is ferromagnetically imperfect when $t'=0$.
However, it is bipartite when $t'=0$ (see Appendix~\ref{sec: biparticity}), a concept useful for strongly correlated systems~\cite{Maier_Betts_2005}. When $|t'|$ is set equal to $|t|$, we recover the triangular lattice and the cluster becomes ferromagnetically perfect, as discussed in Appendix~\ref{sec: ferromagnetic_imperfection}.
In this section, we compare the normal-state phase diagram obtained using the $N_c=6$ cluster with other cluster choices, summarized in Fig.~\ref{fig: cluster_conv}.

Since previous Widom lines were only found in small clusters~\cite{park:2008, Sordi:2012, hebert_superconducting_2015, fratino2016organizing,WalshSordiEntanglement:2019}, one could argue that they are an artifact. For all the clusters in Fig.~\ref{fig: cluster_conv}, the Widom line has been found and there is a clear trend suggesting that it survives in the thermodynamic limit. 
Although it is clearly seen for the largest $N_c=16$ cluster at high temperature, we could not look for it below $T=1/6$ in that cluster due to the long computation time and to the uncertainty coming from the sign problem (see Appendix \ref{sec: biparticity} for further comments on the sign problem).
We claim that these results strongly suggest that the double occupancy Widom line exists in the thermodynamic limit~\footnote{Note that the Widom line for $N_c=1$ seems to be different from the ``quantum Widom line'' in  Ref.~\cite{vucicevic:2013,Eisenlohr_Lee_Vojta_2019}.}.

The shape of the Widom line, which is mostly consistent for every cluster except $N_c=1$ (Fig.~\ref{fig: cluster_conv}), is interesting. 
The Widom line starts at the critical point ($T\sim 0.1$) with a slope whose sign is the same as that of $U_{c1}$ just below the critical point.
In such clusters, the formation of singlets in the insulator leads to smaller entropy in the insulator than in the metal~\cite{park:2008}. 
The Clausius-Clapeyron equation~[\eref{Clausius}] in turn imposes a positive slope.  
In all cases, at sufficiently high temperature the slope becomes negative, as in single-site DMFT, or more generally when there is an odd number of sites in the cluster~\cite{Dang:2015,reymbautMottTransition2020a}. 
The odd number of sites leaves a net spin one-half in the insulator, which yields higher entropy in the insulator than in the metal thus yielding a negative Clausius-Clapeyron slope.
Remarkably, the slopes of every Widom line associated with double occupancy do become negative at the highest temperatures, suggesting that short-range superexchange interaction is not important anymore and that we have almost-free spin one-half degrees of freedom giving the same slope as in single-site DMFT. 
We do not understand why $U_W$ for single-site DMFT is so large. 
If the Widom line for $N_c=1$ joins the other ones, it would be at very high temperature. 

Surprisingly, the position of the transition line for the $N_c=12$ cluster is away from the one of the $N_c=16$ and $N_c=6$ clusters. Although no thorough work has been done on the matter, observations from dynamical variational Monte Carlo (DVMC) led to the proposition that the cluster shape plays a role in the onset of the PG~\cite{rosenberg_fermi_2022}. For example, for a same value of $U$, a rectangular cluster acts more like a FL than a square cluster. This could be the reason why the MT happens at larger $U$ for $N_c=6$ compared to $N_c=4$, and $N_c=12$ compared to $N_c=16$. 

Another observation on the consistency between different clusters is that most of the features of the $N_c=6$ cluster can be observed in the larger $N_c=12$ cluster. Further details are given in Appendix~\ref{sec: Nc=12}. In summary, not only does $U_{W}$ exist and have a similar shape for both cluster sizes, the crossover lines $U_{inc}$, $U_{gap}$ and $T_A$ also exist. We also observe a $\textbf{K}$-dependent loss of spectral weight.

The differences between the clusters are a manifestation of the approximate nature of the technique we are using. However, the qualitative description of the various states is consistent across the different choices of clusters.


We did not seek to work on the detailed $U_{c1}$ and $U_{c2}$ for other clusters than $N_c=6$ because previous papers already did~\cite{Dang:2015}. Although patch and cluster definitions can impact final results~\cite{SakaiSize:2012, GullFerrero:2010}, our values of $U_{c1}$ and $U_{c2}$ are consistent with those of Ref.~\cite{Dang:2015}. Neither did we feel that the $N_c=6^*$ nonbipartite cluster would reveal additional important details. Using the ladder dual fermion approximation, Ref.~\cite{Li_Gull_2020} finds that the system is insulating at $U=10$. Additional results for the latter value of $U$ and a discussion of the disappearance of finite-size effects as a function of temperature may be found in Ref.~\cite{Vranic_Vucicevic_2020}.



\section{Discussion}

The MT in Fig.~\ref{fig: cluster_conv} is found for a range of $U$ and $T$ consistent with that found previously in Ref.~\cite{Dang:2015} for clusters of size $N_c=1,3,4,6$ and, for $N_c=6^*$, with a different tiling of the Brillouin zone. On a $2\times2$ cluster, CDMFT finds comparable values of $U$ for the MT~\cite{Parcollet:2004,kyung:2006} while DCA with $N_c=4$ clusters finds the slightly smaller value $U_c=6.7$ at $T=0.05$~\cite{LeeDualFermions:2008,Vranic_Vucicevic_2020}.

At the lowest temperature, $T=0.1$, studied with CDMFT in Ref.~\citenum{wietek_mott_2021}, there is a sharp maximum in the derivative of the kinetic energy for $N_c=7$ around $U=8$-$9$, but no jump indicative of a first-order MT.
We argue that $T=0.1$ is close to where we find the critical point. In addition, clusters with an odd number of sites are rather different from even-numbered clusters, probably because the $S=1/2$ cluster state yields to an increased entropy. Such artificial degeneracy does not exist in even numbered clusters. This is the reason why the MT in $N_c=1$ and $N_c=3$ clusters~\cite{Dang:2015} has a negative slope. 
As we discuss in Appendix~\ref{sec: Tc}, it might also be the reason why Ref.~\citenum{wietek_mott_2021} may not see the MT in the $N_c=7$ cluster at $T=0.1$.
The other methods used in Ref.~\citenum{wietek_mott_2021} did not show a MT either, but the authors leave open the possibility that they cannot be conclusive on this point. 

Results for lattices with slight anisotropy, $|t'/t|=0.8$, give similar ranges of values of $U$ and $T$ for the MT~\cite{ohashi:2008,Semon:2012}. The PG observed in weakly frustrated organics, $|t'/t| \ll 1$, seems to originate from the long-wavelength spin fluctuation mechanism mentioned in the Introduction~\cite{powell_spin_2009}. This is different from triangular lattice organic compounds~\cite{Shimizu_Tokura_Kanoda_Saito_2006, Pustogow_Bories_2018} that are more strongly correlated and where the PG physics may be more closely related to the one discussed here. Our work shows that the PG in the frustrated systems at $U$ is larger than the MT, contrary to previous work on slightly frustrated systems~\cite{merinoPseudogapSingletFormation2014a}. This suggests that the physics of the PG that we found comes from strong correlations instead of long-wavelength fluctuations. 

All clusters for which low-temperature calculations could be done show a Widom line that ends at a critical point, followed by a MT. 
Hence, the Widom line is clearly a precursor to the MT. The inflection point in double occupancy is also seen in Ref.~\cite{wietek_mott_2021} at high temperature.
The fact that the shape of the Widom line and its position in parameter space stay approximately the same with cluster shape and size suggests that the MT also survives in the thermodynamic limit.

For anisotropic triangular lattices, long-range order may hide the MT at the lowest temperatures as seen experimentally in organics~\cite{Lefebvre:2000,Limelette:2003}. Nevertheless, a metal-insulator transition is seen above the long-range-ordered states. Furthermore, in agreement with theoretical results on the triangular lattice for values of $U$ in the range where we see the MT~\cite{morita_nonmagnetic_2002, kyung:2006, sahebsara_hubbard_2008, laubachPhaseDiagramHubbard2015a, Misumi_Mott_triangular:2017, Tocchio_backflow_Mott:2008, Yoshioka_triangular:2009, yang_effective_2010, szasz_chiral_2020, chenQuantumSpinLiquid2022,wietek_mott_2021}, experiments on organics show a MT separating a Fermi liquid and a spin liquid that is not hidden by long-range order~\cite{Shimizu:2003,Pustogow_Bories_2018,Pustogow_2022,Yesil_Imajo_Yamashita_Akutsu_Saito_Pustogow_Kawamoto_Nakazawa_2023}.

\section{Conclusion}
\label{sec:conclusion}

We used DCA for the half-filled Hubbard model on the triangular lattice to study the Mott transition and how it influences the normal-state phase diagram in its vicinity. Our important contributions are as follows.
First, we gave strong arguments for the existence of a Widom line in the thermodynamic limit by comparing results for different cluster sizes and boundary conditions. Second, we argued that the presence of the Widom line in the thermodynamic limit strongly suggests that the Mott transition also exists in that limit and that it would be observable at finite temperature above long-range-ordered states. 
Indeed, the Widom line always emerges at the end of the critical point of the first-order Mott transition. 
Third, we found that the Mott transition not only separates a correlated Fermi liquid from a Mott insulator, as found experimentally, it also separates the Fermi liquid from a region where a pseudogap appears.
Finally, the pseudogap occurs on the large-$U$ side of the Mott transition, instead of beginning on the small-$U$ side, as is the case for the square lattice. 

This paper motivates further research on the observability of the Mott transition at finite temperature in the triangular lattice Hubbard model. 
Having shown the existence of a pseudogap caused by strong interactions on the triangular lattice at half filling, the next step should be to try to better characterize its underlying mechanisms and to look into the doped regime to find out if a similar pseudogap exists. 
Having shown that, on the triangular lattice, the Widom line, the Mott transition and the pseudogap are deeply interconnected emergent phenomena for strongly correlated electrons, experimental studies should now look for the concomitant presence of all four phenomena.


\section*{Acknowlegments}

We thank Claude Bourbonnais, Michel Ferrero, Chloé-Aminata Gauvin-Ndiaye, Antoine Georges, Emanuel Gull, Marcel Klett, Mario Malcoms, Michael Meixner, Thomas Schäfer, David Sénéchal, Giovanni Sordi, Hanna Terletska, Roser Valent\'i, Caitlin Walsh and Philipp Werner for interesting discussions that contributed to the article. This work has been supported by the Natural Sciences and Engineering Research Council of Canada (NSERC) under Grant No. RGPIN-2019-05312, the Canada First Research Excellence Fund and by the Research Chair in the Theory of Quantum Materials. 
Simulations were performed on computers provided by the Canadian Foundation for Innovation, the Minist\`ere de l'\'Education des Loisirs et du Sport (Qu\'ebec), Calcul Qu\'ebec, and Compute Canada.
The Flatiron Institute is a division of the Simons Foundation.




\makeatletter




\begin{appendix}

\section{\label{sec: biparticity}Biparticity and sign problem.}

In the case of the one-band Hubbard model with nearest-neighbor hopping, any infinite square lattice is bipartite which leads to particle-hole symmetry. Under that symmetry, one finds that $n\to2-n$ and $\mu\to U-\mu$ maps the results for the particle-doped lattices to those for the hole-doped ones. Thus, this symmetry implies that $n=1$ when $\mu=U/2$.
When the lattice is frustrated, this symmetry does not hold but a particle-hole transformation is still possible. 
In the computational basis [Fig.~\ref{fig: lattice}(a)], the infinite triangular lattice can be divided into two sublattices with $t$ the intersublattice and $t'$ (that we eventually set to $-t$) the intrasublattice hopping terms. 
The particle-hole transformation then leads to the following statements about the clusters: 

\begin{enumerate}
    \item Changing $-t'\to t'$ is equivalent to changing $n\to2-n$, $\mu\to U-\mu$, and $\textbf{k}\to\textbf{K}+(\pi,\pi)$.
    \item Changing $t\to-t$ is equivalent to changing $\textbf{k}\to\textbf{K}+(\pi,\pi)$.
\end{enumerate}


Within DCA, we verified thoroughly that this holds true for the $N_c=4$, $6$, and $16$ clusters in Fig.~\ref{fig: cluster_conv}, and not for the $N_c=6^*$ cluster. 
This is a manifestation of the fact that the $N_c=6^*$ cluster does not satisfy biparticity.
Nonbipartite lattices have the attribute of enhancing frustration, thus inhibiting AFM order. 

\begin{figure}
    \centering
    \includegraphics[width=\linewidth]{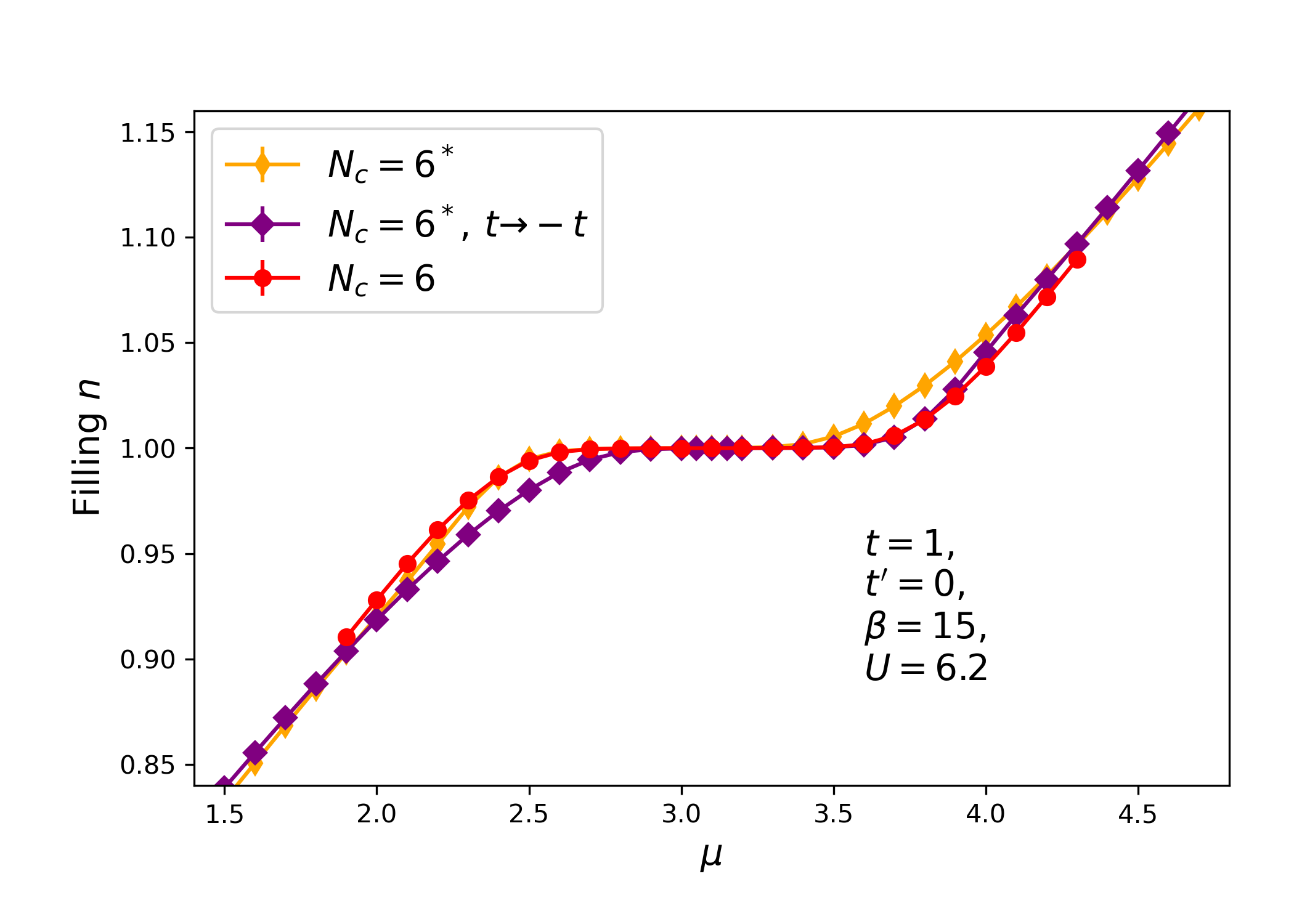}
    \caption{(Color online) This figure displays three curves with $t'=0$ where particle-hole symmetry should hold. The red curve is for the bipartite $N_c=6$ lattice while the two others are for the $N_c=6^*$ nonbipartite lattice with and without the particle-hole transformation. The last two curves do not overlap, showing the lack of particle-hole symmetry of the $N_c=6^*$ cluster.}
    \label{fig: MI-METAL_hole_3x2}
\end{figure}

For $t'=0$, we verified that our bipartite lattices are half-filled when the condition $\mu/2=U$ is satisfied while it is not for the nonbipartite lattice (for $\beta=20$, $U=5.25$ and $\mu=2.625$, the filling is $0.995$). 
In Fig.~\ref{fig: MI-METAL_hole_3x2}, we compare the filling as a function of the chemical potential for both clusters with six sites. The nonbipartite $N_c=6^*$ lattice does not display particle-hole symmetry. Although the large doping behavior is very similar on both sides, near half filling it deviates from particle-hole symmetry. Using such a nonbipartite cluster might be interesting to understand the general behavior, but is not as trustworthy as bipartite ones.

Pushing further the analysis, we found some common behavior for all the clusters. Applying the particle-hole transformation by changing the sign of $t'$, we noticed an important reduction, by a factor of 2, in the average Monte Carlo sign. In order to recover a better average sign, the hopping terms of a nonbipartite cluster needed to satisfy $t\cdot t' = |t\cdot t'|$. 
In cluster DMFT, there are analogous choices of gauge that can increase the average sign~\cite{Semon:2012}.


\section{\label{sec: bistability}Bistability.} 

\begin{figure}
    \centering
    \includegraphics[width=\linewidth]{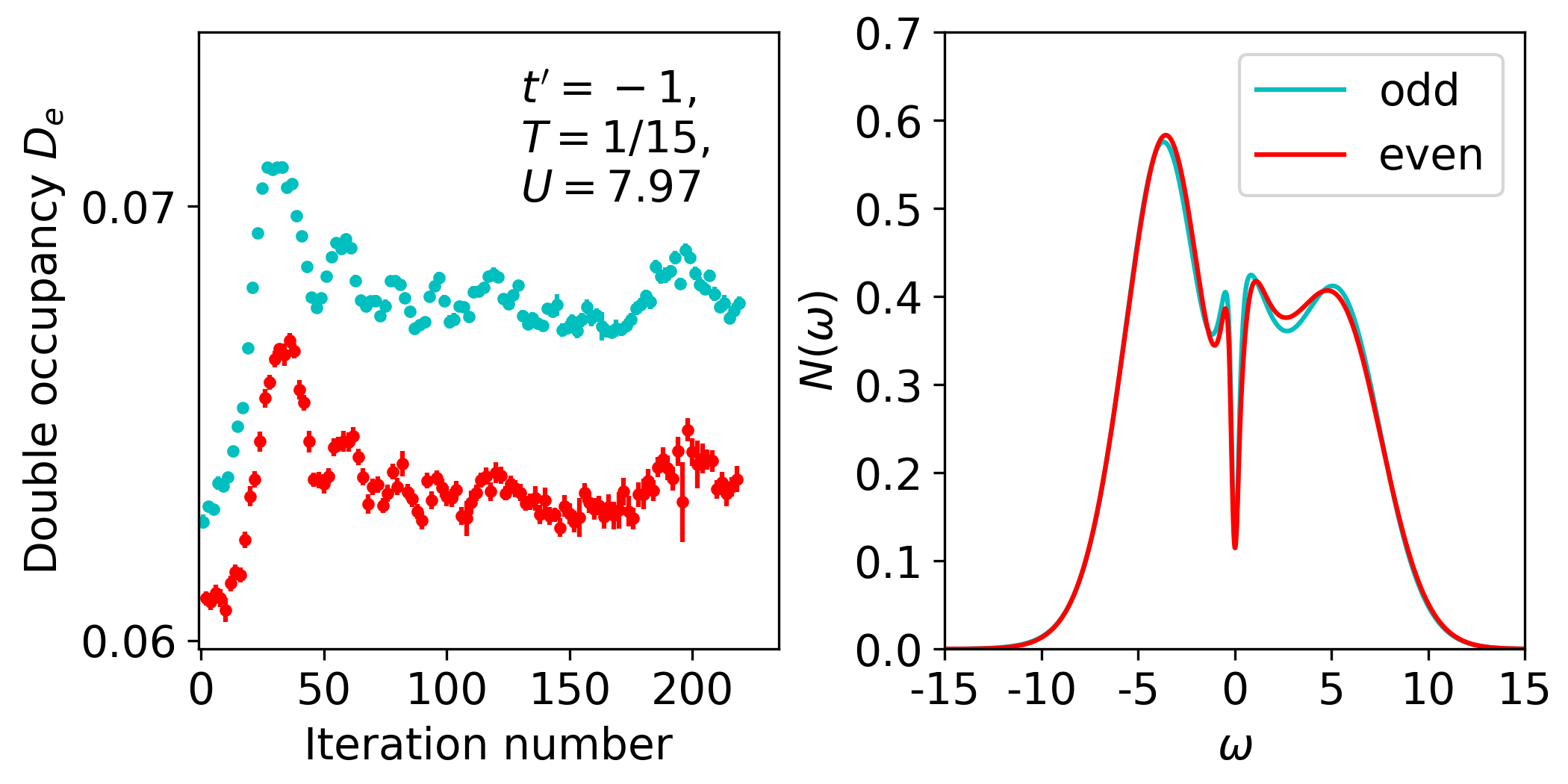}
    \caption{ (Color online) Left : Double occupancy as a function of iteration number for $U=7.97$ and $T=1/15$. The different colors give the parity of every iteration. The error bars are the Monte Carlo errors. Right : Density of states of the two alternating solutions of the left-hand panel. This was made using the analytic continuation algorithm of Ref.~\onlinecite{MaxEntBergeron}. The color coding is identical for both figures.}
    \label{fig: docc_bistable}
\end{figure}

In the coexistence region indicated by the hatched zone in Fig.~\ref{fig: nc6_half_filled_phase_diag}(a), coming from the insulating phase we have noticed a smaller zone where two solutions alternate.
This is a typical behavior of the logistic map and it seems to make the insulating solution in the $U_{c1}<U<U_{c3}$ range extremely unstable. 

To understand $U_{c3}$, consider the simulation performed in the $U_{c1}<U<U_{c3}$ region and illustrated in Fig.~\ref{fig: docc_bistable}. On the left-hand panel, we show the double occupancy as a function of the iteration number, where the even (odd) iteration numbers are plotted in cyan (red). We clearly notice that the bistable phase is a feature that survives under DCA iterations and that it strongly affects double occupancy. Even if we add damping to the iterations by keeping a fraction of the last imaginary and real parts of the hybridization function for the next iteration, it does not change anything to the bistability gap. On the right-hand panel, we plot the spectral weights obtained from analytic continuation of the Green's function for the even and odd numbered iterations. Taking into account that analytic continuation is usually error prone, the DOS tells us that there ought to be little to no differences between the two cases. Hence, this may not be of importance for the pseudogap and does not cast doubts on our results outside this coexistence region.


\section{Data compilation algorithm.} \label{sec: data_compile_algo}



In order to compute the numerous average values and uncertainties presented in this work, we used the following algorithm. 
An observable $X$ calculated for $n$ iterations is given as an input. Its values and Monte Carlo errors at the $i$th iteration are labeled $x_i$ and $\sigma_{M,i}$. 
The result of each iteration is weighted by 
\begin{align}
    w_i = 1/\sigma_{M,i}^2.
\end{align}

With those, we calculate the weighted mean of the last $m$ iterations $\bar x_m$ using
\begin{align}
    \bar x_m = \frac{1}{w_{T,m}}\sum_{i=1+n-m}^n w_i \cdot x_i ,
\end{align}
where we defined $w_{T,m}=\sum_{i=1+n-m}^n w_i$ as the total weight from the $m$th iteration to the last one. We calculate the weighted uncorrelated standard deviation of the simulation as
\begin{align}
    \sigma_m = \sqrt{\frac{\sum_{i=1+n-m}^n w_i^2}{w_{T,m}^2}\Bigg(\frac{1}{w_{T,m}}\sum_{i=1+n-m}^n x_i^2 w_i - \bar x_m^2\Bigg)} .
\end{align}
Using the confidence interval factor $\text{CIF}(0.9999, m)$, we ensure that, in an uncorrelated simulation, we are $99.99\%$ sure that $\bar x_m$ is inside the uncertainty. We define
\begin{align}
    \sigma_{\text{CIF},m} = \sigma_m \cdot \text{CIF}(0.9999, m).
\end{align}
From this, our algorithm aims to find the optimal $m$ ($m_{op}$) such that
\begin{align}
    m_{op} = \min_m \big( \sigma_{\text{CIF},m}).
\end{align}
We selected $m\ge 15$, an arbitrary limit made necessary by the correlations between iterations and by thermalization.

The output of our algorithm is the result $\bar x_{m_{op}}$ and we use $\sigma_{\text{CIF},m_{op}}$ as the uncertainty.

\begin{figure*}
    \centering
    \includegraphics[width=0.8\linewidth]{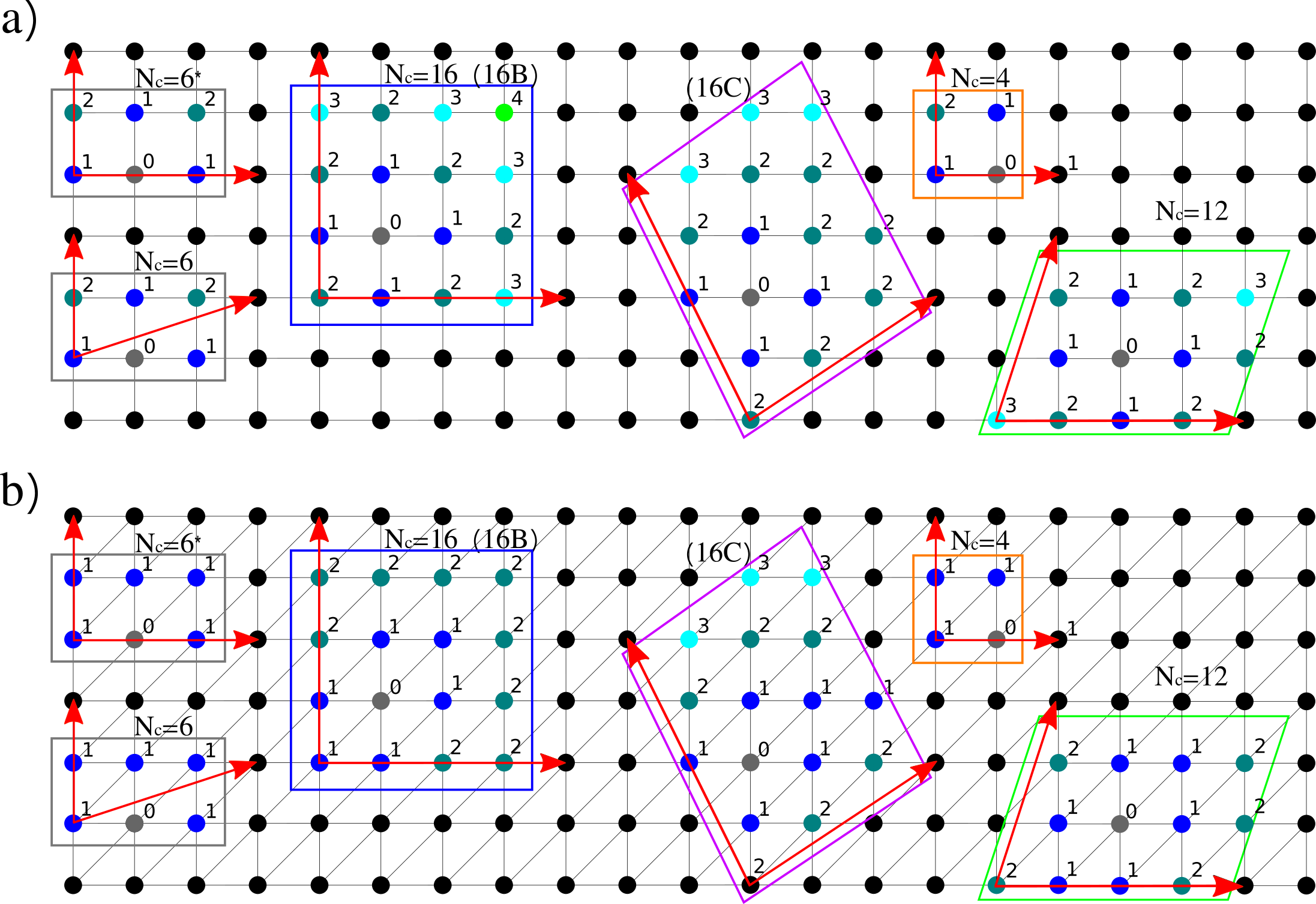}
    \caption{ (Color online) 
   Labelling of the intra-cluster topological neighbors of the site 0 for all clusters used in this work along with the cluster 16C from Ref.~\citenum{betts:1999} for the (a) square and (b) triangular lattices. For each cluster, the super-lattice vectors are drawn in red.
    }
    \label{fig: betts_lattice}
\end{figure*}

\section{\label{sec: ferromagnetic_imperfection}Ferromagnetic imperfections.}



Typically encountered in the context of ferromagnetism, the number of ferromagnetic imperfections $\mathcal{I}_F$ relies on the concept of topological neighbors to predict the quality or goodness of a cluster with periodic boundary conditions~\cite{betts:1996, betts:1999}. The importance of this concept in the context of strongly correlated electrons was emphasized in Ref.~\cite{Maier_Betts_2005}.  

The Betts~\cite{betts:1996, betts:1999} definition of perfect or imperfect finite-size cluster is as follows. First note that in an infinitely large square cluster, a site has 4 topological nearest neighbors, 8 topological next-nearest neighbors, 12 third-nearest neighbors, and so on, with $4n$ $n$th-nearest neighbors. For a given finite size cluster, all the $4n$ of the $n$th-nearest neighbors of a given site may not be present on the cluster. As long as the $4(n-1)$ neighbors are all present, the cluster is perfect, unless further neighbors, such as those in the $(n+1)$th-neighbor category, appear on the cluster. 

The number of ferromagnetic imperfections $\mathcal{I}_F$ of a cluster is defined as the number of displacements, or changes in the shape of the cluster, that are needed to make it perfect.
Let us detail how $\mathcal{I}_F$ can be counted on the square lattice ($\mathcal{I}_{Fs}$) for several clusters shown in Fig.~\ref{fig: betts_lattice}(a).
They are the same as those defined in Fig.~\ref{fig: cluster_conv}(a), with the addition of cluster 16C from Ref.~\cite{betts:1999}.

In the case of cluster 16C, starting from the site zero, we have four first neighbors, eight second neighbors and three third neighbors, thus $\mathcal{I}_{Fs}(\text{16C})=0$. It is important to notice that we assume periodic boundary conditions modulo displacements along the super-lattice vectors. 

For the $N_c=16$ cluster (which is the same as 16B in Ref.~\citenum{betts:1999}), we have four first, six second, four third and one fourth neighbor, denoted $(4,6,4,1)$. The number of ferromagnetic imperfections is given by the number of site displacements needed to build a perfect lattice. 
One can make a perfect lattice by doing the following displacements : $(4,6,4,1)\to(4,6,5)\to(4,7,4)\to(4,8,3)$. Since we did three displacements, $\mathcal{I}_{Fs}(N_c=16)=3$. For each of the square lattice clusters, we find

\begin{align}
    \mathcal{I}_{Fs}(N_c=4)&=1, \nonumber \\
    \mathcal{I}_{Fs}(N_c=6^*)&=2, \nonumber \\
    \mathcal{I}_{Fs}(N_c=6)&=2,\\
    \mathcal{I}_{Fs}(N_c=12)&=2, \nonumber \\
    \mathcal{I}_{Fs}(N_c=16)&=3. \nonumber
\end{align}

In Fig.~\ref{fig: betts_lattice}(b), we illustrate how to extend the concept of ferromagnetic imperfections to triangular lattices, labeled $\mathcal{I}_{Ft}$. In this configuration, a site in the infinite lattice will have $6n$ $n$th nearest neighbors.
We find $\mathcal{I}_{Ft}(\text{16C})=3$ while the clusters used in this work are all ferromagnetically perfect with

\begin{align}
    \mathcal{I}_{Ft}(N_c=4)&=0, \nonumber \\
    \mathcal{I}_{Ft}(N_c=6^*)&=0, \nonumber \\
    \mathcal{I}_{Ft}(N_c=6)&=0,\\
    \mathcal{I}_{Ft}(N_c=12)&=0, \nonumber \\
    \mathcal{I}_{Ft}(N_c=16)&=0. \nonumber
\end{align}

\section{Defining \texorpdfstring{$U_{inc}$ and $U_{gap}$}{Uinc and Ugap}.}\label{sec: Ugap and Uinc}

Defining $U_{inc}$ and $U_{gap}$ is not straightforward, but the idea is that after analytic continuation with the maximum entropy technique~\cite{Jarrell:1996,MaxEntBergeron}, we inspect the low frequency behavior to determine the limit between two qualitatively different behaviors. 

\begin{figure}[h!]
    \centering
    \includegraphics[width=\linewidth]{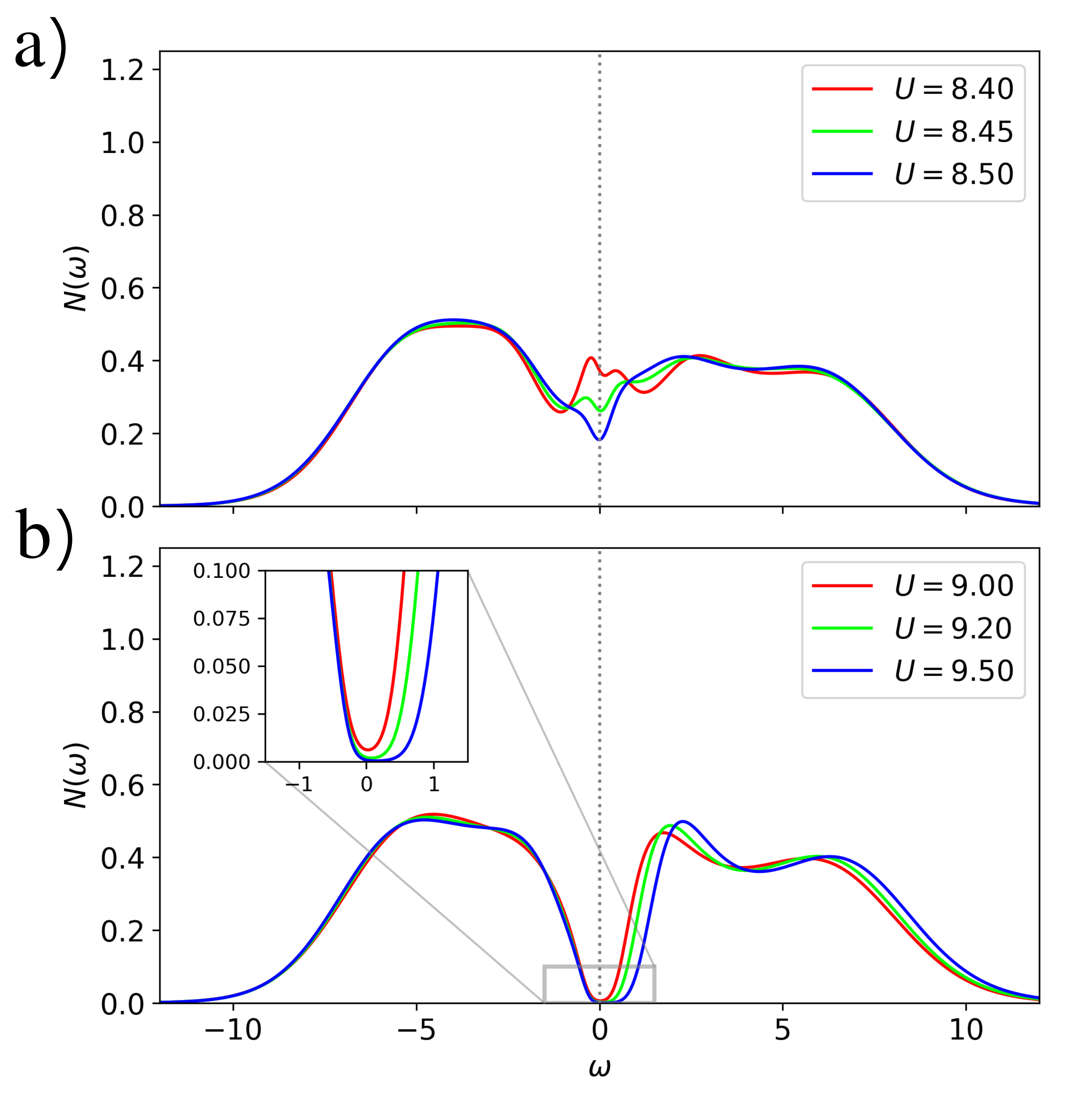}
    \caption{ (color online) Local DOS for many values of $U$ at $\beta=7$. (a) The DOS indicates that $U_{inc}=8.475\pm 0.025$. (b) The DOS indicates that $U_{gap}=9.35\pm0.15$.}
    \label{fig: cFL_PG_MI_lim}
\end{figure}

The $U_{inc}$ line separates cases where a quasiparticle peak exists in the density of states and when it does not. Although it is usually very clear, it is not always the case, particularly when there seems to exist two quasiparticle peaks. In Fig.~\ref{fig: cFL_PG_MI_lim}(a), we have an example of how we choose the value. Once there is no longer a local maximum {\it near} the Fermi level, we consider that we have crossed the $U_{inc}$ line.

The $U_{gap}$ line separates cases where there is a gap at the Fermi level of the local density of states from cases where there is not. The subjectivity of this criterion comes from the fact that the spectral weight never exactly vanishes at $\omega=0$. For this reason, we inspect visually, judging if there is a gap or not. In Fig.~\ref{fig: cFL_PG_MI_lim}(b), we have an example of how we choose the value.

\section{\texorpdfstring{$N_c=12$}{Nc=12} cluster benchmark.}\label{sec: Nc=12}
In order to verify whether the results obtained with the $N_c=6$ cluster were reliable, we performed calculations for the larger $N_c=12$ and $N_c=16$ clusters and studied in more detail the 12-site cluster. This appendix focuses on this cluster.

\begin{figure}[h!]
    \centering
    \includegraphics[width=0.5\linewidth]{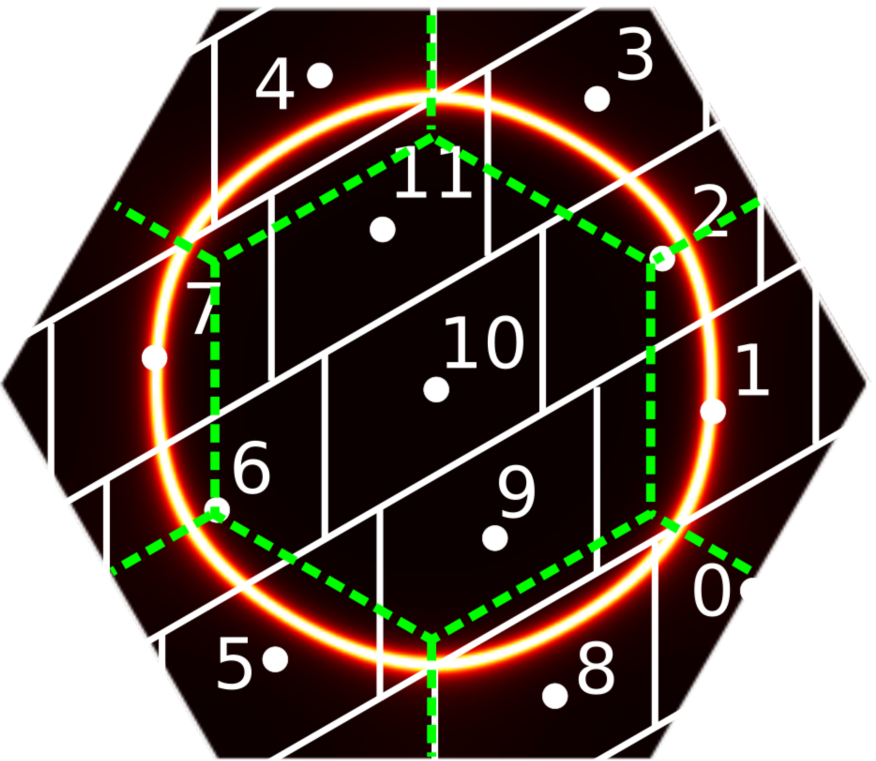}
    \caption{ (Color online) 
    Brillouin zone and patch arrangement of the $N_c=12$ cluster, along with the spectral weight at $U=0$ showing the Fermi surface. The green dashed lines delimit the three sub-lattice reduced Brillouin zone, associated to 120$^\circ$ AFM.
    }
    \label{fig: Nc12_Brillouin_patch}
\end{figure}

\begin{figure}
    \centering
    \includegraphics[width=\linewidth]{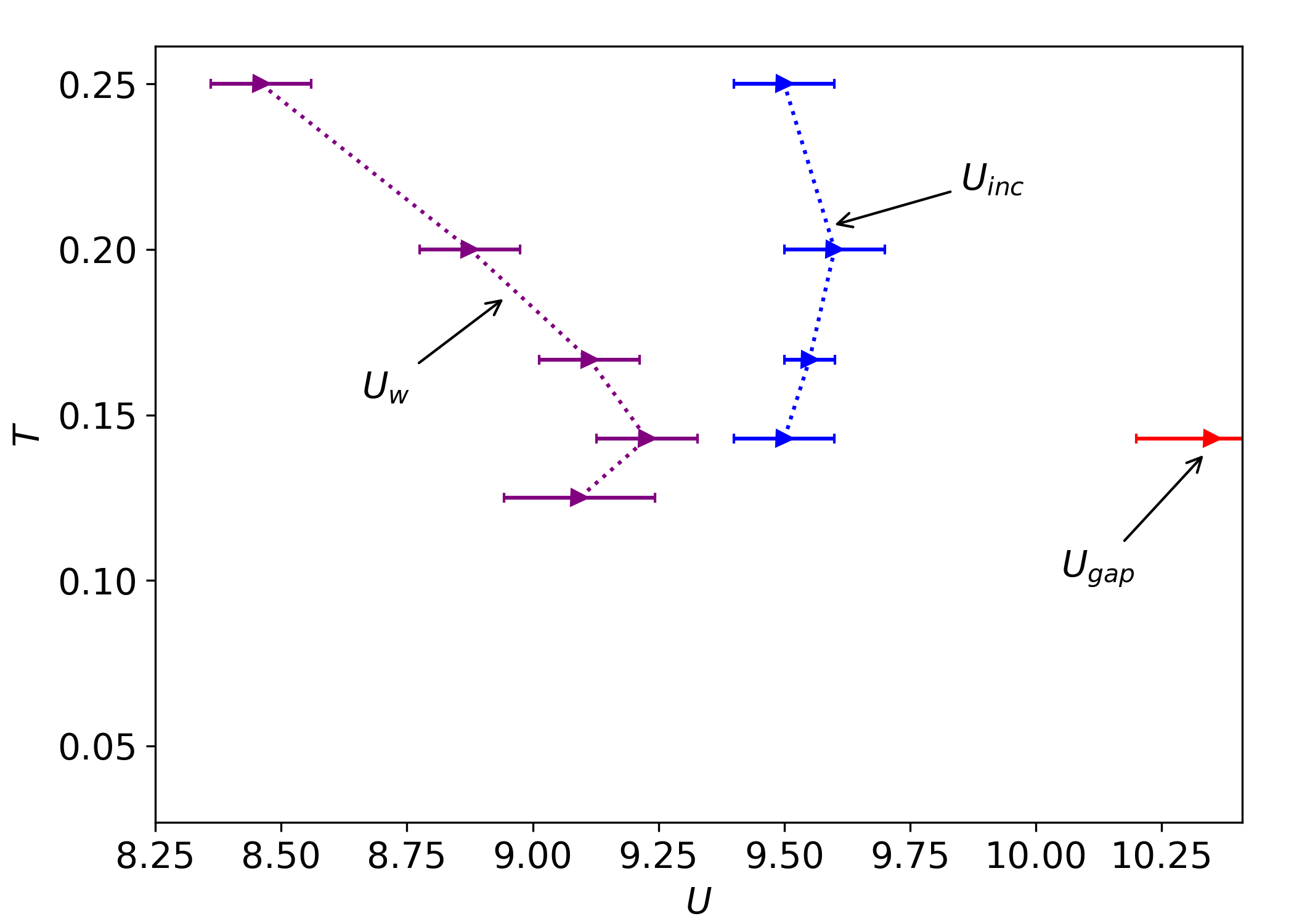}
    \caption{ (Color online)
    $U$-$T$ phase diagram for the $N_c=12$ cluster. This figure shows the same important features as Fig.~\ref{fig: nc6_half_filled_phase_diag}(a), but for a larger cluster. At $T\leq0.125$, the sign problem becomes too important for $U>U_W$.
    For the low-temperature part of the Widom line, some points were discarded because they were outliers, namely $(U=9.5, T=1/7)$, $(U=9.2, T=1/8)$ and $(U=9.25, T=1/8)$.
    }
    \label{fig: nc12_half_filled_phase_diag}
\end{figure}

\begin{figure}
    \centering
    \includegraphics[width=\linewidth]{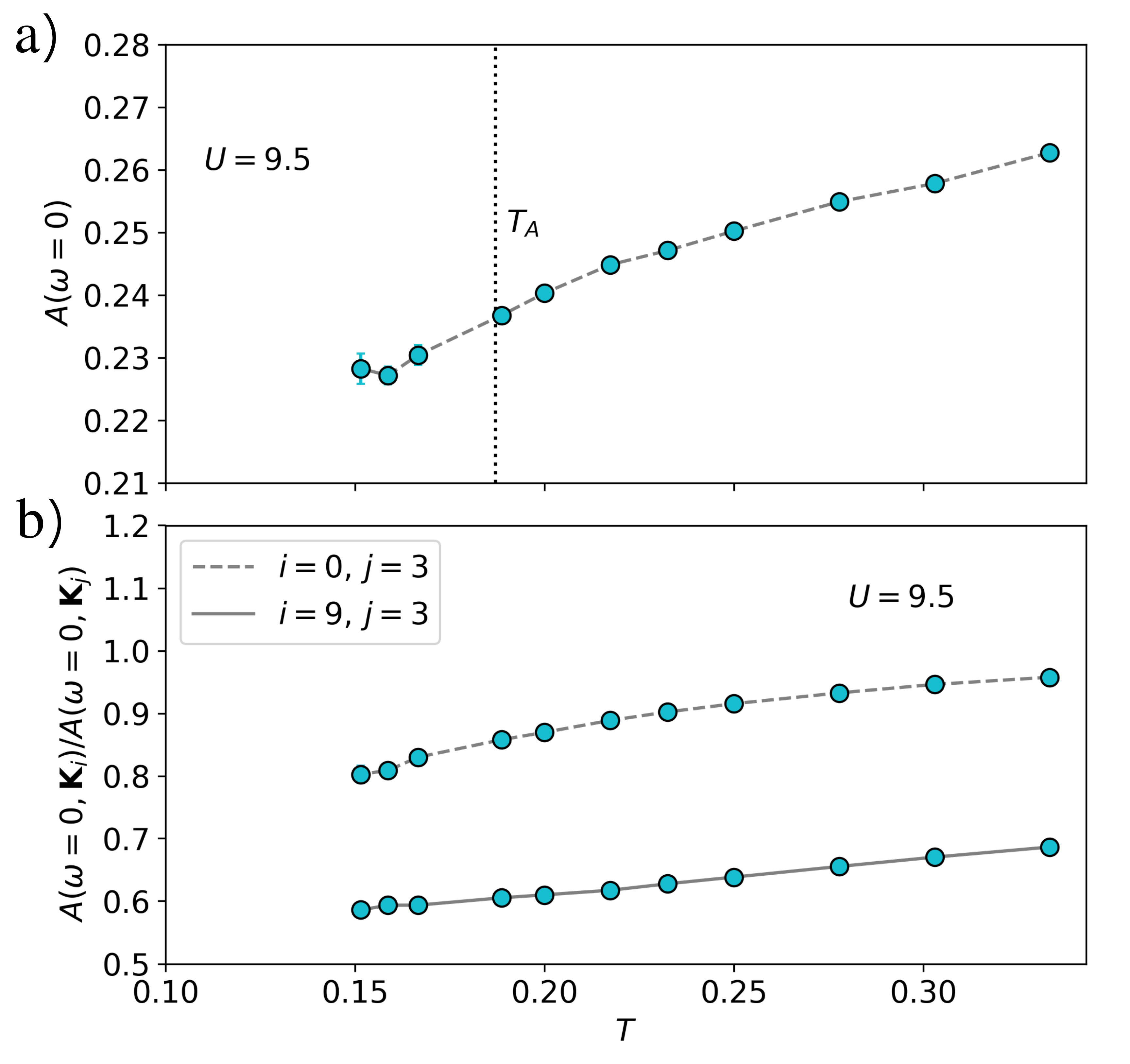}
    \caption{ (Color online) 
    (a) Temperature dependence of the spectral weight at the Fermi level for the $N_c=12$ cluster at $U=9.5$. Lower temperatures suffer from a bad sign problem, increasing their uncertainty. This uncertainty is hard to calculate and is probably underestimated. 
    (b) Temperature dependence of the ratio of the spectral weight between patches $i$ and $j$, $A(\omega=0,\mathbf{K}_i)/A(\omega=0,\mathbf{K}_j)$.}
    \label{fig: nc12_A_omega_K_dep}
\end{figure}

The real-space cluster is depicted in Fig.~\ref{fig: betts_lattice}(b) and the Brillouin zone in momentum space in Fig.~\ref{fig: Nc12_Brillouin_patch}.
The main results obtained using the $N_c=12$ cluster are presented in Fig.~\ref{fig: nc12_half_filled_phase_diag}. Comparing with those from the $N_c=6$ cluster shown in Fig.~\ref{fig: nc6_half_filled_phase_diag}(a), we observe that the most important features are preserved: the lines $U_W$, $U_{inc}$ and $U_{gap}$ are present and spaced out by a similar distance. We also note that the shape of $U_W$ is consistent between the two clusters.
The $U_{inc}$ line however has a different shape, but for the most part, the slope is slightly positive in the bigger cluster similarly to the smaller cluster.
In addition, the important $T_A$ line presented in \fref{fig: nc6_K_dependant_loses_and_hotspots}(a) for the 6-site cluster is also present in the 12-site one, as shown in Fig.~\ref{fig: nc12_A_omega_K_dep}(a).

On the other hand, the comparison between these clusters highlights important differences, the main ones being the shape of $U_{inc}$ and the absence of $T_{A,max}$ in the $N_c = 12$ cluster shown in Fig.~\ref{fig: nc12_A_omega_K_dep}(a). This maximum of the spectral weight at the Fermi level as a function of temperature is a typical feature of the onset of a pseudogap. There are a few possible explanations for its absence. First, $T_{A,max}$ could be present at higher temperature, but comparing with the $N_c=6$ cluster [Fig.~\ref{fig: nc6_K_dependant_loses_and_hotspots}(a)], we do not expect it to appear at much higher temperatures. Another explanation points  to the quasivertical $U_{inc}$ line. The transition from having a quasiparticle peak to not having one, which defines $U_{inc}$, should be closely related to the crossover in the density of states close to the Fermi level, like $T_A$. But since $U_{inc}$ is quasivertical in temperature, any temperature crossover related to $U_{inc}$ should become very blurry, as we observe for $T_A$. The very small derivative of $A(\omega=0,T)$ at $T_A$ in Fig.~\ref{fig: nc12_A_omega_K_dep}(a) should then not come as a surprise. A consequence of such a blurry crossover might be to push $T_{A,max}$ to a much higher temperature.

Another important difference is the small size of the double occupancy's derivative, used to track $U_W$. At higher temperatures (e.g. $T\ge0.167$), the derivative of the double occupancy is similar for the $N_c=6$ and the $N_c=12$ clusters (not shown). At $T\le0.167$, the minimum in the derivative for $N_c=12$ does not become deeper as is the case for $N_c=6$, giving no evidence of a possible critical point close to $T=0.111$. One could infer that the Mott transition happens at much lower temperature or that it does not occur at all, but another possibility is that magnetic ordering could be hiding the Mott transition for this specific cluster. Indeed for the large values of $U$ that are important here, many works find magnetic ordering~\cite{sahebsara_hubbard_2008, laubachPhaseDiagramHubbard2015a, Misumi_Mott_triangular:2017, Tocchio_backflow_Mott:2008, Yoshioka_triangular:2009, wietek_mott_2021, Yu_Li_Iskakov_Gull_2023} in this temperature range.

One last comparison with the $N_c=6$ cluster is the $\textbf{K}$ dependence of the losses of spectral weight, as displayed in Fig.~\ref{fig: nc12_A_omega_K_dep}(b). Although we always observe a quasiparticle peak close to the Fermi level, the patches losing spectral weight the fastest are consistent with 120$^\circ$ spiral AFM order hotspots, shown in Fig.~\ref{fig: Nc12_Brillouin_patch} for the noninteracting Fermi surface. For patch 3 the noninteracting Fermi surface never crosses the three sub-lattice reduced Brillouin zone (green dashed lines in Fig.\ref{fig: Nc12_Brillouin_patch}). In patches 0 and 9, however, the Fermi surface crosses the three sublattice reduced Brillouin zone. Hence patches 0 and 9 lose spectral weight faster than patch 3 with decreasing temperature. Note that patch 9 contains less Fermi surface than patch 0 in the noninteracting case, hence the ratio of patch 9 to patch 3 is overall smaller than the ratio of patch 0 to patch 3. In a Fermi liquid however, the temperature dependence of the ratio would be negligible. 

The mechanism behind the formation of the PG here might be similar to the $N_c=6$ cluster, where we had seen that $\textbf{K}_0$ was losing spectral weight the fastest. But it is important to note that, although it seems to follow the same trend, the change in ratio here is small, probably because we never actually enter the PG phase in in Fig.~\ref{fig: nc12_A_omega_K_dep}(b)
.

\begin{figure}
    \centering
    \includegraphics[width=\linewidth]{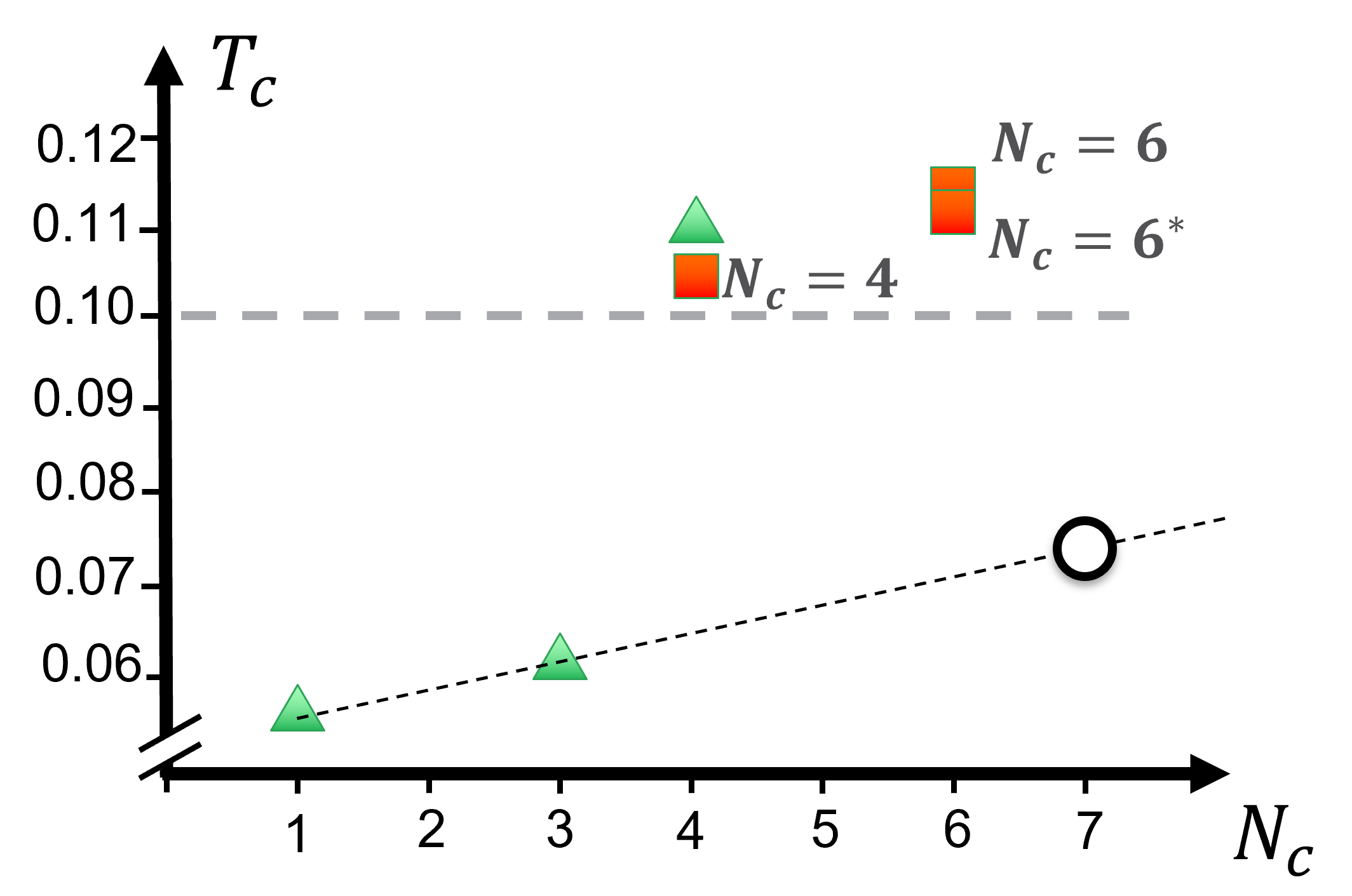}
    \caption{ (Color online) The Mott critical temperature $T_c$ for different cluster sizes $N_c$. The orange squares are data from the present paper. The green triangles are extrapolated data from Ref.~\citenum{Dang:2015}. The horizontal gray dashed line represents the lowest temperature achieved using CDMFT with $N_c = 7$ in Ref.~\citenum{wietek_mott_2021}, $T=0.1$. The dashed black line is the linear extrapolation for the $T_c$ of clusters with an odd number of sites, with the white circle being the predicted $T_c$ for the seven sites cluster.}
    \label{fig: Tc_dang_po}
\end{figure}

\section{Parity dependence of the Mott \texorpdfstring{$T_c$}{Tc}.} \label{sec: Tc}
The first-order transition was not observed in the 7-site cluster used in Ref.~\citenum{wietek_mott_2021}. We believe it is attributable to the parity of the cluster size. Further studies will be needed to confirm this claim.

In Fig.~\ref{fig: Tc_dang_po}, we gathered from this work and the literature the estimated Mott critical temperature $T_c$ for different cluster sizes $N_c$ for a triangular lattice. Note that $T_c$ grows with $N_c$, with a big jump at $N_c=4$. Notice we can also observe a pattern related to the parity of $N_c$. For odd numbers of sites, $T_c$ is much lower than for even numbers. We attribute this behavior to an intrinsic difference in entropy between even and odd number of sites. 
For even $N_c$, every electron at half filling can pair with another one.
This is not true for odd $N_c$ that have larger entropy coming from unpaired electrons. A rough argument from $F=E-TS$ for the free energy is that large entropy favors lower $T_c$ for phase transitions that can also occur at larger $U_c$, i.e. for larger potential energy. This phenomenon can already be deduced from the results of Ref.~\citenum{Dang:2015}, reported in Fig.~\ref{fig: Tc_dang_po}, for various cluster parities.

%
%
%
%
%
%
%
%
%
%
%
%
%
%
%
%
%
%
%
%
%
%
%
%
%
%
%
%
%
%
%
%
%
%
%

\end{appendix}


\begin{thebibliography}{114}%
    \makeatletter
    \providecommand \@ifxundefined [1]{%
     \@ifx{#1\undefined}
    }%
    \providecommand \@ifnum [1]{%
     \ifnum #1\expandafter \@firstoftwo
     \else \expandafter \@secondoftwo
     \fi
    }%
    \providecommand \@ifx [1]{%
     \ifx #1\expandafter \@firstoftwo
     \else \expandafter \@secondoftwo
     \fi
    }%
    \providecommand \natexlab [1]{#1}%
    \providecommand \enquote  [1]{``#1''}%
    \providecommand \bibnamefont  [1]{#1}%
    \providecommand \bibfnamefont [1]{#1}%
    \providecommand \citenamefont [1]{#1}%
    \providecommand \href@noop [0]{\@secondoftwo}%
    \providecommand \href [0]{\begingroup \@sanitize@url \@href}%
    \providecommand \@href[1]{\@@startlink{#1}\@@href}%
    \providecommand \@@href[1]{\endgroup#1\@@endlink}%
    \providecommand \@sanitize@url [0]{\catcode `\\12\catcode `\$12\catcode
      `\&12\catcode `\#12\catcode `\^12\catcode `\_12\catcode `\%12\relax}%
    \providecommand \@@startlink[1]{}%
    \providecommand \@@endlink[0]{}%
    \providecommand \url  [0]{\begingroup\@sanitize@url \@url }%
    \providecommand \@url [1]{\endgroup\@href {#1}{\urlprefix }}%
    \providecommand \urlprefix  [0]{URL }%
    \providecommand \Eprint [0]{\href }%
    \providecommand \doibase [0]{https://doi.org/}%
    \providecommand \selectlanguage [0]{\@gobble}%
    \providecommand \bibinfo  [0]{\@secondoftwo}%
    \providecommand \bibfield  [0]{\@secondoftwo}%
    \providecommand \translation [1]{[#1]}%
    \providecommand \BibitemOpen [0]{}%
    \providecommand \bibitemStop [0]{}%
    \providecommand \bibitemNoStop [0]{.\EOS\space}%
    \providecommand \EOS [0]{\spacefactor3000\relax}%
    \providecommand \BibitemShut  [1]{\csname bibitem#1\endcsname}%
    \let\auto@bib@innerbib\@empty
    \bibitem [{\citenamefont {Hubbard}(1963)}]{hubbard1963electron}%
      \BibitemOpen
      \bibfield  {author} {\bibinfo {author} {\bibfnamefont {J.}~\bibnamefont
      {Hubbard}},\ }\bibfield  {title} {\bibinfo {title} {Electron correlations in
      narrow energy bands},\ }\href
      {https://royalsocietypublishing.org/doi/abs/10.1098/rspa.1963.0204}
      {\bibfield  {journal} {\bibinfo  {journal} {Proceedings of the Royal Society
      of London. Series A. Mathematical and Physical Sciences}\ }\textbf {\bibinfo
      {volume} {276}},\ \bibinfo {pages} {238} (\bibinfo {year}
      {1963})}\BibitemShut {NoStop}%
    \bibitem [{\citenamefont {Hubbard}(1964{\natexlab{a}})}]{hubbard1964electron2}%
      \BibitemOpen
      \bibfield  {author} {\bibinfo {author} {\bibfnamefont {J.}~\bibnamefont
      {Hubbard}},\ }\bibfield  {title} {\bibinfo {title} {Electron correlations in
      narrow energy bands. {II}. {The} degenerate band case},\ }\href
      {https://royalsocietypublishing.org/doi/abs/10.1098/rspa.1964.0019}
      {\bibfield  {journal} {\bibinfo  {journal} {Proceedings of the Royal Society
      of London. Series A. Mathematical and Physical Sciences}\ }\textbf {\bibinfo
      {volume} {277}},\ \bibinfo {pages} {237} (\bibinfo {year}
      {1964}{\natexlab{a}})}\BibitemShut {NoStop}%
    \bibitem [{\citenamefont {Hubbard}(1964{\natexlab{b}})}]{hubbard1964electron3}%
      \BibitemOpen
      \bibfield  {author} {\bibinfo {author} {\bibfnamefont {J.}~\bibnamefont
      {Hubbard}},\ }\bibfield  {title} {\bibinfo {title} {Electron correlations in
      narrow energy bands {III}. {An} improved solution},\ }\href
      {https://royalsocietypublishing.org/doi/abs/10.1098/rspa.1964.0190}
      {\bibfield  {journal} {\bibinfo  {journal} {Proceedings of the Royal Society
      of London. Series A. Mathematical and Physical Sciences}\ }\textbf {\bibinfo
      {volume} {281}},\ \bibinfo {pages} {401} (\bibinfo {year}
      {1964}{\natexlab{b}})}\BibitemShut {NoStop}%
    \bibitem [{\citenamefont {Gutzwiller}(1963)}]{Gutzwiller:1963}%
      \BibitemOpen
      \bibfield  {author} {\bibinfo {author} {\bibfnamefont {M.~C.}\ \bibnamefont
      {Gutzwiller}},\ }\bibfield  {title} {\bibinfo {title} {Effect of correlation
      on the ferromagnetism of transition metals},\ }\href
      {https://doi.org/10.1103/PhysRevLett.10.159} {\bibfield  {journal} {\bibinfo
      {journal} {Phys. Rev. Lett.}\ }\textbf {\bibinfo {volume} {10}},\ \bibinfo
      {pages} {159} (\bibinfo {year} {1963})}\BibitemShut {NoStop}%
    \bibitem [{\citenamefont {Kanamori}(1963)}]{kanamori1963electron}%
      \BibitemOpen
      \bibfield  {author} {\bibinfo {author} {\bibfnamefont {J.}~\bibnamefont
      {Kanamori}},\ }\bibfield  {title} {\bibinfo {title} {Electron correlation and
      ferromagnetism of transition metals},\ }\href
      {https://academic.oup.com/ptp/article/30/3/275/1865799} {\bibfield  {journal}
      {\bibinfo  {journal} {Progress of Theoretical Physics}\ }\textbf {\bibinfo
      {volume} {30}},\ \bibinfo {pages} {275} (\bibinfo {year} {1963})}\BibitemShut
      {NoStop}%
    \bibitem [{\citenamefont {Qin}\ \emph {et~al.}(2022)\citenamefont {Qin},
      \citenamefont {Schäfer}, \citenamefont {Andergassen}, \citenamefont
      {Corboz},\ and\ \citenamefont {Gull}}]{qinHubbardModelComputational2022a}%
      \BibitemOpen
      \bibfield  {author} {\bibinfo {author} {\bibfnamefont {M.}~\bibnamefont
      {Qin}}, \bibinfo {author} {\bibfnamefont {T.}~\bibnamefont {Schäfer}},
      \bibinfo {author} {\bibfnamefont {S.}~\bibnamefont {Andergassen}}, \bibinfo
      {author} {\bibfnamefont {P.}~\bibnamefont {Corboz}},\ and\ \bibinfo {author}
      {\bibfnamefont {E.}~\bibnamefont {Gull}},\ }\bibfield  {title} {\bibinfo
      {title} {The {{Hubbard Model}}: {{A Computational Perspective}}},\ }\href
      {https://doi.org/10.1146/annurev-conmatphys-090921-033948} {\bibfield
      {journal} {\bibinfo  {journal} {Annual Review of Condensed Matter Physics}\
      }\textbf {\bibinfo {volume} {13}},\ \bibinfo {pages} {275} (\bibinfo {year}
      {2022})}\BibitemShut {NoStop}%
    \bibitem [{\citenamefont {Arovas}\ \emph {et~al.}(2022)\citenamefont {Arovas},
      \citenamefont {Berg}, \citenamefont {Kivelson},\ and\ \citenamefont
      {Raghu}}]{Arovas_Berg_Kivelson_Raghu_2022}%
      \BibitemOpen
      \bibfield  {author} {\bibinfo {author} {\bibfnamefont {D.~P.}\ \bibnamefont
      {Arovas}}, \bibinfo {author} {\bibfnamefont {E.}~\bibnamefont {Berg}},
      \bibinfo {author} {\bibfnamefont {S.}~\bibnamefont {Kivelson}},\ and\
      \bibinfo {author} {\bibfnamefont {S.}~\bibnamefont {Raghu}},\ }\bibfield
      {title} {\bibinfo {title} {The hubbard model},\ }\href
      {https://doi.org/10.1146/annurev-conmatphys-031620-102024} {\bibfield
      {journal} {\bibinfo  {journal} {Annual Review of Condensed Matter Physics}\
      }\textbf {\bibinfo {volume} {13}},\ \bibinfo {pages} {239–274} (\bibinfo
      {year} {2022})}\BibitemShut {NoStop}%
    \bibitem [{\citenamefont {Georges}\ and\ \citenamefont
      {Kotliar}(1992)}]{Georges:1992}%
      \BibitemOpen
      \bibfield  {author} {\bibinfo {author} {\bibfnamefont {A.}~\bibnamefont
      {Georges}}\ and\ \bibinfo {author} {\bibfnamefont {G.}~\bibnamefont
      {Kotliar}},\ }\bibfield  {title} {\bibinfo {title} {Hubbard model in infinite
      dimensions},\ }\href
      {https://journals.aps.org/prb/abstract/10.1103/PhysRevB.45.6479} {\bibfield
      {journal} {\bibinfo  {journal} {Phys. Rev. B}\ }\textbf {\bibinfo {volume}
      {45}},\ \bibinfo {pages} {6479} (\bibinfo {year} {1992})}\BibitemShut
      {NoStop}%
    \bibitem [{\citenamefont {Jarrell}(1992)}]{Jarrell:1992}%
      \BibitemOpen
      \bibfield  {author} {\bibinfo {author} {\bibfnamefont {M.}~\bibnamefont
      {Jarrell}},\ }\bibfield  {title} {\bibinfo {title} {Hubbard model in infinite
      dimensions: A quantum monte carlo study},\ }\href
      {https://journals.aps.org/prl/abstract/10.1103/PhysRevLett.69.168} {\bibfield
       {journal} {\bibinfo  {journal} {Phys. Rev. Lett.}\ }\textbf {\bibinfo
      {volume} {69}},\ \bibinfo {pages} {168} (\bibinfo {year} {1992})}\BibitemShut
      {NoStop}%
    \bibitem [{\citenamefont {Georges}\ \emph {et~al.}(1996)\citenamefont
      {Georges}, \citenamefont {Kotliar}, \citenamefont {Krauth},\ and\
      \citenamefont {Rozenberg}}]{Georges:1996}%
      \BibitemOpen
      \bibfield  {author} {\bibinfo {author} {\bibfnamefont {A.}~\bibnamefont
      {Georges}}, \bibinfo {author} {\bibfnamefont {G.}~\bibnamefont {Kotliar}},
      \bibinfo {author} {\bibfnamefont {W.}~\bibnamefont {Krauth}},\ and\ \bibinfo
      {author} {\bibfnamefont {M.~J.}\ \bibnamefont {Rozenberg}},\ }\bibfield
      {title} {\bibinfo {title} {Dynamical mean-field theory of strongly correlated
      fermion systems and the limit of infinite dimensions},\ }\href
      {https://journals.aps.org/rmp/abstract/10.1103/RevModPhys.68.13} {\bibfield
      {journal} {\bibinfo  {journal} {Rev. Mod. Phys.}\ }\textbf {\bibinfo {volume}
      {68}},\ \bibinfo {pages} {13 } (\bibinfo {year} {1996})}\BibitemShut
      {NoStop}%
    \bibitem [{\citenamefont {Maier}\ \emph
      {et~al.}(2005{\natexlab{a}})\citenamefont {Maier}, \citenamefont {Jarrell},
      \citenamefont {Pruschke},\ and\ \citenamefont
      {Hettler}}]{maier_quantum_2005}%
      \BibitemOpen
      \bibfield  {author} {\bibinfo {author} {\bibfnamefont {T.}~\bibnamefont
      {Maier}}, \bibinfo {author} {\bibfnamefont {M.}~\bibnamefont {Jarrell}},
      \bibinfo {author} {\bibfnamefont {T.}~\bibnamefont {Pruschke}},\ and\
      \bibinfo {author} {\bibfnamefont {M.~H.}\ \bibnamefont {Hettler}},\
      }\bibfield  {title} {\bibinfo {title} {Quantum cluster theories},\ }\href
      {https://doi.org/10.1103/RevModPhys.77.1027} {\bibfield  {journal} {\bibinfo
      {journal} {Rev. Mod. Phys.}\ }\textbf {\bibinfo {volume} {77}},\ \bibinfo
      {pages} {1027} (\bibinfo {year} {2005}{\natexlab{a}})}\BibitemShut {NoStop}%
    \bibitem [{\citenamefont {Kotliar}\ \emph {et~al.}(2006)\citenamefont
      {Kotliar}, \citenamefont {Savrasov}, \citenamefont {Haule}, \citenamefont
      {Oudovenko}, \citenamefont {Parcollet},\ and\ \citenamefont
      {Marianetti}}]{KotliarRMP:2006}%
      \BibitemOpen
      \bibfield  {author} {\bibinfo {author} {\bibfnamefont {G.}~\bibnamefont
      {Kotliar}}, \bibinfo {author} {\bibfnamefont {S.~Y.}\ \bibnamefont
      {Savrasov}}, \bibinfo {author} {\bibfnamefont {K.}~\bibnamefont {Haule}},
      \bibinfo {author} {\bibfnamefont {V.~S.}\ \bibnamefont {Oudovenko}}, \bibinfo
      {author} {\bibfnamefont {O.}~\bibnamefont {Parcollet}},\ and\ \bibinfo
      {author} {\bibfnamefont {C.~A.}\ \bibnamefont {Marianetti}},\ }\bibfield
      {title} {\bibinfo {title} {Electronic structure calculations with dynamical
      mean-field theory},\ }\href {https://doi.org/10.1103/RevModPhys.78.865}
      {\bibfield  {journal} {\bibinfo  {journal} {Reviews of Modern Physics}\
      }\textbf {\bibinfo {volume} {78}},\ \bibinfo {eid} {865} (\bibinfo {year}
      {2006})}\BibitemShut {NoStop}%
    \bibitem [{\citenamefont {Tremblay}\ \emph {et~al.}(2006)\citenamefont
      {Tremblay}, \citenamefont {Kyung},\ and\ \citenamefont
      {S\'en\'echal}}]{LTP:2006}%
      \BibitemOpen
      \bibfield  {author} {\bibinfo {author} {\bibfnamefont {A.~M.~S.}\
      \bibnamefont {Tremblay}}, \bibinfo {author} {\bibfnamefont {B.}~\bibnamefont
      {Kyung}},\ and\ \bibinfo {author} {\bibfnamefont {D.}~\bibnamefont
      {S\'en\'echal}},\ }\bibfield  {title} {\bibinfo {title} {Pseudogap and
      high-temperature superconductivity from weak to strong coupling. towards a
      quantitative theory},\ }\href {http://dx.doi.org/10.1063/1.2199446}
      {\bibfield  {journal} {\bibinfo  {journal} {Low Temp. Phys.}\ }\textbf
      {\bibinfo {volume} {32}},\ \bibinfo {pages} {424} (\bibinfo {year}
      {2006})}\BibitemShut {NoStop}%
    \bibitem [{\citenamefont {Metzner}\ and\ \citenamefont
      {Vollhardt}(1989)}]{Metzner:1989}%
      \BibitemOpen
      \bibfield  {author} {\bibinfo {author} {\bibfnamefont {W.}~\bibnamefont
      {Metzner}}\ and\ \bibinfo {author} {\bibfnamefont {D.}~\bibnamefont
      {Vollhardt}},\ }\bibfield  {title} {\bibinfo {title} {Correlated lattice
      fermions in $d=\ensuremath{\infty}$ dimensions},\ }\href
      {https://doi.org/10.1103/PhysRevLett.62.324} {\bibfield  {journal} {\bibinfo
      {journal} {Phys. Rev. Lett.}\ }\textbf {\bibinfo {volume} {62}},\ \bibinfo
      {pages} {324} (\bibinfo {year} {1989})}\BibitemShut {NoStop}%
    \bibitem [{\citenamefont {Kyung}\ and\ \citenamefont
      {Tremblay}(2006)}]{kyung:2006}%
      \BibitemOpen
      \bibfield  {author} {\bibinfo {author} {\bibfnamefont {B.}~\bibnamefont
      {Kyung}}\ and\ \bibinfo {author} {\bibfnamefont {A.-M.~S.}\ \bibnamefont
      {Tremblay}},\ }\bibfield  {title} {\bibinfo {title} {{Mott} {Transition},
      {Antiferromagnetism}, and {d-Wave} {Superconductivity} in {Two-Dimensional}
      {Organic} {Conductors}},\ }\href
      {https://doi.org/10.1103/PhysRevLett.97.046402} {\bibfield  {journal}
      {\bibinfo  {journal} {Physical Review Letters}\ }\textbf {\bibinfo {volume}
      {97}},\ \bibinfo {eid} {046402} (\bibinfo {year} {2006})}\BibitemShut
      {NoStop}%
    \bibitem [{\citenamefont {Park}\ \emph {et~al.}(2008)\citenamefont {Park},
      \citenamefont {Haule},\ and\ \citenamefont {Kotliar}}]{park:2008}%
      \BibitemOpen
      \bibfield  {author} {\bibinfo {author} {\bibfnamefont {H.}~\bibnamefont
      {Park}}, \bibinfo {author} {\bibfnamefont {K.}~\bibnamefont {Haule}},\ and\
      \bibinfo {author} {\bibfnamefont {G.}~\bibnamefont {Kotliar}},\ }\bibfield
      {title} {\bibinfo {title} {Cluster dynamical mean field theory of the mott
      transition},\ }\href {https://doi.org/10.1103/PhysRevLett.101.186403}
      {\bibfield  {journal} {\bibinfo  {journal} {Physical Review Letters}\
      }\textbf {\bibinfo {volume} {101}},\ \bibinfo {eid} {186403} (\bibinfo {year}
      {2008})}\BibitemShut {NoStop}%
    \bibitem [{\citenamefont {Balzer}\ \emph {et~al.}(2009)\citenamefont {Balzer},
      \citenamefont {Kyung}, \citenamefont {S\'en\'echal}, \citenamefont
      {Tremblay},\ and\ \citenamefont {Potthoff}}]{Balzer:2009}%
      \BibitemOpen
      \bibfield  {author} {\bibinfo {author} {\bibfnamefont {M.}~\bibnamefont
      {Balzer}}, \bibinfo {author} {\bibfnamefont {B.}~\bibnamefont {Kyung}},
      \bibinfo {author} {\bibfnamefont {D.}~\bibnamefont {S\'en\'echal}}, \bibinfo
      {author} {\bibfnamefont {A.-M.~S.}\ \bibnamefont {Tremblay}},\ and\ \bibinfo
      {author} {\bibfnamefont {M.}~\bibnamefont {Potthoff}},\ }\bibfield  {title}
      {\bibinfo {title} {First-order mott transition at zero temperature in two
      dimensions: Variational plaquette study},\ }\href
      {http://stacks.iop.org/0295-5075/85/i=1/a=17002} {\bibfield  {journal}
      {\bibinfo  {journal} {Euro. Phys. Lett.}\ }\textbf {\bibinfo {volume} {85}},\
      \bibinfo {pages} {17002} (\bibinfo {year} {2009})}\BibitemShut {NoStop}%
    \bibitem [{\citenamefont {Dang}\ \emph {et~al.}(2015)\citenamefont {Dang},
      \citenamefont {Xu}, \citenamefont {Chen}, \citenamefont {Meng},\ and\
      \citenamefont {Wessel}}]{Dang:2015}%
      \BibitemOpen
      \bibfield  {author} {\bibinfo {author} {\bibfnamefont {H.~T.}\ \bibnamefont
      {Dang}}, \bibinfo {author} {\bibfnamefont {X.~Y.}\ \bibnamefont {Xu}},
      \bibinfo {author} {\bibfnamefont {K.-S.}\ \bibnamefont {Chen}}, \bibinfo
      {author} {\bibfnamefont {Z.~Y.}\ \bibnamefont {Meng}},\ and\ \bibinfo
      {author} {\bibfnamefont {S.}~\bibnamefont {Wessel}},\ }\bibfield  {title}
      {\bibinfo {title} {Mott transition in the triangular lattice hubbard model: A
      dynamical cluster approximation study},\ }\href
      {https://doi.org/10.1103/PhysRevB.91.155101} {\bibfield  {journal} {\bibinfo
      {journal} {Physical Review B}\ }\textbf {\bibinfo {volume} {91}},\ \bibinfo
      {pages} {155101} (\bibinfo {year} {2015})}\BibitemShut {NoStop}%
    \bibitem [{\citenamefont {Walsh}\ \emph {et~al.}(2019)\citenamefont {Walsh},
      \citenamefont {S\'emon}, \citenamefont {Poulin}, \citenamefont {Sordi},\ and\
      \citenamefont {Tremblay}}]{WalshSordiEntanglement:2019}%
      \BibitemOpen
      \bibfield  {author} {\bibinfo {author} {\bibfnamefont {C.}~\bibnamefont
      {Walsh}}, \bibinfo {author} {\bibfnamefont {P.}~\bibnamefont {S\'emon}},
      \bibinfo {author} {\bibfnamefont {D.}~\bibnamefont {Poulin}}, \bibinfo
      {author} {\bibfnamefont {G.}~\bibnamefont {Sordi}},\ and\ \bibinfo {author}
      {\bibfnamefont {A.-M.~S.}\ \bibnamefont {Tremblay}},\ }\bibfield  {title}
      {\bibinfo {title} {Local entanglement entropy and mutual information across
      the mott transition in the two-dimensional hubbard model},\ }\href
      {https://doi.org/10.1103/PhysRevLett.122.067203} {\bibfield  {journal}
      {\bibinfo  {journal} {Phys. Rev. Lett.}\ }\textbf {\bibinfo {volume} {122}},\
      \bibinfo {pages} {067203} (\bibinfo {year} {2019})}\BibitemShut {NoStop}%
    \bibitem [{\citenamefont {McWhan}\ \emph {et~al.}(1973)\citenamefont {McWhan},
      \citenamefont {Menth}, \citenamefont {Remeika}, \citenamefont {Brinkman},\
      and\ \citenamefont {Rice}}]{mcwhan_metal-insulator_1973}%
      \BibitemOpen
      \bibfield  {author} {\bibinfo {author} {\bibfnamefont {D.~B.}\ \bibnamefont
      {McWhan}}, \bibinfo {author} {\bibfnamefont {A.}~\bibnamefont {Menth}},
      \bibinfo {author} {\bibfnamefont {J.~P.}\ \bibnamefont {Remeika}}, \bibinfo
      {author} {\bibfnamefont {W.~F.}\ \bibnamefont {Brinkman}},\ and\ \bibinfo
      {author} {\bibfnamefont {T.~M.}\ \bibnamefont {Rice}},\ }\bibfield  {title}
      {\bibinfo {title} {{Metal-Insulator} {Transitions} in {Pure} and {Doped}
      {V$_2$O$_3$}},\ }\href {https://doi.org/10.1103/PhysRevB.7.1920} {\bibfield
      {journal} {\bibinfo  {journal} {Phys. Rev. B}\ }\textbf {\bibinfo {volume}
      {7}},\ \bibinfo {pages} {1920} (\bibinfo {year} {1973})}\BibitemShut
      {NoStop}%
    \bibitem [{\citenamefont {Granados}\ \emph {et~al.}(1993)\citenamefont
      {Granados}, \citenamefont {Fontcuberta}, \citenamefont {Obradors},
      \citenamefont {Mañosa},\ and\ \citenamefont {Torrance}}]{granados1993}%
      \BibitemOpen
      \bibfield  {author} {\bibinfo {author} {\bibfnamefont {X.}~\bibnamefont
      {Granados}}, \bibinfo {author} {\bibfnamefont {J.}~\bibnamefont
      {Fontcuberta}}, \bibinfo {author} {\bibfnamefont {X.}~\bibnamefont
      {Obradors}}, \bibinfo {author} {\bibfnamefont {L.}~\bibnamefont {Mañosa}},\
      and\ \bibinfo {author} {\bibfnamefont {J.~B.}\ \bibnamefont {Torrance}},\
      }\bibfield  {title} {\bibinfo {title} {Metallic state and the metal-insulator
      transition of {{NdNiO}} 3},\ }\href
      {https://doi.org/10.1103/PhysRevB.48.11666} {\bibfield  {journal} {\bibinfo
      {journal} {Physical Review B}\ }\textbf {\bibinfo {volume} {48}},\ \bibinfo
      {pages} {11666} (\bibinfo {year} {1993})}\BibitemShut {NoStop}%
    \bibitem [{\citenamefont {Lefebvre}\ \emph {et~al.}(2000)\citenamefont
      {Lefebvre}, \citenamefont {Wzietek}, \citenamefont {Brown}, \citenamefont
      {Bourbonnais}, \citenamefont {J\'erome}, \citenamefont {M\'ezi\`ere},
      \citenamefont {Fourmigu\'e},\ and\ \citenamefont {Batail}}]{Lefebvre:2000}%
      \BibitemOpen
      \bibfield  {author} {\bibinfo {author} {\bibfnamefont {S.}~\bibnamefont
      {Lefebvre}}, \bibinfo {author} {\bibfnamefont {P.}~\bibnamefont {Wzietek}},
      \bibinfo {author} {\bibfnamefont {S.}~\bibnamefont {Brown}}, \bibinfo
      {author} {\bibfnamefont {C.}~\bibnamefont {Bourbonnais}}, \bibinfo {author}
      {\bibfnamefont {D.}~\bibnamefont {J\'erome}}, \bibinfo {author}
      {\bibfnamefont {C.}~\bibnamefont {M\'ezi\`ere}}, \bibinfo {author}
      {\bibfnamefont {M.}~\bibnamefont {Fourmigu\'e}},\ and\ \bibinfo {author}
      {\bibfnamefont {P.}~\bibnamefont {Batail}},\ }\bibfield  {title} {\bibinfo
      {title} {Mott transition, antiferromagnetism, and unconventional
      superconductivity in layered organic superconductors},\ }\href
      {https://doi.org/10.1103/PhysRevLett.85.5420} {\bibfield  {journal} {\bibinfo
       {journal} {Phys. Rev. Lett.}\ }\textbf {\bibinfo {volume} {85}},\ \bibinfo
      {pages} {5420} (\bibinfo {year} {2000})}\BibitemShut {NoStop}%
    \bibitem [{\citenamefont {Dumm}\ \emph {et~al.}(2009)\citenamefont {Dumm},
      \citenamefont {Faltermeier}, \citenamefont {Drichko}, \citenamefont
      {Dressel}, \citenamefont {M\'ezi\`ere},\ and\ \citenamefont
      {Batail}}]{dumm:2009}%
      \BibitemOpen
      \bibfield  {author} {\bibinfo {author} {\bibfnamefont {M.}~\bibnamefont
      {Dumm}}, \bibinfo {author} {\bibfnamefont {D.}~\bibnamefont {Faltermeier}},
      \bibinfo {author} {\bibfnamefont {N.}~\bibnamefont {Drichko}}, \bibinfo
      {author} {\bibfnamefont {M.}~\bibnamefont {Dressel}}, \bibinfo {author}
      {\bibfnamefont {C.}~\bibnamefont {M\'ezi\`ere}},\ and\ \bibinfo {author}
      {\bibfnamefont {P.}~\bibnamefont {Batail}},\ }\bibfield  {title} {\bibinfo
      {title} {Bandwidth-controlled {Mott} transition in
      $\kappa-{(\mathrm{BEDT}-\mathrm{TTF})}_{2}\mathrm{Cu}[\mathrm{N}{(\mathrm{CN})}_{2}]\mathrm{Br}_{x}\mathrm{Cl}_{1-x}$:
      Optical studies of correlated carriers},\ }\href
      {https://doi.org/10.1103/PhysRevB.79.195106} {\bibfield  {journal} {\bibinfo
      {journal} {Phys. Rev. B}\ }\textbf {\bibinfo {volume} {79}},\ \bibinfo
      {pages} {195106} (\bibinfo {year} {2009})}\BibitemShut {NoStop}%
    \bibitem [{\citenamefont {Kanoda}\ and\ \citenamefont
      {Kato}(2011)}]{kanoda_mott_2011}%
      \BibitemOpen
      \bibfield  {author} {\bibinfo {author} {\bibfnamefont {K.}~\bibnamefont
      {Kanoda}}\ and\ \bibinfo {author} {\bibfnamefont {R.}~\bibnamefont {Kato}},\
      }\bibfield  {title} {\bibinfo {title} {Mott physics in organic conductors
      with triangular lattices},\ }\href
      {https://doi.org/10.1146/annurev-conmatphys-062910-140521} {\bibfield
      {journal} {\bibinfo  {journal} {Annual Review of Condensed Matter Physics}\
      }\textbf {\bibinfo {volume} {2}},\ \bibinfo {pages} {167} (\bibinfo {year}
      {2011})},\ \Eprint
      {https://arxiv.org/abs/https://doi.org/10.1146/annurev-conmatphys-062910-140521}
      {https://doi.org/10.1146/annurev-conmatphys-062910-140521} \BibitemShut
      {NoStop}%
    \bibitem [{\citenamefont {Pustogow}\ \emph {et~al.}(2018)\citenamefont
      {Pustogow}, \citenamefont {Bories}, \citenamefont {Löhle}, \citenamefont
      {Rösslhuber}, \citenamefont {Zhukova}, \citenamefont {Gorshunov},
      \citenamefont {Tomić}, \citenamefont {Schlueter}, \citenamefont {Hübner},
      \citenamefont {Hiramatsu}, \citenamefont {Yoshida}, \citenamefont {Saito},
      \citenamefont {Kato}, \citenamefont {Lee}, \citenamefont {Dobrosavljević},
      \citenamefont {Fratini},\ and\ \citenamefont
      {Dressel}}]{Pustogow_Bories_2018}%
      \BibitemOpen
      \bibfield  {author} {\bibinfo {author} {\bibfnamefont {A.}~\bibnamefont
      {Pustogow}}, \bibinfo {author} {\bibfnamefont {M.}~\bibnamefont {Bories}},
      \bibinfo {author} {\bibfnamefont {A.}~\bibnamefont {Löhle}}, \bibinfo
      {author} {\bibfnamefont {R.}~\bibnamefont {Rösslhuber}}, \bibinfo {author}
      {\bibfnamefont {E.}~\bibnamefont {Zhukova}}, \bibinfo {author} {\bibfnamefont
      {B.}~\bibnamefont {Gorshunov}}, \bibinfo {author} {\bibfnamefont
      {S.}~\bibnamefont {Tomić}}, \bibinfo {author} {\bibfnamefont {J.~A.}\
      \bibnamefont {Schlueter}}, \bibinfo {author} {\bibfnamefont {R.}~\bibnamefont
      {Hübner}}, \bibinfo {author} {\bibfnamefont {T.}~\bibnamefont {Hiramatsu}},
      \bibinfo {author} {\bibfnamefont {Y.}~\bibnamefont {Yoshida}}, \bibinfo
      {author} {\bibfnamefont {G.}~\bibnamefont {Saito}}, \bibinfo {author}
      {\bibfnamefont {R.}~\bibnamefont {Kato}}, \bibinfo {author} {\bibfnamefont
      {T.-H.}\ \bibnamefont {Lee}}, \bibinfo {author} {\bibfnamefont
      {V.}~\bibnamefont {Dobrosavljević}}, \bibinfo {author} {\bibfnamefont
      {S.}~\bibnamefont {Fratini}},\ and\ \bibinfo {author} {\bibfnamefont
      {M.}~\bibnamefont {Dressel}},\ }\bibfield  {title} {\bibinfo {title} {Quantum
      spin liquids unveil the genuine mott state},\ }\href
      {https://doi.org/10.1038/s41563-018-0140-3} {\bibfield  {journal} {\bibinfo
      {journal} {Nature Materials}\ ,\ \bibinfo {pages} {1}} (\bibinfo {year}
      {2018})}\BibitemShut {NoStop}%
    \bibitem [{\citenamefont {Sordi}\ \emph {et~al.}(2011)\citenamefont {Sordi},
      \citenamefont {Haule},\ and\ \citenamefont {Tremblay}}]{Sordi:2011}%
      \BibitemOpen
      \bibfield  {author} {\bibinfo {author} {\bibfnamefont {G.}~\bibnamefont
      {Sordi}}, \bibinfo {author} {\bibfnamefont {K.}~\bibnamefont {Haule}},\ and\
      \bibinfo {author} {\bibfnamefont {A.-M.~S.}\ \bibnamefont {Tremblay}},\
      }\bibfield  {title} {\bibinfo {title} {Mott physics and first-order
      transition between two metals in the normal-state phase diagram of the
      two-dimensional hubbard model},\ }\href
      {https://doi.org/10.1103/PhysRevB.84.075161} {\bibfield  {journal} {\bibinfo
      {journal} {Phys. Rev. B}\ }\textbf {\bibinfo {volume} {84}},\ \bibinfo
      {pages} {075161} (\bibinfo {year} {2011})}\BibitemShut {NoStop}%
    \bibitem [{\citenamefont {S\'emon}\ and\ \citenamefont
      {Tremblay}(2012)}]{Semon:2012}%
      \BibitemOpen
      \bibfield  {author} {\bibinfo {author} {\bibfnamefont {P.}~\bibnamefont
      {S\'emon}}\ and\ \bibinfo {author} {\bibfnamefont {A.-M.~S.}\ \bibnamefont
      {Tremblay}},\ }\bibfield  {title} {\bibinfo {title} {Importance of subleading
      corrections for the mott critical point},\ }\href
      {https://doi.org/10.1103/PhysRevB.85.201101} {\bibfield  {journal} {\bibinfo
      {journal} {Phys. Rev. B}\ }\textbf {\bibinfo {volume} {85}},\ \bibinfo
      {pages} {201101} (\bibinfo {year} {2012})}\BibitemShut {NoStop}%
    \bibitem [{\citenamefont {Vučičević}\ \emph {et~al.}(2013)\citenamefont
      {Vučičević}, \citenamefont {Terletska}, \citenamefont {Tanasković},\ and\
      \citenamefont {Dobrosavljević}}]{vucicevic:2013}%
      \BibitemOpen
      \bibfield  {author} {\bibinfo {author} {\bibfnamefont {J.}~\bibnamefont
      {Vučičević}}, \bibinfo {author} {\bibfnamefont {H.}~\bibnamefont
      {Terletska}}, \bibinfo {author} {\bibfnamefont {D.}~\bibnamefont
      {Tanasković}},\ and\ \bibinfo {author} {\bibfnamefont {V.}~\bibnamefont
      {Dobrosavljević}},\ }\bibfield  {title} {\bibinfo {title}
      {Finite-temperature crossover and the quantum {{Widom}} line near the
      {{Mott}} transition},\ }\href {https://doi.org/10.1103/PhysRevB.88.075143}
      {\bibfield  {journal} {\bibinfo  {journal} {Physical Review B}\ }\textbf
      {\bibinfo {volume} {88}},\ \bibinfo {pages} {075143} (\bibinfo {year}
      {2013})}\BibitemShut {NoStop}%
    \bibitem [{\citenamefont {H\'ebert}\ \emph {et~al.}(2015)\citenamefont
      {H\'ebert}, \citenamefont {S\'emon},\ and\ \citenamefont
      {Tremblay}}]{hebert_superconducting_2015}%
      \BibitemOpen
      \bibfield  {author} {\bibinfo {author} {\bibfnamefont {C.-D.}\ \bibnamefont
      {H\'ebert}}, \bibinfo {author} {\bibfnamefont {P.}~\bibnamefont {S\'emon}},\
      and\ \bibinfo {author} {\bibfnamefont {A.-M.~S.}\ \bibnamefont {Tremblay}},\
      }\bibfield  {title} {\bibinfo {title} {Superconducting dome in doped
      quasi-two-dimensional organic {Mott} insulators: {A} paradigm for strongly
      correlated superconductivity},\ }\href
      {https://doi.org/10.1103/PhysRevB.92.195112} {\bibfield  {journal} {\bibinfo
      {journal} {Physical Review B}\ }\textbf {\bibinfo {volume} {92}},\ \bibinfo
      {pages} {195112} (\bibinfo {year} {2015})}\BibitemShut {NoStop}%
    \bibitem [{\citenamefont {Xu}\ \emph {et~al.}(2005)\citenamefont {Xu},
      \citenamefont {Kumar}, \citenamefont {Buldyrev}, \citenamefont {Chen},
      \citenamefont {Poole}, \citenamefont {Sciortino},\ and\ \citenamefont
      {Stanley}}]{XuStanleyWidom:2005}%
      \BibitemOpen
      \bibfield  {author} {\bibinfo {author} {\bibfnamefont {L.}~\bibnamefont
      {Xu}}, \bibinfo {author} {\bibfnamefont {P.}~\bibnamefont {Kumar}}, \bibinfo
      {author} {\bibfnamefont {S.~V.}\ \bibnamefont {Buldyrev}}, \bibinfo {author}
      {\bibfnamefont {S.-H.}\ \bibnamefont {Chen}}, \bibinfo {author}
      {\bibfnamefont {P.~H.}\ \bibnamefont {Poole}}, \bibinfo {author}
      {\bibfnamefont {F.}~\bibnamefont {Sciortino}},\ and\ \bibinfo {author}
      {\bibfnamefont {H.~E.}\ \bibnamefont {Stanley}},\ }\bibfield  {title}
      {\bibinfo {title} {Relation between the widom line and the dynamic crossover
      in systems with a liquid–liquid phase transition},\ }\href
      {https://doi.org/10.1073/pnas.0507870102} {\bibfield  {journal} {\bibinfo
      {journal} {Proc. Nat. Acad. Sci. USA}\ }\textbf {\bibinfo {volume} {102}},\
      \bibinfo {pages} {16558} (\bibinfo {year} {2005})}\BibitemShut {NoStop}%
    \bibitem [{\citenamefont {McMillan}\ and\ \citenamefont
      {Stanley}(2010)}]{McMillan_Stanley_2010}%
      \BibitemOpen
      \bibfield  {author} {\bibinfo {author} {\bibfnamefont {P.~F.}\ \bibnamefont
      {McMillan}}\ and\ \bibinfo {author} {\bibfnamefont {H.~E.}\ \bibnamefont
      {Stanley}},\ }\bibfield  {title} {\bibinfo {title} {Going supercritical},\
      }\href {https://doi.org/10.1038/nphys1711} {\bibfield  {journal} {\bibinfo
      {journal} {Nature Physics}\ }\textbf {\bibinfo {volume} {6}},\ \bibinfo
      {pages} {479–480} (\bibinfo {year} {2010})}\BibitemShut {NoStop}%
    \bibitem [{\citenamefont {Fratino}\ \emph {et~al.}(2016)\citenamefont
      {Fratino}, \citenamefont {S{\'e}mon}, \citenamefont {Sordi},\ and\
      \citenamefont {Tremblay}}]{fratino2016organizing}%
      \BibitemOpen
      \bibfield  {author} {\bibinfo {author} {\bibfnamefont {L.}~\bibnamefont
      {Fratino}}, \bibinfo {author} {\bibfnamefont {P.}~\bibnamefont {S{\'e}mon}},
      \bibinfo {author} {\bibfnamefont {G.}~\bibnamefont {Sordi}},\ and\ \bibinfo
      {author} {\bibfnamefont {A.-M.}\ \bibnamefont {Tremblay}},\ }\bibfield
      {title} {\bibinfo {title} {An organizing principle for two-dimensional
      strongly correlated superconductivity},\ }\href
      {https://www.nature.com/articles/srep22715} {\bibfield  {journal} {\bibinfo
      {journal} {Scientific reports}\ }\textbf {\bibinfo {volume} {6}},\ \bibinfo
      {pages} {22715} (\bibinfo {year} {2016})}\BibitemShut {NoStop}%
    \bibitem [{\citenamefont {Fratino}\ \emph {et~al.}(2017)\citenamefont
      {Fratino}, \citenamefont {Sémon}, \citenamefont {Charlebois}, \citenamefont
      {Sordi},\ and\ \citenamefont
      {Tremblay}}]{Fratino_Semon_Charlebois_Sordi_Tremblay_2017}%
      \BibitemOpen
      \bibfield  {author} {\bibinfo {author} {\bibfnamefont {L.}~\bibnamefont
      {Fratino}}, \bibinfo {author} {\bibfnamefont {P.}~\bibnamefont {Sémon}},
      \bibinfo {author} {\bibfnamefont {M.}~\bibnamefont {Charlebois}}, \bibinfo
      {author} {\bibfnamefont {G.}~\bibnamefont {Sordi}},\ and\ \bibinfo {author}
      {\bibfnamefont {A.-M.~S.}\ \bibnamefont {Tremblay}},\ }\bibfield  {title}
      {\bibinfo {title} {Signatures of the mott transition in the antiferromagnetic
      state of the two-dimensional hubbard model},\ }\href
      {https://doi.org/10.1103/PhysRevB.95.235109} {\bibfield  {journal} {\bibinfo
      {journal} {Physical Review B}\ }\textbf {\bibinfo {volume} {95}},\ \bibinfo
      {pages} {235109} (\bibinfo {year} {2017})}\BibitemShut {NoStop}%
    \bibitem [{\citenamefont {Vilk}\ and\ \citenamefont
      {Tremblay}(1995)}]{Vilk:1995}%
      \BibitemOpen
      \bibfield  {author} {\bibinfo {author} {\bibfnamefont {Y.~M.}\ \bibnamefont
      {Vilk}}\ and\ \bibinfo {author} {\bibfnamefont {A.-M.~S.}\ \bibnamefont
      {Tremblay}},\ }\bibfield  {title} {\bibinfo {title} {Destruction of the fermi
      liquid by spin fluctuations in two dimensions},\ }\href
      {https://www.sciencedirect.com/science/article/abs/pii/0022369795001689}
      {\bibfield  {journal} {\bibinfo  {journal} {J. Phys. Chem. Solids (UK)}\
      }\textbf {\bibinfo {volume} {56}},\ \bibinfo {pages} {1769 } (\bibinfo {year}
      {1995})}\BibitemShut {NoStop}%
    \bibitem [{\citenamefont {{Y.M. Vilk}}\ and\ \citenamefont {{A.-M.S.
      Tremblay}}(1997)}]{Vilk:1997}%
      \BibitemOpen
      \bibfield  {author} {\bibinfo {author} {\bibnamefont {{Y.M. Vilk}}}\ and\
      \bibinfo {author} {\bibnamefont {{A.-M.S. Tremblay}}},\ }\bibfield  {title}
      {\bibinfo {title} {Non-perturbative many-body approach to the hubbard model
      and single-particle pseudogap},\ }\href {https://doi.org/10.1051/jp1:1997135}
      {\bibfield  {journal} {\bibinfo  {journal} {J. Phys. I France}\ }\textbf
      {\bibinfo {volume} {7}},\ \bibinfo {pages} {1309} (\bibinfo {year}
      {1997})}\BibitemShut {NoStop}%
    \bibitem [{\citenamefont {Huscroft}\ \emph {et~al.}(2001)\citenamefont
      {Huscroft}, \citenamefont {Jarrell}, \citenamefont {Maier}, \citenamefont
      {Moukouri},\ and\ \citenamefont {Tahvildarzadeh}}]{Huscroft:2001}%
      \BibitemOpen
      \bibfield  {author} {\bibinfo {author} {\bibfnamefont {C.}~\bibnamefont
      {Huscroft}}, \bibinfo {author} {\bibfnamefont {M.}~\bibnamefont {Jarrell}},
      \bibinfo {author} {\bibfnamefont {T.}~\bibnamefont {Maier}}, \bibinfo
      {author} {\bibfnamefont {S.}~\bibnamefont {Moukouri}},\ and\ \bibinfo
      {author} {\bibfnamefont {A.~N.}\ \bibnamefont {Tahvildarzadeh}},\ }\bibfield
      {title} {\bibinfo {title} {Pseudogaps in the 2d hubbard model},\ }\href
      {https://journals.aps.org/prl/abstract/10.1103/PhysRevLett.86.139} {\bibfield
       {journal} {\bibinfo  {journal} {Phys. Rev. Lett.}\ }\textbf {\bibinfo
      {volume} {86}},\ \bibinfo {pages} {139 } (\bibinfo {year}
      {2001})}\BibitemShut {NoStop}%
    \bibitem [{\citenamefont {Rost}\ \emph {et~al.}(2012)\citenamefont {Rost},
      \citenamefont {Gorelik}, \citenamefont {Assaad},\ and\ \citenamefont
      {Bl\"umer}}]{rost:2012}%
      \BibitemOpen
      \bibfield  {author} {\bibinfo {author} {\bibfnamefont {D.}~\bibnamefont
      {Rost}}, \bibinfo {author} {\bibfnamefont {E.~V.}\ \bibnamefont {Gorelik}},
      \bibinfo {author} {\bibfnamefont {F.}~\bibnamefont {Assaad}},\ and\ \bibinfo
      {author} {\bibfnamefont {N.}~\bibnamefont {Bl\"umer}},\ }\bibfield  {title}
      {\bibinfo {title} {Momentum-dependent pseudogaps in the half-filled
      two-dimensional {Hubbard} model},\ }\href
      {https://doi.org/10.1103/PhysRevB.86.155109} {\bibfield  {journal} {\bibinfo
      {journal} {Phys. Rev. B}\ }\textbf {\bibinfo {volume} {86}},\ \bibinfo
      {pages} {155109} (\bibinfo {year} {2012})}\BibitemShut {NoStop}%
    \bibitem [{\citenamefont {Merino}\ and\ \citenamefont
      {Gunnarsson}(2014)}]{merinoPseudogapSingletFormation2014a}%
      \BibitemOpen
      \bibfield  {author} {\bibinfo {author} {\bibfnamefont {J.}~\bibnamefont
      {Merino}}\ and\ \bibinfo {author} {\bibfnamefont {O.}~\bibnamefont
      {Gunnarsson}},\ }\bibfield  {title} {\bibinfo {title} {Pseudogap and singlet
      formation in organic and cuprate superconductors},\ }\href
      {https://doi.org/10.1103/PhysRevB.89.245130} {\bibfield  {journal} {\bibinfo
      {journal} {Physical Review B}\ }\textbf {\bibinfo {volume} {89}},\ \bibinfo
      {pages} {245130} (\bibinfo {year} {2014})}\BibitemShut {NoStop}%
    \bibitem [{\citenamefont {Ye}\ and\ \citenamefont
      {Chubukov}(2019)}]{Ye_Chubukov_2019}%
      \BibitemOpen
      \bibfield  {author} {\bibinfo {author} {\bibfnamefont {M.}~\bibnamefont
      {Ye}}\ and\ \bibinfo {author} {\bibfnamefont {A.~V.}\ \bibnamefont
      {Chubukov}},\ }\bibfield  {title} {\bibinfo {title} {Hubbard model on a
      triangular lattice: Pseudogap due to spin density wave fluctuations},\ }\href
      {https://doi.org/10.1103/PhysRevB.100.035135} {\bibfield  {journal} {\bibinfo
       {journal} {Physical Review B}\ }\textbf {\bibinfo {volume} {100}},\ \bibinfo
      {pages} {035135} (\bibinfo {year} {2019})}\BibitemShut {NoStop}%
    \bibitem [{\citenamefont {Hankevych}\ \emph {et~al.}(2006)\citenamefont
      {Hankevych}, \citenamefont {Kyung}, \citenamefont {Dar\'e}, \citenamefont
      {S\'en\'echal},\ and\ \citenamefont {Tremblay}}]{Hankevych:2006}%
      \BibitemOpen
      \bibfield  {author} {\bibinfo {author} {\bibfnamefont {V.}~\bibnamefont
      {Hankevych}}, \bibinfo {author} {\bibfnamefont {B.}~\bibnamefont {Kyung}},
      \bibinfo {author} {\bibfnamefont {A.-M.}\ \bibnamefont {Dar\'e}}, \bibinfo
      {author} {\bibfnamefont {D.}~\bibnamefont {S\'en\'echal}},\ and\ \bibinfo
      {author} {\bibfnamefont {A.-M.~S.}\ \bibnamefont {Tremblay}},\ }\bibfield
      {title} {\bibinfo {title} {Strong- and weak-coupling mechanisms for pseudogap
      in electron-doped cuprates},\ }\href
      {https://doi.org/10.1016/j.jpcs.2005.10.121} {\bibfield  {journal} {\bibinfo
      {journal} {Journal of Physics and Chemistry of Solids}\ }\textbf {\bibinfo
      {volume} {67}},\ \bibinfo {pages} {189 } (\bibinfo {year} {2006})},\ \bibinfo
      {note} {spectroscopies in Novel Superconductors 2004}\BibitemShut {NoStop}%
    \bibitem [{\citenamefont {S\'en\'echal}\ and\ \citenamefont
      {Tremblay}(2004)}]{Senechal:2004}%
      \BibitemOpen
      \bibfield  {author} {\bibinfo {author} {\bibfnamefont {D.}~\bibnamefont
      {S\'en\'echal}}\ and\ \bibinfo {author} {\bibfnamefont {A.-M.~S.}\
      \bibnamefont {Tremblay}},\ }\bibfield  {title} {\bibinfo {title} {Hot spots
      and pseudogaps for hole- and electron-doped high-temperature
      superconductors},\ }\href
      {http://link.aps.org/doi/10.1103/PhysRevLett.92.126401} {\bibfield  {journal}
      {\bibinfo  {journal} {Phys. Rev. Lett.}\ }\textbf {\bibinfo {volume} {92}},\
      \bibinfo {pages} {126401} (\bibinfo {year} {2004})}\BibitemShut {NoStop}%
    \bibitem [{\citenamefont {Powell}\ \emph {et~al.}(2009)\citenamefont {Powell},
      \citenamefont {Yusuf},\ and\ \citenamefont {{McKenzie}}}]{powell_spin_2009}%
      \BibitemOpen
      \bibfield  {author} {\bibinfo {author} {\bibfnamefont {B.~J.}\ \bibnamefont
      {Powell}}, \bibinfo {author} {\bibfnamefont {E.}~\bibnamefont {Yusuf}},\ and\
      \bibinfo {author} {\bibfnamefont {R.~H.}\ \bibnamefont {{McKenzie}}},\
      }\bibfield  {title} {\bibinfo {title} {Spin fluctuations and the pseudogap in
      organic superconductors},\ }\href
      {http://link.aps.org/doi/10.1103/PhysRevB.80.054505} {\bibfield  {journal}
      {\bibinfo  {journal} {Physical Review B}\ }\textbf {\bibinfo {volume} {80}},\
      \bibinfo {pages} {054505} (\bibinfo {year} {2009})}\BibitemShut {NoStop}%
    \bibitem [{\citenamefont {Powell}\ and\ \citenamefont
      {McKenzie}(2011)}]{PowellMcKenzieReview:2011}%
      \BibitemOpen
      \bibfield  {author} {\bibinfo {author} {\bibfnamefont {B.~J.}\ \bibnamefont
      {Powell}}\ and\ \bibinfo {author} {\bibfnamefont {R.~H.}\ \bibnamefont
      {McKenzie}},\ }\bibfield  {title} {\bibinfo {title} {Quantum frustration in
      organic mott insulators: from spin liquids to unconventional
      superconductors},\ }\href {http://stacks.iop.org/0034-4885/74/i=5/a=056501}
      {\bibfield  {journal} {\bibinfo  {journal} {Reports on Progress in Physics}\
      }\textbf {\bibinfo {volume} {74}},\ \bibinfo {pages} {056501} (\bibinfo
      {year} {2011})}\BibitemShut {NoStop}%
    \bibitem [{\citenamefont {Maegawa}\ \emph {et~al.}(2011)\citenamefont
      {Maegawa}, \citenamefont {Itou}, \citenamefont {Oyamada},\ and\ \citenamefont
      {Kato}}]{Maegawa_Itou_Oyamada_Kato_2011}%
      \BibitemOpen
      \bibfield  {author} {\bibinfo {author} {\bibfnamefont {S.}~\bibnamefont
      {Maegawa}}, \bibinfo {author} {\bibfnamefont {T.}~\bibnamefont {Itou}},
      \bibinfo {author} {\bibfnamefont {A.}~\bibnamefont {Oyamada}},\ and\ \bibinfo
      {author} {\bibfnamefont {R.}~\bibnamefont {Kato}},\ }\bibfield  {title}
      {\bibinfo {title} {Nmr study of quantum spin liquid and its phase transition
      in the organic spin-1/2 triangular lattice antiferromagnet
      {EtMe$_3$Sb[Pd(dmit)$_2$]$_2$}},\ }\href
      {https://doi.org/10.1088/1742-6596/320/1/012032} {\bibfield  {journal}
      {\bibinfo  {journal} {Journal of Physics: Conference Series}\ }\textbf
      {\bibinfo {volume} {320}},\ \bibinfo {pages} {012032} (\bibinfo {year}
      {2011})}\BibitemShut {NoStop}%
    \bibitem [{\citenamefont {Limelette}\ \emph {et~al.}(2003)\citenamefont
      {Limelette}, \citenamefont {Georges}, \citenamefont {Jerome}, \citenamefont
      {Wzietek}, \citenamefont {Metcalf},\ and\ \citenamefont
      {Honig}}]{Limelette:2003}%
      \BibitemOpen
      \bibfield  {author} {\bibinfo {author} {\bibfnamefont {P.}~\bibnamefont
      {Limelette}}, \bibinfo {author} {\bibfnamefont {A.}~\bibnamefont {Georges}},
      \bibinfo {author} {\bibfnamefont {D.}~\bibnamefont {Jerome}}, \bibinfo
      {author} {\bibfnamefont {P.}~\bibnamefont {Wzietek}}, \bibinfo {author}
      {\bibfnamefont {P.}~\bibnamefont {Metcalf}},\ and\ \bibinfo {author}
      {\bibfnamefont {J.~M.}\ \bibnamefont {Honig}},\ }\bibfield  {title} {\bibinfo
      {title} {{Universality and Critical Behavior at the Mott Transition}},\
      }\href {https://doi.org/10.1126/science.1088386} {\bibfield  {journal}
      {\bibinfo  {journal} {Science}\ }\textbf {\bibinfo {volume} {302}},\ \bibinfo
      {pages} {89} (\bibinfo {year} {2003})},\ \Eprint
      {https://arxiv.org/abs/http://www.sciencemag.org/cgi/reprint/302/5642/89.pdf}
      {http://www.sciencemag.org/cgi/reprint/302/5642/89.pdf} \BibitemShut
      {NoStop}%
    \bibitem [{\citenamefont {Shimizu}\ \emph {et~al.}(2003)\citenamefont
      {Shimizu}, \citenamefont {Miyagawa}, \citenamefont {Kanoda}, \citenamefont
      {Maesato},\ and\ \citenamefont {Saito}}]{Shimizu:2003}%
      \BibitemOpen
      \bibfield  {author} {\bibinfo {author} {\bibfnamefont {Y.}~\bibnamefont
      {Shimizu}}, \bibinfo {author} {\bibfnamefont {K.}~\bibnamefont {Miyagawa}},
      \bibinfo {author} {\bibfnamefont {K.}~\bibnamefont {Kanoda}}, \bibinfo
      {author} {\bibfnamefont {M.}~\bibnamefont {Maesato}},\ and\ \bibinfo {author}
      {\bibfnamefont {G.}~\bibnamefont {Saito}},\ }\bibfield  {title} {\bibinfo
      {title} {Spin liquid state in an organic mott insulator with a triangular
      lattice},\ }\href {https://doi.org/10.1103/PhysRevLett.91.107001} {\bibfield
      {journal} {\bibinfo  {journal} {Phys. Rev. Lett.}\ }\textbf {\bibinfo
      {volume} {91}},\ \bibinfo {pages} {107001} (\bibinfo {year}
      {2003})}\BibitemShut {NoStop}%
    \bibitem [{\citenamefont {Kurosaki}\ \emph {et~al.}(2005)\citenamefont
      {Kurosaki}, \citenamefont {Shimizu}, \citenamefont {Miyagawa}, \citenamefont
      {Kanoda},\ and\ \citenamefont {Saito}}]{Kurosaki:2005}%
      \BibitemOpen
      \bibfield  {author} {\bibinfo {author} {\bibfnamefont {Y.}~\bibnamefont
      {Kurosaki}}, \bibinfo {author} {\bibfnamefont {Y.}~\bibnamefont {Shimizu}},
      \bibinfo {author} {\bibfnamefont {K.}~\bibnamefont {Miyagawa}}, \bibinfo
      {author} {\bibfnamefont {K.}~\bibnamefont {Kanoda}},\ and\ \bibinfo {author}
      {\bibfnamefont {G.}~\bibnamefont {Saito}},\ }\bibfield  {title} {\bibinfo
      {title} {Mott transition from a spin liquid to a fermi liquid in the
      spin-frustrated organic conductor
      $\ensuremath{\kappa}\mathrm{\text{\ensuremath{-}}}(\mathrm{ET}{)}_{2}{\mathrm{cu}}_{2}(\mathrm{CN}{)}_{3}$},\
      }\href {https://doi.org/10.1103/PhysRevLett.95.177001} {\bibfield  {journal}
      {\bibinfo  {journal} {Phys. Rev. Lett.}\ }\textbf {\bibinfo {volume} {95}},\
      \bibinfo {pages} {177001} (\bibinfo {year} {2005})}\BibitemShut {NoStop}%
    \bibitem [{\citenamefont {Itou}\ \emph {et~al.}(2007)\citenamefont {Itou},
      \citenamefont {Oyamada}, \citenamefont {Maegawa}, \citenamefont {Tamura},\
      and\ \citenamefont {Kato}}]{Itou_Oyamada_Maegawa_Tamura_Kato_2007}%
      \BibitemOpen
      \bibfield  {author} {\bibinfo {author} {\bibfnamefont {T.}~\bibnamefont
      {Itou}}, \bibinfo {author} {\bibfnamefont {A.}~\bibnamefont {Oyamada}},
      \bibinfo {author} {\bibfnamefont {S.}~\bibnamefont {Maegawa}}, \bibinfo
      {author} {\bibfnamefont {M.}~\bibnamefont {Tamura}},\ and\ \bibinfo {author}
      {\bibfnamefont {R.}~\bibnamefont {Kato}},\ }\bibfield  {title} {\bibinfo
      {title} {Spin-liquid state in an organic spin-1/2 system on a triangular
      lattice, {EtMe$_3$Sb[Pd(dmit)$_2$]$_2$}},\ }\href
      {https://doi.org/10.1088/0953-8984/19/14/145247} {\bibfield  {journal}
      {\bibinfo  {journal} {Journal of Physics: Condensed Matter}\ }\textbf
      {\bibinfo {volume} {19}},\ \bibinfo {pages} {145247} (\bibinfo {year}
      {2007})}\BibitemShut {NoStop}%
    \bibitem [{\citenamefont {Kandpal}\ \emph {et~al.}(2009)\citenamefont
      {Kandpal}, \citenamefont {Opahle}, \citenamefont {Zhang}, \citenamefont
      {Jeschke},\ and\ \citenamefont {Valent\'\i}}]{Kandpal:2009}%
      \BibitemOpen
      \bibfield  {author} {\bibinfo {author} {\bibfnamefont {H.~C.}\ \bibnamefont
      {Kandpal}}, \bibinfo {author} {\bibfnamefont {I.}~\bibnamefont {Opahle}},
      \bibinfo {author} {\bibfnamefont {Y.-Z.}\ \bibnamefont {Zhang}}, \bibinfo
      {author} {\bibfnamefont {H.~O.}\ \bibnamefont {Jeschke}},\ and\ \bibinfo
      {author} {\bibfnamefont {R.}~\bibnamefont {Valent\'\i}},\ }\bibfield  {title}
      {\bibinfo {title} {Revision of model parameters for $\kappa{}$-type charge
      transfer salts: An \textit{Ab~Initio} study},\ }\href
      {https://doi.org/10.1103/PhysRevLett.103.067004} {\bibfield  {journal}
      {\bibinfo  {journal} {Phys. Rev. Lett.}\ }\textbf {\bibinfo {volume} {103}},\
      \bibinfo {pages} {067004} (\bibinfo {year} {2009})}\BibitemShut {NoStop}%
    \bibitem [{\citenamefont {Isono}\ \emph {et~al.}(2014)\citenamefont {Isono},
      \citenamefont {Kamo}, \citenamefont {Ueda}, \citenamefont {Takahashi},
      \citenamefont {Kimata}, \citenamefont {Tajima}, \citenamefont {Tsuchiya},
      \citenamefont {Terashima}, \citenamefont {Uji},\ and\ \citenamefont
      {Mori}}]{Isono:2014}%
      \BibitemOpen
      \bibfield  {author} {\bibinfo {author} {\bibfnamefont {T.}~\bibnamefont
      {Isono}}, \bibinfo {author} {\bibfnamefont {H.}~\bibnamefont {Kamo}},
      \bibinfo {author} {\bibfnamefont {A.}~\bibnamefont {Ueda}}, \bibinfo {author}
      {\bibfnamefont {K.}~\bibnamefont {Takahashi}}, \bibinfo {author}
      {\bibfnamefont {M.}~\bibnamefont {Kimata}}, \bibinfo {author} {\bibfnamefont
      {H.}~\bibnamefont {Tajima}}, \bibinfo {author} {\bibfnamefont
      {S.}~\bibnamefont {Tsuchiya}}, \bibinfo {author} {\bibfnamefont
      {T.}~\bibnamefont {Terashima}}, \bibinfo {author} {\bibfnamefont
      {S.}~\bibnamefont {Uji}},\ and\ \bibinfo {author} {\bibfnamefont
      {H.}~\bibnamefont {Mori}},\ }\bibfield  {title} {\bibinfo {title} {Gapless
      quantum spin liquid in an organic spin-1/2 triangular-lattice
      $\ensuremath{\kappa}\ensuremath{-}{\mathrm{h}}_{3}(\mathbf{\text{cat-edt-ttf}}{)}_{2}$},\
      }\href {https://doi.org/10.1103/PhysRevLett.112.177201} {\bibfield  {journal}
      {\bibinfo  {journal} {Phys. Rev. Lett.}\ }\textbf {\bibinfo {volume} {112}},\
      \bibinfo {pages} {177201} (\bibinfo {year} {2014})}\BibitemShut {NoStop}%
    \bibitem [{\citenamefont {Lee}\ \emph {et~al.}(2007)\citenamefont {Lee},
      \citenamefont {Kune\ifmmode~\check{s}\else \v{s}\fi{}}, \citenamefont
      {Scalettar},\ and\ \citenamefont {Pickett}}]{Lee_LiNbO:2007}%
      \BibitemOpen
      \bibfield  {author} {\bibinfo {author} {\bibfnamefont {K.-W.}\ \bibnamefont
      {Lee}}, \bibinfo {author} {\bibfnamefont {J.}~\bibnamefont
      {Kune\ifmmode~\check{s}\else \v{s}\fi{}}}, \bibinfo {author} {\bibfnamefont
      {R.~T.}\ \bibnamefont {Scalettar}},\ and\ \bibinfo {author} {\bibfnamefont
      {W.~E.}\ \bibnamefont {Pickett}},\ }\bibfield  {title} {\bibinfo {title}
      {Correlation effects in the triangular lattice single-band system
      {Li$_x$NbO$_2$}},\ }\href {https://doi.org/10.1103/PhysRevB.76.144513}
      {\bibfield  {journal} {\bibinfo  {journal} {Phys. Rev. B}\ }\textbf {\bibinfo
      {volume} {76}},\ \bibinfo {pages} {144513} (\bibinfo {year}
      {2007})}\BibitemShut {NoStop}%
    \bibitem [{\citenamefont {Soma}\ \emph {et~al.}(2020)\citenamefont {Soma},
      \citenamefont {Yoshimatsu},\ and\ \citenamefont
      {Ohtomo}}]{Soma_Yoshimatsu_Ohtomo_2020}%
      \BibitemOpen
      \bibfield  {author} {\bibinfo {author} {\bibfnamefont {T.}~\bibnamefont
      {Soma}}, \bibinfo {author} {\bibfnamefont {K.}~\bibnamefont {Yoshimatsu}},\
      and\ \bibinfo {author} {\bibfnamefont {A.}~\bibnamefont {Ohtomo}},\
      }\bibfield  {title} {\bibinfo {title} {p-type transparent superconductivity
      in a layered oxide},\ }\href {https://doi.org/10.1126/sciadv.abb8570}
      {\bibfield  {journal} {\bibinfo  {journal} {Science Advances}\ }\textbf
      {\bibinfo {volume} {6}},\ \bibinfo {pages} {eabb8570} (\bibinfo {year}
      {2020})}\BibitemShut {NoStop}%
    \bibitem [{\citenamefont {Rawl}\ \emph {et~al.}(2017)\citenamefont {Rawl},
      \citenamefont {Ge}, \citenamefont {Agrawal}, \citenamefont {Kamiya},
      \citenamefont {Dela~Cruz}, \citenamefont {Butch}, \citenamefont {Sun},
      \citenamefont {Lee}, \citenamefont {Choi}, \citenamefont {Oitmaa},
      \citenamefont {Batista}, \citenamefont {Mourigal}, \citenamefont {Zhou},\
      and\ \citenamefont {Ma}}]{Rawl_BaCoNbO:2017}%
      \BibitemOpen
      \bibfield  {author} {\bibinfo {author} {\bibfnamefont {R.}~\bibnamefont
      {Rawl}}, \bibinfo {author} {\bibfnamefont {L.}~\bibnamefont {Ge}}, \bibinfo
      {author} {\bibfnamefont {H.}~\bibnamefont {Agrawal}}, \bibinfo {author}
      {\bibfnamefont {Y.}~\bibnamefont {Kamiya}}, \bibinfo {author} {\bibfnamefont
      {C.~R.}\ \bibnamefont {Dela~Cruz}}, \bibinfo {author} {\bibfnamefont {N.~P.}\
      \bibnamefont {Butch}}, \bibinfo {author} {\bibfnamefont {X.~F.}\ \bibnamefont
      {Sun}}, \bibinfo {author} {\bibfnamefont {M.}~\bibnamefont {Lee}}, \bibinfo
      {author} {\bibfnamefont {E.~S.}\ \bibnamefont {Choi}}, \bibinfo {author}
      {\bibfnamefont {J.}~\bibnamefont {Oitmaa}}, \bibinfo {author} {\bibfnamefont
      {C.~D.}\ \bibnamefont {Batista}}, \bibinfo {author} {\bibfnamefont
      {M.}~\bibnamefont {Mourigal}}, \bibinfo {author} {\bibfnamefont {H.~D.}\
      \bibnamefont {Zhou}},\ and\ \bibinfo {author} {\bibfnamefont
      {J.}~\bibnamefont {Ma}},\ }\bibfield  {title} {\bibinfo {title}
      {${\mathrm{ba}}_{8}{\mathrm{conb}}_{6}{\mathrm{o}}_{24}$: A
      spin-$\frac{1}{2}$ triangular-lattice heisenberg antiferromagnet in the
      two-dimensional limit},\ }\href {https://doi.org/10.1103/PhysRevB.95.060412}
      {\bibfield  {journal} {\bibinfo  {journal} {Phys. Rev. B}\ }\textbf {\bibinfo
      {volume} {95}},\ \bibinfo {pages} {060412} (\bibinfo {year}
      {2017})}\BibitemShut {NoStop}%
    \bibitem [{\citenamefont {Guratinder}\ \emph {et~al.}(2021)\citenamefont
      {Guratinder}, \citenamefont {Schmidt}, \citenamefont {Walker}, \citenamefont
      {Bewley}, \citenamefont {W\"orle}, \citenamefont {Cabra}, \citenamefont
      {Osorio}, \citenamefont {Villalba}, \citenamefont {Madsen}, \citenamefont
      {Keller}, \citenamefont {Wildes}, \citenamefont {Puphal}, \citenamefont
      {Cervellino}, \citenamefont {R\"uegg},\ and\ \citenamefont
      {Zaharko}}]{Guratinder_FeGaS}%
      \BibitemOpen
      \bibfield  {author} {\bibinfo {author} {\bibfnamefont {K.}~\bibnamefont
      {Guratinder}}, \bibinfo {author} {\bibfnamefont {M.}~\bibnamefont {Schmidt}},
      \bibinfo {author} {\bibfnamefont {H.~C.}\ \bibnamefont {Walker}}, \bibinfo
      {author} {\bibfnamefont {R.}~\bibnamefont {Bewley}}, \bibinfo {author}
      {\bibfnamefont {M.}~\bibnamefont {W\"orle}}, \bibinfo {author} {\bibfnamefont
      {D.}~\bibnamefont {Cabra}}, \bibinfo {author} {\bibfnamefont {S.~A.}\
      \bibnamefont {Osorio}}, \bibinfo {author} {\bibfnamefont {M.}~\bibnamefont
      {Villalba}}, \bibinfo {author} {\bibfnamefont {A.~K.}\ \bibnamefont
      {Madsen}}, \bibinfo {author} {\bibfnamefont {L.}~\bibnamefont {Keller}},
      \bibinfo {author} {\bibfnamefont {A.}~\bibnamefont {Wildes}}, \bibinfo
      {author} {\bibfnamefont {P.}~\bibnamefont {Puphal}}, \bibinfo {author}
      {\bibfnamefont {A.}~\bibnamefont {Cervellino}}, \bibinfo {author}
      {\bibfnamefont {C.}~\bibnamefont {R\"uegg}},\ and\ \bibinfo {author}
      {\bibfnamefont {O.}~\bibnamefont {Zaharko}},\ }\bibfield  {title} {\bibinfo
      {title} {Magnetic correlations in the triangular antiferromagnet
      {FeGa$_2$S$_4$}},\ }\href {https://doi.org/10.1103/PhysRevB.104.064412}
      {\bibfield  {journal} {\bibinfo  {journal} {Phys. Rev. B}\ }\textbf {\bibinfo
      {volume} {104}},\ \bibinfo {pages} {064412} (\bibinfo {year}
      {2021})}\BibitemShut {NoStop}%
    \bibitem [{\citenamefont {Foo}\ \emph {et~al.}(2004)\citenamefont {Foo},
      \citenamefont {Wang}, \citenamefont {Watauchi}, \citenamefont {Zandbergen},
      \citenamefont {He}, \citenamefont {Cava},\ and\ \citenamefont
      {Ong}}]{Cobaltates:2004}%
      \BibitemOpen
      \bibfield  {author} {\bibinfo {author} {\bibfnamefont {M.~L.}\ \bibnamefont
      {Foo}}, \bibinfo {author} {\bibfnamefont {Y.}~\bibnamefont {Wang}}, \bibinfo
      {author} {\bibfnamefont {S.}~\bibnamefont {Watauchi}}, \bibinfo {author}
      {\bibfnamefont {H.~W.}\ \bibnamefont {Zandbergen}}, \bibinfo {author}
      {\bibfnamefont {T.}~\bibnamefont {He}}, \bibinfo {author} {\bibfnamefont
      {R.~J.}\ \bibnamefont {Cava}},\ and\ \bibinfo {author} {\bibfnamefont
      {N.~P.}\ \bibnamefont {Ong}},\ }\bibfield  {title} {\bibinfo {title} {Charge
      ordering, commensurability, and metallicity in the phase diagram of the
      layered {Na}$_x${CoO}$_2$},\ }\href
      {https://doi.org/10.1103/PhysRevLett.92.247001} {\bibfield  {journal}
      {\bibinfo  {journal} {Phys. Rev. Lett.}\ }\textbf {\bibinfo {volume} {92}},\
      \bibinfo {pages} {247001} (\bibinfo {year} {2004})}\BibitemShut {NoStop}%
    \bibitem [{\citenamefont {Ming}\ \emph {et~al.}(2017)\citenamefont {Ming},
      \citenamefont {Johnston}, \citenamefont {Mulugeta}, \citenamefont {Smith},
      \citenamefont {Vilmercati}, \citenamefont {Lee}, \citenamefont {Maier},
      \citenamefont {Snijders},\ and\ \citenamefont
      {Weitering}}]{Ming_Johnston_Sn_Si:2017}%
      \BibitemOpen
      \bibfield  {author} {\bibinfo {author} {\bibfnamefont {F.}~\bibnamefont
      {Ming}}, \bibinfo {author} {\bibfnamefont {S.}~\bibnamefont {Johnston}},
      \bibinfo {author} {\bibfnamefont {D.}~\bibnamefont {Mulugeta}}, \bibinfo
      {author} {\bibfnamefont {T.~S.}\ \bibnamefont {Smith}}, \bibinfo {author}
      {\bibfnamefont {P.}~\bibnamefont {Vilmercati}}, \bibinfo {author}
      {\bibfnamefont {G.}~\bibnamefont {Lee}}, \bibinfo {author} {\bibfnamefont
      {T.~A.}\ \bibnamefont {Maier}}, \bibinfo {author} {\bibfnamefont {P.~C.}\
      \bibnamefont {Snijders}},\ and\ \bibinfo {author} {\bibfnamefont {H.~H.}\
      \bibnamefont {Weitering}},\ }\bibfield  {title} {\bibinfo {title}
      {Realization of a hole-doped mott insulator on a triangular silicon
      lattice},\ }\href {https://doi.org/10.1103/PhysRevLett.119.266802} {\bibfield
       {journal} {\bibinfo  {journal} {Phys. Rev. Lett.}\ }\textbf {\bibinfo
      {volume} {119}},\ \bibinfo {pages} {266802} (\bibinfo {year}
      {2017})}\BibitemShut {NoStop}%
    \bibitem [{\citenamefont {Yang}\ \emph {et~al.}(2021)\citenamefont {Yang},
      \citenamefont {Liu}, \citenamefont {Mongkolkiattichai},\ and\ \citenamefont
      {Schauss}}]{Yang_Liu_Mongkolkiattichai_Schauss_2021}%
      \BibitemOpen
      \bibfield  {author} {\bibinfo {author} {\bibfnamefont {J.}~\bibnamefont
      {Yang}}, \bibinfo {author} {\bibfnamefont {L.}~\bibnamefont {Liu}}, \bibinfo
      {author} {\bibfnamefont {J.}~\bibnamefont {Mongkolkiattichai}},\ and\
      \bibinfo {author} {\bibfnamefont {P.}~\bibnamefont {Schauss}},\ }\bibfield
      {title} {\bibinfo {title} {Site-resolved imaging of ultracold fermions in a
      triangular-lattice quantum gas microscope},\ }\href
      {https://doi.org/10.1103/PRXQuantum.2.020344} {\bibfield  {journal} {\bibinfo
       {journal} {PRX Quantum}\ }\textbf {\bibinfo {volume} {2}},\ \bibinfo {pages}
      {020344} (\bibinfo {year} {2021})}\BibitemShut {NoStop}%
    \bibitem [{\citenamefont {Cao}\ \emph {et~al.}(2018{\natexlab{a}})\citenamefont
      {Cao}, \citenamefont {Fatemi}, \citenamefont {Fang}, \citenamefont
      {Watanabe}, \citenamefont {Taniguchi}, \citenamefont {Kaxiras},\ and\
      \citenamefont {Jarillo-Herrero}}]{cao_unconventional_2018}%
      \BibitemOpen
      \bibfield  {author} {\bibinfo {author} {\bibfnamefont {Y.}~\bibnamefont
      {Cao}}, \bibinfo {author} {\bibfnamefont {V.}~\bibnamefont {Fatemi}},
      \bibinfo {author} {\bibfnamefont {S.}~\bibnamefont {Fang}}, \bibinfo {author}
      {\bibfnamefont {K.}~\bibnamefont {Watanabe}}, \bibinfo {author}
      {\bibfnamefont {T.}~\bibnamefont {Taniguchi}}, \bibinfo {author}
      {\bibfnamefont {E.}~\bibnamefont {Kaxiras}},\ and\ \bibinfo {author}
      {\bibfnamefont {P.}~\bibnamefont {Jarillo-Herrero}},\ }\bibfield  {title}
      {\bibinfo {title} {Unconventional superconductivity in magic-angle graphene
      superlattices},\ }\href {https://doi.org/10.1038/nature26160} {\bibfield
      {journal} {\bibinfo  {journal} {Nature}\ }\textbf {\bibinfo {volume} {556}},\
      \bibinfo {pages} {43–50} (\bibinfo {year}
      {2018}{\natexlab{a}})}\BibitemShut {NoStop}%
    \bibitem [{\citenamefont {Cao}\ \emph {et~al.}(2018{\natexlab{b}})\citenamefont
      {Cao}, \citenamefont {Fatemi}, \citenamefont {Demir}, \citenamefont {Fang},
      \citenamefont {Tomarken}, \citenamefont {Luo}, \citenamefont
      {Sanchez-Yamagishi}, \citenamefont {Watanabe}, \citenamefont {Taniguchi},
      \citenamefont {Kaxiras}, \citenamefont {Ashoori},\ and\ \citenamefont
      {Jarillo-Herrero}}]{cao_correlated_2018}%
      \BibitemOpen
      \bibfield  {author} {\bibinfo {author} {\bibfnamefont {Y.}~\bibnamefont
      {Cao}}, \bibinfo {author} {\bibfnamefont {V.}~\bibnamefont {Fatemi}},
      \bibinfo {author} {\bibfnamefont {A.}~\bibnamefont {Demir}}, \bibinfo
      {author} {\bibfnamefont {S.}~\bibnamefont {Fang}}, \bibinfo {author}
      {\bibfnamefont {S.~L.}\ \bibnamefont {Tomarken}}, \bibinfo {author}
      {\bibfnamefont {J.~Y.}\ \bibnamefont {Luo}}, \bibinfo {author} {\bibfnamefont
      {J.~D.}\ \bibnamefont {Sanchez-Yamagishi}}, \bibinfo {author} {\bibfnamefont
      {K.}~\bibnamefont {Watanabe}}, \bibinfo {author} {\bibfnamefont
      {T.}~\bibnamefont {Taniguchi}}, \bibinfo {author} {\bibfnamefont
      {E.}~\bibnamefont {Kaxiras}}, \bibinfo {author} {\bibfnamefont {R.~C.}\
      \bibnamefont {Ashoori}},\ and\ \bibinfo {author} {\bibfnamefont
      {P.}~\bibnamefont {Jarillo-Herrero}},\ }\bibfield  {title} {\bibinfo {title}
      {Correlated insulator behaviour at half-filling in magic-angle graphene
      superlattices},\ }\href {https://doi.org/10.1038/nature26154} {\bibfield
      {journal} {\bibinfo  {journal} {Nature}\ }\textbf {\bibinfo {volume} {556}},\
      \bibinfo {pages} {80–84} (\bibinfo {year}
      {2018}{\natexlab{b}})}\BibitemShut {NoStop}%
    \bibitem [{\citenamefont {Yankowitz}\ \emph {et~al.}(2019)\citenamefont
      {Yankowitz}, \citenamefont {Chen}, \citenamefont {Polshyn}, \citenamefont
      {Zhang}, \citenamefont {Watanabe}, \citenamefont {Taniguchi}, \citenamefont
      {Graf}, \citenamefont {Young},\ and\ \citenamefont
      {Dean}}]{yankowitz_tuning_2019}%
      \BibitemOpen
      \bibfield  {author} {\bibinfo {author} {\bibfnamefont {M.}~\bibnamefont
      {Yankowitz}}, \bibinfo {author} {\bibfnamefont {S.}~\bibnamefont {Chen}},
      \bibinfo {author} {\bibfnamefont {H.}~\bibnamefont {Polshyn}}, \bibinfo
      {author} {\bibfnamefont {Y.}~\bibnamefont {Zhang}}, \bibinfo {author}
      {\bibfnamefont {K.}~\bibnamefont {Watanabe}}, \bibinfo {author}
      {\bibfnamefont {T.}~\bibnamefont {Taniguchi}}, \bibinfo {author}
      {\bibfnamefont {D.}~\bibnamefont {Graf}}, \bibinfo {author} {\bibfnamefont
      {A.~F.}\ \bibnamefont {Young}},\ and\ \bibinfo {author} {\bibfnamefont
      {C.~R.}\ \bibnamefont {Dean}},\ }\bibfield  {title} {\bibinfo {title} {Tuning
      superconductivity in twisted bilayer graphene},\ }\href
      {https://doi.org/10.1126/science.aav1910} {\bibfield  {journal} {\bibinfo
      {journal} {Science}\ }\textbf {\bibinfo {volume} {363}},\ \bibinfo {pages}
      {1059} (\bibinfo {year} {2019})}\BibitemShut {NoStop}%
    \bibitem [{\citenamefont {Wu}\ \emph {et~al.}(2018)\citenamefont {Wu},
      \citenamefont {Lovorn}, \citenamefont {Tutuc},\ and\ \citenamefont
      {MacDonald}}]{wu2018hubbard}%
      \BibitemOpen
      \bibfield  {author} {\bibinfo {author} {\bibfnamefont {F.}~\bibnamefont
      {Wu}}, \bibinfo {author} {\bibfnamefont {T.}~\bibnamefont {Lovorn}}, \bibinfo
      {author} {\bibfnamefont {E.}~\bibnamefont {Tutuc}},\ and\ \bibinfo {author}
      {\bibfnamefont {A.~H.}\ \bibnamefont {MacDonald}},\ }\bibfield  {title}
      {\bibinfo {title} {Hubbard {Model} {Physics} in {Transition} {Metal}
      {Dichalcogenide} {Moir\'e} {Bands}},\ }\href
      {https://doi.org/10.1103/PhysRevLett.121.026402} {\bibfield  {journal}
      {\bibinfo  {journal} {Phys. Rev. Lett.}\ }\textbf {\bibinfo {volume} {121}},\
      \bibinfo {pages} {026402} (\bibinfo {year} {2018})}\BibitemShut {NoStop}%
    \bibitem [{\citenamefont {Wu}\ \emph {et~al.}(2019)\citenamefont {Wu},
      \citenamefont {Lovorn}, \citenamefont {Tutuc}, \citenamefont {Martin},\ and\
      \citenamefont {MacDonald}}]{wu2019topological}%
      \BibitemOpen
      \bibfield  {author} {\bibinfo {author} {\bibfnamefont {F.}~\bibnamefont
      {Wu}}, \bibinfo {author} {\bibfnamefont {T.}~\bibnamefont {Lovorn}}, \bibinfo
      {author} {\bibfnamefont {E.}~\bibnamefont {Tutuc}}, \bibinfo {author}
      {\bibfnamefont {I.}~\bibnamefont {Martin}},\ and\ \bibinfo {author}
      {\bibfnamefont {A.~H.}\ \bibnamefont {MacDonald}},\ }\bibfield  {title}
      {\bibinfo {title} {Topological {Insulators} in {Twisted} {Transition} {Metal}
      {Dichalcogenide} {Homobilayers}},\ }\href
      {https://doi.org/10.1103/PhysRevLett.122.086402} {\bibfield  {journal}
      {\bibinfo  {journal} {Phys. Rev. Lett.}\ }\textbf {\bibinfo {volume} {122}},\
      \bibinfo {pages} {086402} (\bibinfo {year} {2019})}\BibitemShut {NoStop}%
    \bibitem [{\citenamefont {Zang}\ \emph {et~al.}(2022)\citenamefont {Zang},
      \citenamefont {Wang}, \citenamefont {Cano}, \citenamefont {Georges},\ and\
      \citenamefont {Millis}}]{Zang_Wang_Cano_Georges_Millis_2022}%
      \BibitemOpen
      \bibfield  {author} {\bibinfo {author} {\bibfnamefont {J.}~\bibnamefont
      {Zang}}, \bibinfo {author} {\bibfnamefont {J.}~\bibnamefont {Wang}}, \bibinfo
      {author} {\bibfnamefont {J.}~\bibnamefont {Cano}}, \bibinfo {author}
      {\bibfnamefont {A.}~\bibnamefont {Georges}},\ and\ \bibinfo {author}
      {\bibfnamefont {A.~J.}\ \bibnamefont {Millis}},\ }\bibfield  {title}
      {\bibinfo {title} {Dynamical mean-field theory of moiré bilayer transition
      metal dichalcogenides: Phase diagram, resistivity, and quantum criticality},\
      }\href {https://doi.org/10.1103/PhysRevX.12.021064} {\bibfield  {journal}
      {\bibinfo  {journal} {Physical Review X}\ }\textbf {\bibinfo {volume} {12}},\
      \bibinfo {pages} {021064} (\bibinfo {year} {2022})}\BibitemShut {NoStop}%
    \bibitem [{\citenamefont {Morita}\ \emph {et~al.}(2002)\citenamefont {Morita},
      \citenamefont {Watanabe},\ and\ \citenamefont
      {Imada}}]{morita_nonmagnetic_2002}%
      \BibitemOpen
      \bibfield  {author} {\bibinfo {author} {\bibfnamefont {H.}~\bibnamefont
      {Morita}}, \bibinfo {author} {\bibfnamefont {S.}~\bibnamefont {Watanabe}},\
      and\ \bibinfo {author} {\bibfnamefont {M.}~\bibnamefont {Imada}},\ }\bibfield
       {title} {\bibinfo {title} {Nonmagnetic {Insulating} {States} near the {Mott}
      {Transitions} on {Lattices} with {Geometrical} {Frustration} and
      {Implications} for $\kappa$-{(ET)$_2$Cu$_2$(CN)$_3$}},\ }\href
      {https://doi.org/10.1143/JPSJ.71.2109} {\bibfield  {journal} {\bibinfo
      {journal} {Journal of the Physical Society of Japan}\ }\textbf {\bibinfo
      {volume} {71}},\ \bibinfo {pages} {2109} (\bibinfo {year}
      {2002})}\BibitemShut {NoStop}%
    \bibitem [{\citenamefont {Sahebsara}\ and\ \citenamefont
      {S\'en\'echal}(2008)}]{sahebsara_hubbard_2008}%
      \BibitemOpen
      \bibfield  {author} {\bibinfo {author} {\bibfnamefont {P.}~\bibnamefont
      {Sahebsara}}\ and\ \bibinfo {author} {\bibfnamefont {D.}~\bibnamefont
      {S\'en\'echal}},\ }\bibfield  {title} {\bibinfo {title} {Hubbard {Model} on
      the {Triangular} {Lattice}: {Spiral} {Order} and {Spin} {Liquid}},\ }\href
      {https://doi.org/10.1103/PhysRevLett.100.136402} {\bibfield  {journal}
      {\bibinfo  {journal} {Phys. Rev. Lett.}\ }\textbf {\bibinfo {volume} {100}},\
      \bibinfo {pages} {136402} (\bibinfo {year} {2008})}\BibitemShut {NoStop}%
    \bibitem [{\citenamefont {Laubach}\ \emph {et~al.}(2015)\citenamefont
      {Laubach}, \citenamefont {Thomale}, \citenamefont {Platt}, \citenamefont
      {Hanke},\ and\ \citenamefont {Li}}]{laubachPhaseDiagramHubbard2015a}%
      \BibitemOpen
      \bibfield  {author} {\bibinfo {author} {\bibfnamefont {M.}~\bibnamefont
      {Laubach}}, \bibinfo {author} {\bibfnamefont {R.}~\bibnamefont {Thomale}},
      \bibinfo {author} {\bibfnamefont {C.}~\bibnamefont {Platt}}, \bibinfo
      {author} {\bibfnamefont {W.}~\bibnamefont {Hanke}},\ and\ \bibinfo {author}
      {\bibfnamefont {G.}~\bibnamefont {Li}},\ }\bibfield  {title} {\bibinfo
      {title} {Phase diagram of the {{Hubbard}} model on the anisotropic triangular
      lattice},\ }\href {https://doi.org/10.1103/PhysRevB.91.245125} {\bibfield
      {journal} {\bibinfo  {journal} {Physical Review B}\ }\textbf {\bibinfo
      {volume} {91}},\ \bibinfo {pages} {245125} (\bibinfo {year}
      {2015})}\BibitemShut {NoStop}%
    \bibitem [{\citenamefont {Misumi}\ \emph {et~al.}(2017)\citenamefont {Misumi},
      \citenamefont {Kaneko},\ and\ \citenamefont
      {Ohta}}]{Misumi_Mott_triangular:2017}%
      \BibitemOpen
      \bibfield  {author} {\bibinfo {author} {\bibfnamefont {K.}~\bibnamefont
      {Misumi}}, \bibinfo {author} {\bibfnamefont {T.}~\bibnamefont {Kaneko}},\
      and\ \bibinfo {author} {\bibfnamefont {Y.}~\bibnamefont {Ohta}},\ }\bibfield
      {title} {\bibinfo {title} {Mott transition and magnetism of the
      triangular-lattice hubbard model with next-nearest-neighbor hopping},\ }\href
      {https://doi.org/10.1103/PhysRevB.95.075124} {\bibfield  {journal} {\bibinfo
      {journal} {Phys. Rev. B}\ }\textbf {\bibinfo {volume} {95}},\ \bibinfo
      {pages} {075124} (\bibinfo {year} {2017})}\BibitemShut {NoStop}%
    \bibitem [{\citenamefont {Tocchio}\ \emph {et~al.}(2008)\citenamefont
      {Tocchio}, \citenamefont {Becca}, \citenamefont {Parola},\ and\ \citenamefont
      {Sorella}}]{Tocchio_backflow_Mott:2008}%
      \BibitemOpen
      \bibfield  {author} {\bibinfo {author} {\bibfnamefont {L.~F.}\ \bibnamefont
      {Tocchio}}, \bibinfo {author} {\bibfnamefont {F.}~\bibnamefont {Becca}},
      \bibinfo {author} {\bibfnamefont {A.}~\bibnamefont {Parola}},\ and\ \bibinfo
      {author} {\bibfnamefont {S.}~\bibnamefont {Sorella}},\ }\bibfield  {title}
      {\bibinfo {title} {Role of backflow correlations for the nonmagnetic phase of
      the $t\text{--}{t}^{\ensuremath{'}}$ hubbard model},\ }\href
      {https://doi.org/10.1103/PhysRevB.78.041101} {\bibfield  {journal} {\bibinfo
      {journal} {Phys. Rev. B}\ }\textbf {\bibinfo {volume} {78}},\ \bibinfo
      {pages} {041101} (\bibinfo {year} {2008})}\BibitemShut {NoStop}%
    \bibitem [{\citenamefont {Yoshioka}\ \emph {et~al.}(2009)\citenamefont
      {Yoshioka}, \citenamefont {Koga},\ and\ \citenamefont
      {Kawakami}}]{Yoshioka_triangular:2009}%
      \BibitemOpen
      \bibfield  {author} {\bibinfo {author} {\bibfnamefont {T.}~\bibnamefont
      {Yoshioka}}, \bibinfo {author} {\bibfnamefont {A.}~\bibnamefont {Koga}},\
      and\ \bibinfo {author} {\bibfnamefont {N.}~\bibnamefont {Kawakami}},\
      }\bibfield  {title} {\bibinfo {title} {Quantum phase transitions in the
      hubbard model on a triangular lattice},\ }\href
      {https://doi.org/10.1103/PhysRevLett.103.036401} {\bibfield  {journal}
      {\bibinfo  {journal} {Phys. Rev. Lett.}\ }\textbf {\bibinfo {volume} {103}},\
      \bibinfo {pages} {036401} (\bibinfo {year} {2009})}\BibitemShut {NoStop}%
    \bibitem [{\citenamefont {Yang}\ \emph {et~al.}(2010)\citenamefont {Yang},
      \citenamefont {L\"auchli}, \citenamefont {Mila},\ and\ \citenamefont
      {Schmidt}}]{yang_effective_2010}%
      \BibitemOpen
      \bibfield  {author} {\bibinfo {author} {\bibfnamefont {H.-Y.}\ \bibnamefont
      {Yang}}, \bibinfo {author} {\bibfnamefont {A.~M.}\ \bibnamefont {L\"auchli}},
      \bibinfo {author} {\bibfnamefont {F.}~\bibnamefont {Mila}},\ and\ \bibinfo
      {author} {\bibfnamefont {K.~P.}\ \bibnamefont {Schmidt}},\ }\bibfield
      {title} {\bibinfo {title} {Effective {Spin} {Model} for the {Spin-Liquid}
      {Phase} of the {Hubbard} {Model} on the {Triangular} {Lattice}},\ }\href
      {https://doi.org/10.1103/PhysRevLett.105.267204} {\bibfield  {journal}
      {\bibinfo  {journal} {Phys. Rev. Lett.}\ }\textbf {\bibinfo {volume} {105}},\
      \bibinfo {pages} {267204} (\bibinfo {year} {2010})}\BibitemShut {NoStop}%
    \bibitem [{\citenamefont {Szasz}\ \emph {et~al.}(2020)\citenamefont {Szasz},
      \citenamefont {Motruk}, \citenamefont {Zaletel},\ and\ \citenamefont
      {Moore}}]{szasz_chiral_2020}%
      \BibitemOpen
      \bibfield  {author} {\bibinfo {author} {\bibfnamefont {A.}~\bibnamefont
      {Szasz}}, \bibinfo {author} {\bibfnamefont {J.}~\bibnamefont {Motruk}},
      \bibinfo {author} {\bibfnamefont {M.~P.}\ \bibnamefont {Zaletel}},\ and\
      \bibinfo {author} {\bibfnamefont {J.~E.}\ \bibnamefont {Moore}},\ }\bibfield
      {title} {\bibinfo {title} {Chiral {Spin} {Liquid} {Phase} of the {Triangular}
      {Lattice} {Hubbard} {Model}: {A} {Density} {Matrix} {Renormalization} {Group}
      {Study}},\ }\href {https://doi.org/10.1103/PhysRevX.10.021042} {\bibfield
      {journal} {\bibinfo  {journal} {Phys. Rev. X}\ }\textbf {\bibinfo {volume}
      {10}},\ \bibinfo {pages} {021042} (\bibinfo {year} {2020})}\BibitemShut
      {NoStop}%
    \bibitem [{\citenamefont {Chen}\ \emph {et~al.}(2022)\citenamefont {Chen},
      \citenamefont {Chen}, \citenamefont {Gong}, \citenamefont {Sheng},
      \citenamefont {Li},\ and\ \citenamefont
      {Weichselbaum}}]{chenQuantumSpinLiquid2022}%
      \BibitemOpen
      \bibfield  {author} {\bibinfo {author} {\bibfnamefont {B.-B.}\ \bibnamefont
      {Chen}}, \bibinfo {author} {\bibfnamefont {Z.}~\bibnamefont {Chen}}, \bibinfo
      {author} {\bibfnamefont {S.-S.}\ \bibnamefont {Gong}}, \bibinfo {author}
      {\bibfnamefont {D.~N.}\ \bibnamefont {Sheng}}, \bibinfo {author}
      {\bibfnamefont {W.}~\bibnamefont {Li}},\ and\ \bibinfo {author}
      {\bibfnamefont {A.}~\bibnamefont {Weichselbaum}},\ }\bibfield  {title}
      {\bibinfo {title} {Quantum {Spin} {Liquid} with {Emergent} {Chiral} {Order}
      in the {Triangular}-{Lattice} {Hubbard} {Model}},\ }\href
      {https://doi.org/10.1103/PhysRevB.106.094420} {\bibfield  {journal} {\bibinfo
       {journal} {Phys. Rev. B}\ }\textbf {\bibinfo {volume} {106}},\ \bibinfo
      {pages} {094420} (\bibinfo {year} {2022})}\BibitemShut {NoStop}%
    \bibitem [{\citenamefont {Wietek}\ \emph {et~al.}(2021)\citenamefont {Wietek},
      \citenamefont {Rossi}, \citenamefont {\ifmmode~\check{S}\else
      \v{S}\fi{}imkovic}, \citenamefont {Klett}, \citenamefont {Hansmann},
      \citenamefont {Ferrero}, \citenamefont {Stoudenmire}, \citenamefont
      {Sch\"afer},\ and\ \citenamefont {Georges}}]{wietek_mott_2021}%
      \BibitemOpen
      \bibfield  {author} {\bibinfo {author} {\bibfnamefont {A.}~\bibnamefont
      {Wietek}}, \bibinfo {author} {\bibfnamefont {R.}~\bibnamefont {Rossi}},
      \bibinfo {author} {\bibfnamefont {F.}~\bibnamefont {\ifmmode~\check{S}\else
      \v{S}\fi{}imkovic}}, \bibinfo {author} {\bibfnamefont {M.}~\bibnamefont
      {Klett}}, \bibinfo {author} {\bibfnamefont {P.}~\bibnamefont {Hansmann}},
      \bibinfo {author} {\bibfnamefont {M.}~\bibnamefont {Ferrero}}, \bibinfo
      {author} {\bibfnamefont {E.~M.}\ \bibnamefont {Stoudenmire}}, \bibinfo
      {author} {\bibfnamefont {T.}~\bibnamefont {Sch\"afer}},\ and\ \bibinfo
      {author} {\bibfnamefont {A.}~\bibnamefont {Georges}},\ }\bibfield  {title}
      {\bibinfo {title} {Mott {Insulating} {States} with {Competing} {Orders} in
      the {Triangular} {Lattice} {Hubbard} {Model}},\ }\href
      {https://doi.org/10.1103/PhysRevX.11.041013} {\bibfield  {journal} {\bibinfo
      {journal} {Phys. Rev. X}\ }\textbf {\bibinfo {volume} {11}},\ \bibinfo
      {pages} {041013} (\bibinfo {year} {2021})}\BibitemShut {NoStop}%
    \bibitem [{\citenamefont {Tocchio}\ \emph {et~al.}(2020)\citenamefont
      {Tocchio}, \citenamefont {Montorsi},\ and\ \citenamefont
      {Becca}}]{Tocchio_Montorsi_Becca_2020}%
      \BibitemOpen
      \bibfield  {author} {\bibinfo {author} {\bibfnamefont {L.~F.}\ \bibnamefont
      {Tocchio}}, \bibinfo {author} {\bibfnamefont {A.}~\bibnamefont {Montorsi}},\
      and\ \bibinfo {author} {\bibfnamefont {F.}~\bibnamefont {Becca}},\ }\bibfield
       {title} {\bibinfo {title} {Magnetic and spin-liquid phases in the frustrated
      $t\ensuremath{-}{t}^{\ensuremath{’}}$ hubbard model on the triangular
      lattice},\ }\href {https://doi.org/10.1103/PhysRevB.102.115150} {\bibfield
      {journal} {\bibinfo  {journal} {Physical Review B}\ }\textbf {\bibinfo
      {volume} {102}},\ \bibinfo {pages} {115150} (\bibinfo {year}
      {2020})}\BibitemShut {NoStop}%
    \bibitem [{\citenamefont {Yu}\ \emph {et~al.}(2023)\citenamefont {Yu},
      \citenamefont {Li}, \citenamefont {Iskakov},\ and\ \citenamefont
      {Gull}}]{Yu_Li_Iskakov_Gull_2023}%
      \BibitemOpen
      \bibfield  {author} {\bibinfo {author} {\bibfnamefont {Y.}~\bibnamefont
      {Yu}}, \bibinfo {author} {\bibfnamefont {S.}~\bibnamefont {Li}}, \bibinfo
      {author} {\bibfnamefont {S.}~\bibnamefont {Iskakov}},\ and\ \bibinfo {author}
      {\bibfnamefont {E.}~\bibnamefont {Gull}},\ }\bibfield  {title} {\bibinfo
      {title} {Magnetic phases of the anisotropic triangular lattice hubbard
      model},\ }\href {https://doi.org/10.1103/PhysRevB.107.075106} {\bibfield
      {journal} {\bibinfo  {journal} {Physical Review B}\ }\textbf {\bibinfo
      {volume} {107}},\ \bibinfo {pages} {075106} (\bibinfo {year}
      {2023})}\BibitemShut {NoStop}%
    \bibitem [{\citenamefont {Hettler}\ \emph {et~al.}(2000)\citenamefont
      {Hettler}, \citenamefont {Mukherjee}, \citenamefont {Jarrell},\ and\
      \citenamefont {Krishnamurthy}}]{hettler_dynamical_2000}%
      \BibitemOpen
      \bibfield  {author} {\bibinfo {author} {\bibfnamefont {M.~H.}\ \bibnamefont
      {Hettler}}, \bibinfo {author} {\bibfnamefont {M.}~\bibnamefont {Mukherjee}},
      \bibinfo {author} {\bibfnamefont {M.}~\bibnamefont {Jarrell}},\ and\ \bibinfo
      {author} {\bibfnamefont {H.~R.}\ \bibnamefont {Krishnamurthy}},\ }\bibfield
      {title} {\bibinfo {title} {Dynamical {cluster} approximation: {Nonlocal}
      dynamics of correlated electron systems},\ }\href
      {https://doi.org/10.1103/PhysRevB.61.12739} {\bibfield  {journal} {\bibinfo
      {journal} {Phys. Rev. B}\ }\textbf {\bibinfo {volume} {61}},\ \bibinfo
      {pages} {12739} (\bibinfo {year} {2000})}\BibitemShut {NoStop}%
    \bibitem [{\citenamefont {on~the Many-Electron~Problem}\ \emph
      {et~al.}(2015)\citenamefont {on~the Many-Electron~Problem}, \citenamefont
      {LeBlanc}, \citenamefont {Antipov}, \citenamefont {Becca}, \citenamefont
      {Bulik}, \citenamefont {Chan}, \citenamefont {Chung}, \citenamefont {Deng},
      \citenamefont {Ferrero}, \citenamefont {Henderson}, \citenamefont
      {Jiménez-Hoyos}, \citenamefont {Kozik}, \citenamefont {Liu}, \citenamefont
      {Millis}, \citenamefont {Prokof’ev}, \citenamefont {Qin}, \citenamefont
      {Scuseria}, \citenamefont {Shi}, \citenamefont {Svistunov}, \citenamefont
      {Tocchio}, \citenamefont {Tupitsyn}, \citenamefont {White}, \citenamefont
      {Zhang}, \citenamefont {Zheng}, \citenamefont {Zhu},\ and\ \citenamefont
      {Gull}}]{LeBlanc_2015}%
      \BibitemOpen
      \bibfield  {author} {\bibinfo {author} {\bibfnamefont {S.~C.}\ \bibnamefont
      {on~the Many-Electron~Problem}}, \bibinfo {author} {\bibfnamefont
      {J.}~\bibnamefont {LeBlanc}}, \bibinfo {author} {\bibfnamefont {A.~E.}\
      \bibnamefont {Antipov}}, \bibinfo {author} {\bibfnamefont {F.}~\bibnamefont
      {Becca}}, \bibinfo {author} {\bibfnamefont {I.~W.}\ \bibnamefont {Bulik}},
      \bibinfo {author} {\bibfnamefont {G.~K.-L.}\ \bibnamefont {Chan}}, \bibinfo
      {author} {\bibfnamefont {C.-M.}\ \bibnamefont {Chung}}, \bibinfo {author}
      {\bibfnamefont {Y.}~\bibnamefont {Deng}}, \bibinfo {author} {\bibfnamefont
      {M.}~\bibnamefont {Ferrero}}, \bibinfo {author} {\bibfnamefont {T.~M.}\
      \bibnamefont {Henderson}}, \bibinfo {author} {\bibfnamefont {C.~A.}\
      \bibnamefont {Jiménez-Hoyos}}, \bibinfo {author} {\bibfnamefont
      {E.}~\bibnamefont {Kozik}}, \bibinfo {author} {\bibfnamefont {X.-W.}\
      \bibnamefont {Liu}}, \bibinfo {author} {\bibfnamefont {A.~J.}\ \bibnamefont
      {Millis}}, \bibinfo {author} {\bibfnamefont {N.}~\bibnamefont {Prokof’ev}},
      \bibinfo {author} {\bibfnamefont {M.}~\bibnamefont {Qin}}, \bibinfo {author}
      {\bibfnamefont {G.~E.}\ \bibnamefont {Scuseria}}, \bibinfo {author}
      {\bibfnamefont {H.}~\bibnamefont {Shi}}, \bibinfo {author} {\bibfnamefont
      {B.}~\bibnamefont {Svistunov}}, \bibinfo {author} {\bibfnamefont {L.~F.}\
      \bibnamefont {Tocchio}}, \bibinfo {author} {\bibfnamefont {I.}~\bibnamefont
      {Tupitsyn}}, \bibinfo {author} {\bibfnamefont {S.~R.}\ \bibnamefont {White}},
      \bibinfo {author} {\bibfnamefont {S.}~\bibnamefont {Zhang}}, \bibinfo
      {author} {\bibfnamefont {B.-X.}\ \bibnamefont {Zheng}}, \bibinfo {author}
      {\bibfnamefont {Z.}~\bibnamefont {Zhu}},\ and\ \bibinfo {author}
      {\bibfnamefont {E.}~\bibnamefont {Gull}},\ }\bibfield  {title} {\bibinfo
      {title} {Solutions of the two-dimensional hubbard model: Benchmarks and
      results from a wide range of numerical algorithms},\ }\href
      {https://doi.org/10.1103/PhysRevX.5.041041} {\bibfield  {journal} {\bibinfo
      {journal} {Physical Review X}\ }\textbf {\bibinfo {volume} {5}},\ \bibinfo
      {pages} {041041} (\bibinfo {year} {2015})}\BibitemShut {NoStop}%
    \bibitem [{\citenamefont {Schäfer}\ \emph {et~al.}(2021)\citenamefont
      {Schäfer}, \citenamefont {Wentzell}, \citenamefont {Šimkovic},
      \citenamefont {He}, \citenamefont {Hille}, \citenamefont {Klett},
      \citenamefont {Eckhardt}, \citenamefont {Arzhang}, \citenamefont {Harkov},
      \citenamefont {Le~Régent}, \citenamefont {Kirsch}, \citenamefont {Wang},
      \citenamefont {Kim}, \citenamefont {Kozik}, \citenamefont {Stepanov},
      \citenamefont {Kauch}, \citenamefont {Andergassen}, \citenamefont {Hansmann},
      \citenamefont {Rohe}, \citenamefont {Vilk}, \citenamefont {LeBlanc},
      \citenamefont {Zhang}, \citenamefont {Tremblay}, \citenamefont {Ferrero},
      \citenamefont {Parcollet},\ and\ \citenamefont
      {Georges}}]{Schaefer_Wentzell_2021}%
      \BibitemOpen
      \bibfield  {author} {\bibinfo {author} {\bibfnamefont {T.}~\bibnamefont
      {Schäfer}}, \bibinfo {author} {\bibfnamefont {N.}~\bibnamefont {Wentzell}},
      \bibinfo {author} {\bibfnamefont {F.}~\bibnamefont {Šimkovic}}, \bibinfo
      {author} {\bibfnamefont {Y.-Y.}\ \bibnamefont {He}}, \bibinfo {author}
      {\bibfnamefont {C.}~\bibnamefont {Hille}}, \bibinfo {author} {\bibfnamefont
      {M.}~\bibnamefont {Klett}}, \bibinfo {author} {\bibfnamefont {C.~J.}\
      \bibnamefont {Eckhardt}}, \bibinfo {author} {\bibfnamefont {B.}~\bibnamefont
      {Arzhang}}, \bibinfo {author} {\bibfnamefont {V.}~\bibnamefont {Harkov}},
      \bibinfo {author} {\bibfnamefont {F.-M.}\ \bibnamefont {Le~Régent}},
      \bibinfo {author} {\bibfnamefont {A.}~\bibnamefont {Kirsch}}, \bibinfo
      {author} {\bibfnamefont {Y.}~\bibnamefont {Wang}}, \bibinfo {author}
      {\bibfnamefont {A.~J.}\ \bibnamefont {Kim}}, \bibinfo {author} {\bibfnamefont
      {E.}~\bibnamefont {Kozik}}, \bibinfo {author} {\bibfnamefont {E.~A.}\
      \bibnamefont {Stepanov}}, \bibinfo {author} {\bibfnamefont {A.}~\bibnamefont
      {Kauch}}, \bibinfo {author} {\bibfnamefont {S.}~\bibnamefont {Andergassen}},
      \bibinfo {author} {\bibfnamefont {P.}~\bibnamefont {Hansmann}}, \bibinfo
      {author} {\bibfnamefont {D.}~\bibnamefont {Rohe}}, \bibinfo {author}
      {\bibfnamefont {Y.~M.}\ \bibnamefont {Vilk}}, \bibinfo {author}
      {\bibfnamefont {J.~P.}\ \bibnamefont {LeBlanc}}, \bibinfo {author}
      {\bibfnamefont {S.}~\bibnamefont {Zhang}}, \bibinfo {author} {\bibfnamefont
      {A.-M.}\ \bibnamefont {Tremblay}}, \bibinfo {author} {\bibfnamefont
      {M.}~\bibnamefont {Ferrero}}, \bibinfo {author} {\bibfnamefont
      {O.}~\bibnamefont {Parcollet}},\ and\ \bibinfo {author} {\bibfnamefont
      {A.}~\bibnamefont {Georges}},\ }\bibfield  {title} {\bibinfo {title}
      {Tracking the footprints of spin fluctuations: A multimethod, multimessenger
      study of the two-dimensional hubbard model},\ }\href
      {https://doi.org/10.1103/PhysRevX.11.011058} {\bibfield  {journal} {\bibinfo
      {journal} {Physical Review X}\ }\textbf {\bibinfo {volume} {11}},\ \bibinfo
      {pages} {011058} (\bibinfo {year} {2021})}\BibitemShut {NoStop}%
    \bibitem [{\citenamefont {Pavarini}\ \emph {et~al.}(2015)\citenamefont
      {Pavarini}, \citenamefont {Koch},\ and\ \citenamefont
      {Coleman}}]{Pavarini_Koch_Coleman:2015}%
      \BibitemOpen
      \bibinfo {editor} {\bibfnamefont {E.}~\bibnamefont {Pavarini}}, \bibinfo
      {editor} {\bibfnamefont {E.}~\bibnamefont {Koch}},\ and\ \bibinfo {editor}
      {\bibfnamefont {P.}~\bibnamefont {Coleman}},\ eds.,\ \href
      {https://juser.fz-juelich.de/record/205123} {\emph {\bibinfo {title}
      {{M}any-{B}ody {P}hysics: {F}rom {K}ondo to {H}ubbard}}},\ \bibinfo {series}
      {Schriften des Forschungszentrums Jülich. Reihe modeling and simulation},
      Vol.~\bibinfo {volume} {5},\ \bibinfo {organization} {Autumn School on
      Correlated Electrons, Jülich (Germany), 21 Sep 2015 - 25 Sep 2015}\
      (\bibinfo  {publisher} {Forschungszentrum Jülich GmbH Zentralbibliothek,
      Verlag},\ \bibinfo {address} {Jülich},\ \bibinfo {year} {2015})\BibitemShut
      {NoStop}%
    \bibitem [{\citenamefont {Gull}\ \emph {et~al.}(2008)\citenamefont {Gull},
      \citenamefont {Werner}, \citenamefont {Parcollet},\ and\ \citenamefont
      {Troyer}}]{Gull_continuous_2008}%
      \BibitemOpen
      \bibfield  {author} {\bibinfo {author} {\bibfnamefont {E.}~\bibnamefont
      {Gull}}, \bibinfo {author} {\bibfnamefont {P.}~\bibnamefont {Werner}},
      \bibinfo {author} {\bibfnamefont {O.}~\bibnamefont {Parcollet}},\ and\
      \bibinfo {author} {\bibfnamefont {M.}~\bibnamefont {Troyer}},\ }\bibfield
      {title} {\bibinfo {title} {Continuous-time auxiliary-field {Monte} {Carlo}
      for quantum impurity models},\ }\href
      {https://doi.org/10.1209/0295-5075/82/57003} {\bibfield  {journal} {\bibinfo
      {journal} {{EPL} (Europhysics Letters)}\ }\textbf {\bibinfo {volume} {82}},\
      \bibinfo {pages} {57003} (\bibinfo {year} {2008})}\BibitemShut {NoStop}%
    \bibitem [{\citenamefont {Gull}\ \emph {et~al.}(2011)\citenamefont {Gull},
      \citenamefont {Millis}, \citenamefont {Lichtenstein}, \citenamefont
      {Rubtsov}, \citenamefont {Troyer},\ and\ \citenamefont {Werner}}]{Gull:2011}%
      \BibitemOpen
      \bibfield  {author} {\bibinfo {author} {\bibfnamefont {E.}~\bibnamefont
      {Gull}}, \bibinfo {author} {\bibfnamefont {A.~J.}\ \bibnamefont {Millis}},
      \bibinfo {author} {\bibfnamefont {A.~I.}\ \bibnamefont {Lichtenstein}},
      \bibinfo {author} {\bibfnamefont {A.~N.}\ \bibnamefont {Rubtsov}}, \bibinfo
      {author} {\bibfnamefont {M.}~\bibnamefont {Troyer}},\ and\ \bibinfo {author}
      {\bibfnamefont {P.}~\bibnamefont {Werner}},\ }\bibfield  {title} {\bibinfo
      {title} {Continuous-time monte~carlo methods for quantum impurity models},\
      }\href {https://doi.org/10.1103/RevModPhys.83.349} {\bibfield  {journal}
      {\bibinfo  {journal} {Rev. Mod. Phys.}\ }\textbf {\bibinfo {volume} {83}},\
      \bibinfo {pages} {349} (\bibinfo {year} {2011})}\BibitemShut {NoStop}%
    \bibitem [{\citenamefont {Troyer}\ and\ \citenamefont
      {Wiese}(2005)}]{troyer2005computational}%
      \BibitemOpen
      \bibfield  {author} {\bibinfo {author} {\bibfnamefont {M.}~\bibnamefont
      {Troyer}}\ and\ \bibinfo {author} {\bibfnamefont {U.-J.}\ \bibnamefont
      {Wiese}},\ }\bibfield  {title} {\bibinfo {title} {Computational complexity
      and fundamental limitations to fermionic quantum monte carlo simulations},\
      }\href {https://journals.aps.org/prl/abstract/10.1103/PhysRevLett.94.170201}
      {\bibfield  {journal} {\bibinfo  {journal} {Physical review letters}\
      }\textbf {\bibinfo {volume} {94}},\ \bibinfo {pages} {170201} (\bibinfo
      {year} {2005})}\BibitemShut {NoStop}%
    \bibitem [{\citenamefont {Gull}\ \emph {et~al.}(2010)\citenamefont {Gull},
      \citenamefont {Ferrero}, \citenamefont {Parcollet}, \citenamefont {Georges},\
      and\ \citenamefont {Millis}}]{GullFerrero:2010}%
      \BibitemOpen
      \bibfield  {author} {\bibinfo {author} {\bibfnamefont {E.}~\bibnamefont
      {Gull}}, \bibinfo {author} {\bibfnamefont {M.}~\bibnamefont {Ferrero}},
      \bibinfo {author} {\bibfnamefont {O.}~\bibnamefont {Parcollet}}, \bibinfo
      {author} {\bibfnamefont {A.}~\bibnamefont {Georges}},\ and\ \bibinfo {author}
      {\bibfnamefont {A.~J.}\ \bibnamefont {Millis}},\ }\bibfield  {title}
      {\bibinfo {title} {Momentum-space anisotropy and pseudogaps: A comparative
      cluster dynamical mean-field analysis of the doping-driven metal-insulator
      transition in the two-dimensional hubbard model},\ }\href
      {https://doi.org/10.1103/PhysRevB.82.155101} {\bibfield  {journal} {\bibinfo
      {journal} {Phys. Rev. B}\ }\textbf {\bibinfo {volume} {82}},\ \bibinfo
      {pages} {155101} (\bibinfo {year} {2010})}\BibitemShut {NoStop}%
    \bibitem [{\citenamefont {Sakai}\ \emph {et~al.}(2012)\citenamefont {Sakai},
      \citenamefont {Sangiovanni}, \citenamefont {Civelli}, \citenamefont {Motome},
      \citenamefont {Held},\ and\ \citenamefont {Imada}}]{SakaiSize:2012}%
      \BibitemOpen
      \bibfield  {author} {\bibinfo {author} {\bibfnamefont {S.}~\bibnamefont
      {Sakai}}, \bibinfo {author} {\bibfnamefont {G.}~\bibnamefont {Sangiovanni}},
      \bibinfo {author} {\bibfnamefont {M.}~\bibnamefont {Civelli}}, \bibinfo
      {author} {\bibfnamefont {Y.}~\bibnamefont {Motome}}, \bibinfo {author}
      {\bibfnamefont {K.}~\bibnamefont {Held}},\ and\ \bibinfo {author}
      {\bibfnamefont {M.}~\bibnamefont {Imada}},\ }\bibfield  {title} {\bibinfo
      {title} {Cluster-size dependence in cellular dynamical mean-field theory},\
      }\href {https://doi.org/10.1103/PhysRevB.85.035102} {\bibfield  {journal}
      {\bibinfo  {journal} {Phys. Rev. B}\ }\textbf {\bibinfo {volume} {85}},\
      \bibinfo {pages} {035102} (\bibinfo {year} {2012})}\BibitemShut {NoStop}%
    \bibitem [{\citenamefont {Bergeron}\ and\ \citenamefont
      {Tremblay}(2016)}]{MaxEntBergeron}%
      \BibitemOpen
      \bibfield  {author} {\bibinfo {author} {\bibfnamefont {D.}~\bibnamefont
      {Bergeron}}\ and\ \bibinfo {author} {\bibfnamefont {A.-M.~S.}\ \bibnamefont
      {Tremblay}},\ }\bibfield  {title} {\bibinfo {title} {Algorithms for optimized
      maximum entropy and diagnostic tools for analytic continuation},\ }\href
      {https://doi.org/10.1103/PhysRevE.94.023303} {\bibfield  {journal} {\bibinfo
      {journal} {Phys. Rev. E}\ }\textbf {\bibinfo {volume} {94}},\ \bibinfo
      {pages} {023303} (\bibinfo {year} {2016})}\BibitemShut {NoStop}%
    \bibitem [{\citenamefont {Deng}\ \emph {et~al.}(2013)\citenamefont {Deng},
      \citenamefont {Mravlje}, \citenamefont {\ifmmode~\check{Z}\else
      \v{Z}\fi{}itko}, \citenamefont {Ferrero}, \citenamefont {Kotliar},\ and\
      \citenamefont {Georges}}]{dengHowBadMetals2013}%
      \BibitemOpen
      \bibfield  {author} {\bibinfo {author} {\bibfnamefont {X.}~\bibnamefont
      {Deng}}, \bibinfo {author} {\bibfnamefont {J.}~\bibnamefont {Mravlje}},
      \bibinfo {author} {\bibfnamefont {R.}~\bibnamefont {\ifmmode~\check{Z}\else
      \v{Z}\fi{}itko}}, \bibinfo {author} {\bibfnamefont {M.}~\bibnamefont
      {Ferrero}}, \bibinfo {author} {\bibfnamefont {G.}~\bibnamefont {Kotliar}},\
      and\ \bibinfo {author} {\bibfnamefont {A.}~\bibnamefont {Georges}},\
      }\bibfield  {title} {\bibinfo {title} {How {Bad} {Metals} {Turn} {Good}:
      {Spectroscopic} {Signatures} of {Resilient} {Quasiparticles}},\ }\href
      {https://doi.org/10.1103/PhysRevLett.110.086401} {\bibfield  {journal}
      {\bibinfo  {journal} {Phys. Rev. Lett.}\ }\textbf {\bibinfo {volume} {110}},\
      \bibinfo {pages} {086401} (\bibinfo {year} {2013})}\BibitemShut {NoStop}%
    \bibitem [{\citenamefont {Jarrell}\ and\ \citenamefont
      {Gubernatis}(1996)}]{Jarrell:1996}%
      \BibitemOpen
      \bibfield  {author} {\bibinfo {author} {\bibfnamefont {M.}~\bibnamefont
      {Jarrell}}\ and\ \bibinfo {author} {\bibfnamefont {J.}~\bibnamefont
      {Gubernatis}},\ }\bibfield  {title} {\bibinfo {title} {Bayesian inference and
      the analytic continuation of imaginary-time quantum monte carlo data},\
      }\href {https://doi.org/10.1016/0370-1573(95)00074-7} {\bibfield  {journal}
      {\bibinfo  {journal} {Physics Reports}\ }\textbf {\bibinfo {volume} {269}},\
      \bibinfo {pages} {133 } (\bibinfo {year} {1996})}\BibitemShut {NoStop}%
    \bibitem [{\citenamefont {Lebigot}(2010)}]{lebigotUncertaintiesPythonPackagea}%
      \BibitemOpen
      \bibfield  {author} {\bibinfo {author} {\bibfnamefont {E.~O.}\ \bibnamefont
      {Lebigot}},\ }\href {https://pythonhosted.org/uncertainties/index.html}
      {\bibinfo {title} {Uncertainties: A {{Python}} package for calculations with
      uncertainties}} (\bibinfo {year} {2010}),\ \bibinfo {note}
      {\url{https://pythonhosted.org/uncertainties/index.html}}\BibitemShut
      {NoStop}%
    \bibitem [{\citenamefont {Sordi}\ \emph {et~al.}(2013)\citenamefont {Sordi},
      \citenamefont {S\'emon}, \citenamefont {Haule},\ and\ \citenamefont
      {Tremblay}}]{SordiResistivity:2013}%
      \BibitemOpen
      \bibfield  {author} {\bibinfo {author} {\bibfnamefont {G.}~\bibnamefont
      {Sordi}}, \bibinfo {author} {\bibfnamefont {P.}~\bibnamefont {S\'emon}},
      \bibinfo {author} {\bibfnamefont {K.}~\bibnamefont {Haule}},\ and\ \bibinfo
      {author} {\bibfnamefont {A.-M.~S.}\ \bibnamefont {Tremblay}},\ }\bibfield
      {title} {\bibinfo {title} {$c$-axis resistivity, pseudogap,
      superconductivity, and widom line in doped mott insulators},\ }\href
      {https://doi.org/10.1103/PhysRevB.87.041101} {\bibfield  {journal} {\bibinfo
      {journal} {Phys. Rev. B}\ }\textbf {\bibinfo {volume} {87}},\ \bibinfo
      {pages} {041101} (\bibinfo {year} {2013})}\BibitemShut {NoStop}%
    \bibitem [{\citenamefont {Parcollet}\ \emph {et~al.}(2004)\citenamefont
      {Parcollet}, \citenamefont {Biroli},\ and\ \citenamefont
      {Kotliar}}]{Parcollet:2004}%
      \BibitemOpen
      \bibfield  {author} {\bibinfo {author} {\bibfnamefont {O.}~\bibnamefont
      {Parcollet}}, \bibinfo {author} {\bibfnamefont {G.}~\bibnamefont {Biroli}},\
      and\ \bibinfo {author} {\bibfnamefont {G.}~\bibnamefont {Kotliar}},\
      }\bibfield  {title} {\bibinfo {title} {Cluster dynamical mean field analysis
      of the mott transition},\ }\href
      {https://doi.org/10.1103/PhysRevLett.92.226402} {\bibfield  {journal}
      {\bibinfo  {journal} {Phys. Rev. Lett.}\ }\textbf {\bibinfo {volume} {92}},\
      \bibinfo {pages} {226402} (\bibinfo {year} {2004})}\BibitemShut {NoStop}%
    \bibitem [{\citenamefont {Kang}\ \emph {et~al.}(2011)\citenamefont {Kang},
      \citenamefont {Yu}, \citenamefont {Xiang},\ and\ \citenamefont
      {Li}}]{Kang_Yu_Xiang_Li_2011}%
      \BibitemOpen
      \bibfield  {author} {\bibinfo {author} {\bibfnamefont {J.}~\bibnamefont
      {Kang}}, \bibinfo {author} {\bibfnamefont {S.-L.}\ \bibnamefont {Yu}},
      \bibinfo {author} {\bibfnamefont {T.}~\bibnamefont {Xiang}},\ and\ \bibinfo
      {author} {\bibfnamefont {J.-X.}\ \bibnamefont {Li}},\ }\bibfield  {title}
      {\bibinfo {title} {Pseudogap and fermi arc in $\ensuremath{\kappa}$-type
      organic superconductors},\ }\href
      {https://doi.org/10.1103/PhysRevB.84.064520} {\bibfield  {journal} {\bibinfo
      {journal} {Physical Review B}\ }\textbf {\bibinfo {volume} {84}},\ \bibinfo
      {pages} {064520} (\bibinfo {year} {2011})}\BibitemShut {NoStop}%
    \bibitem [{\citenamefont {Vilk}(1997)}]{VilkShadow:1997}%
      \BibitemOpen
      \bibfield  {author} {\bibinfo {author} {\bibfnamefont {Y.~M.}\ \bibnamefont
      {Vilk}},\ }\bibfield  {title} {\bibinfo {title} {Shadow features and shadow
      bands in the paramagnetic state of cuprate superconductors},\ }\href
      {https://doi.org/10.1103/PhysRevB.55.3870} {\bibfield  {journal} {\bibinfo
      {journal} {Phys. Rev. B}\ }\textbf {\bibinfo {volume} {55}},\ \bibinfo
      {pages} {3870} (\bibinfo {year} {1997})}\BibitemShut {NoStop}%
    \bibitem [{\citenamefont {Kyung}\ \emph {et~al.}(2004)\citenamefont {Kyung},
      \citenamefont {Hankevych}, \citenamefont {Dar\'e},\ and\ \citenamefont
      {Tremblay}}]{Kyung:2004}%
      \BibitemOpen
      \bibfield  {author} {\bibinfo {author} {\bibfnamefont {B.}~\bibnamefont
      {Kyung}}, \bibinfo {author} {\bibfnamefont {V.}~\bibnamefont {Hankevych}},
      \bibinfo {author} {\bibfnamefont {A.-M.}\ \bibnamefont {Dar\'e}},\ and\
      \bibinfo {author} {\bibfnamefont {A.-M.~S.}\ \bibnamefont {Tremblay}},\
      }\bibfield  {title} {\bibinfo {title} {Pseudogap and spin fluctuations in the
      normal state of the electron-doped cuprates},\ }\href
      {https://doi.org/10.1103/PhysRevLett.93.147004} {\bibfield  {journal}
      {\bibinfo  {journal} {Phys. Rev. Lett.}\ }\textbf {\bibinfo {volume} {93}},\
      \bibinfo {pages} {147004} (\bibinfo {year} {2004})}\BibitemShut {NoStop}%
    \bibitem [{\citenamefont {Motoyama}\ \emph {et~al.}(2007)\citenamefont
      {Motoyama}, \citenamefont {Yu}, \citenamefont {Vishik}, \citenamefont {Vajk},
      \citenamefont {Mang},\ and\ \citenamefont {Greven}}]{Motoyama:2007}%
      \BibitemOpen
      \bibfield  {author} {\bibinfo {author} {\bibfnamefont {E.~M.}\ \bibnamefont
      {Motoyama}}, \bibinfo {author} {\bibfnamefont {G.}~\bibnamefont {Yu}},
      \bibinfo {author} {\bibfnamefont {I.~M.}\ \bibnamefont {Vishik}}, \bibinfo
      {author} {\bibfnamefont {O.~P.}\ \bibnamefont {Vajk}}, \bibinfo {author}
      {\bibfnamefont {P.~K.}\ \bibnamefont {Mang}},\ and\ \bibinfo {author}
      {\bibfnamefont {M.}~\bibnamefont {Greven}},\ }\bibfield  {title} {\bibinfo
      {title} {Spin correlations in the electron-doped high-transition-temperature
      superconductor {NCCO}},\ }\href
      {https://doi.org/https://doi.org/10.1038/nature05437} {\bibfield  {journal}
      {\bibinfo  {journal} {Nature}\ }\textbf {\bibinfo {volume} {445}},\ \bibinfo
      {pages} {186} (\bibinfo {year} {2007})}\BibitemShut {NoStop}%
    \bibitem [{\citenamefont {Sahebsara}\ and\ \citenamefont
      {S\'en\'echal}(2006)}]{Sahebsara:2006}%
      \BibitemOpen
      \bibfield  {author} {\bibinfo {author} {\bibfnamefont {P.}~\bibnamefont
      {Sahebsara}}\ and\ \bibinfo {author} {\bibfnamefont {D.}~\bibnamefont
      {S\'en\'echal}},\ }\bibfield  {title} {\bibinfo {title} {Antiferromagnetism
      and {Superconductivity} in {Layered} {Organic} {Conductors}: {Variational}
      {Cluster} {Approach}},\ }\href
      {https://doi.org/10.1103/PhysRevLett.97.257004} {\bibfield  {journal}
      {\bibinfo  {journal} {Phys. Rev. Lett.}\ }\textbf {\bibinfo {volume} {97}},\
      \bibinfo {pages} {257004} (\bibinfo {year} {2006})}\BibitemShut {NoStop}%
    \bibitem [{\citenamefont {Sordi}\ \emph {et~al.}(2012)\citenamefont {Sordi},
      \citenamefont {S{\'e}mon}, \citenamefont {Haule},\ and\ \citenamefont
      {Tremblay}}]{Sordi:2012}%
      \BibitemOpen
      \bibfield  {author} {\bibinfo {author} {\bibfnamefont {G.}~\bibnamefont
      {Sordi}}, \bibinfo {author} {\bibfnamefont {P.}~\bibnamefont {S{\'e}mon}},
      \bibinfo {author} {\bibfnamefont {K.}~\bibnamefont {Haule}},\ and\ \bibinfo
      {author} {\bibfnamefont {A.-M.~S.}\ \bibnamefont {Tremblay}},\ }\bibfield
      {title} {\bibinfo {title} {Pseudogap temperature as a widom line in doped
      mott insulators},\ }\bibfield  {journal} {\bibinfo  {journal} {Scientific
      Reports}\ }\textbf {\bibinfo {volume} {2}},\ \href
      {https://doi.org/10.1038/srep00547} {10.1038/srep00547} (\bibinfo {year}
      {2012})\BibitemShut {NoStop}%
    \bibitem [{\citenamefont {Kotliar}\ \emph {et~al.}(2000)\citenamefont
      {Kotliar}, \citenamefont {Lange},\ and\ \citenamefont
      {Rozenberg}}]{KotliarLange:2000}%
      \BibitemOpen
      \bibfield  {author} {\bibinfo {author} {\bibfnamefont {G.}~\bibnamefont
      {Kotliar}}, \bibinfo {author} {\bibfnamefont {E.}~\bibnamefont {Lange}},\
      and\ \bibinfo {author} {\bibfnamefont {M.~J.}\ \bibnamefont {Rozenberg}},\
      }\bibfield  {title} {\bibinfo {title} {Landau theory of the finite
      temperature mott transition},\ }\href
      {https://doi.org/10.1103/PhysRevLett.84.5180} {\bibfield  {journal} {\bibinfo
       {journal} {Phys. Rev. Lett.}\ }\textbf {\bibinfo {volume} {84}},\ \bibinfo
      {pages} {5180} (\bibinfo {year} {2000})}\BibitemShut {NoStop}%
    \bibitem [{\citenamefont {Imada}(2005)}]{imadaQuantumMottTransition2005}%
      \BibitemOpen
      \bibfield  {author} {\bibinfo {author} {\bibfnamefont {M.}~\bibnamefont
      {Imada}},\ }\bibfield  {title} {\bibinfo {title} {Quantum {{Mott Transition}}
      and {{Superconductivity}}},\ }\href {https://doi.org/10.1143/JPSJ.74.859}
      {\bibfield  {journal} {\bibinfo  {journal} {J. Phys. Soc. Jpn.}\ }\textbf
      {\bibinfo {volume} {74}},\ \bibinfo {pages} {859} (\bibinfo {year}
      {2005})}\BibitemShut {NoStop}%
    \bibitem [{\citenamefont {Szasz}\ and\ \citenamefont
      {Motruk}(2021)}]{szaszPhaseDiagramAnisotropic2021}%
      \BibitemOpen
      \bibfield  {author} {\bibinfo {author} {\bibfnamefont {A.}~\bibnamefont
      {Szasz}}\ and\ \bibinfo {author} {\bibfnamefont {J.}~\bibnamefont {Motruk}},\
      }\bibfield  {title} {\bibinfo {title} {Phase diagram of the anisotropic
      triangular lattice {{Hubbard}} model},\ }\href
      {https://doi.org/10.1103/PhysRevB.103.235132} {\bibfield  {journal} {\bibinfo
       {journal} {Physical Review B}\ }\textbf {\bibinfo {volume} {103}},\ \bibinfo
      {pages} {235132} (\bibinfo {year} {2021})}\BibitemShut {NoStop}%
    \bibitem [{\citenamefont {Reymbaut}\ \emph {et~al.}(2020)\citenamefont
      {Reymbaut}, \citenamefont {Boulay}, \citenamefont {Fratino}, \citenamefont
      {Sémon}, \citenamefont {Wu}, \citenamefont {Sordi},\ and\ \citenamefont
      {Tremblay}}]{reymbautMottTransition2020a}%
      \BibitemOpen
      \bibfield  {author} {\bibinfo {author} {\bibfnamefont {A.}~\bibnamefont
      {Reymbaut}}, \bibinfo {author} {\bibfnamefont {M.}~\bibnamefont {Boulay}},
      \bibinfo {author} {\bibfnamefont {L.}~\bibnamefont {Fratino}}, \bibinfo
      {author} {\bibfnamefont {P.}~\bibnamefont {Sémon}}, \bibinfo {author}
      {\bibfnamefont {W.}~\bibnamefont {Wu}}, \bibinfo {author} {\bibfnamefont
      {G.}~\bibnamefont {Sordi}},\ and\ \bibinfo {author} {\bibfnamefont
      {A.~M.~S.}\ \bibnamefont {Tremblay}},\ }\href
      {https://doi.org/10.48550/ARXIV.2004.02302} {\bibinfo {title} {Mott
      transition and high-temperature crossovers at half-filling}} (\bibinfo {year}
      {2020})\BibitemShut {NoStop}%
    \bibitem [{\citenamefont {Sordi}\ \emph {et~al.}(2010)\citenamefont {Sordi},
      \citenamefont {Haule},\ and\ \citenamefont {Tremblay}}]{Sordi:2010}%
      \BibitemOpen
      \bibfield  {author} {\bibinfo {author} {\bibfnamefont {G.}~\bibnamefont
      {Sordi}}, \bibinfo {author} {\bibfnamefont {K.}~\bibnamefont {Haule}},\ and\
      \bibinfo {author} {\bibfnamefont {A.~M.~S.}\ \bibnamefont {Tremblay}},\
      }\bibfield  {title} {\bibinfo {title} {Finite doping signatures of the mott
      transition in the two-dimensional hubbard model},\ }\href
      {https://doi.org/10.1103/PhysRevLett.104.226402} {\bibfield  {journal}
      {\bibinfo  {journal} {Phys. Rev. Lett.}\ }\textbf {\bibinfo {volume} {104}},\
      \bibinfo {pages} {226402} (\bibinfo {year} {2010})}\BibitemShut {NoStop}%
    \bibitem [{\citenamefont {Betts}\ \emph {et~al.}(1999)\citenamefont {Betts},
      \citenamefont {Lin},\ and\ \citenamefont {Flynn}}]{betts:1999}%
      \BibitemOpen
      \bibfield  {author} {\bibinfo {author} {\bibfnamefont {D.~D.}\ \bibnamefont
      {Betts}}, \bibinfo {author} {\bibfnamefont {H.~Q.}\ \bibnamefont {Lin}},\
      and\ \bibinfo {author} {\bibfnamefont {J.~S.}\ \bibnamefont {Flynn}},\
      }\bibfield  {title} {\bibinfo {title} {Improved finite-lattice estimates of
      the properties of two quantum spin models on the infinite square lattice},\
      }\href {https://doi.org/10.1139/p99-041} {\bibfield  {journal} {\bibinfo
      {journal} {Canadian Journal of Physics}\ }\textbf {\bibinfo {volume} {77}},\
      \bibinfo {pages} {353} (\bibinfo {year} {1999})}\BibitemShut {NoStop}%
    \bibitem [{\citenamefont {Maier}\ \emph
      {et~al.}(2005{\natexlab{b}})\citenamefont {Maier}, \citenamefont {Jarrell},
      \citenamefont {Schulthess}, \citenamefont {Kent},\ and\ \citenamefont
      {White}}]{Maier_Betts_2005}%
      \BibitemOpen
      \bibfield  {author} {\bibinfo {author} {\bibfnamefont {T.~A.}\ \bibnamefont
      {Maier}}, \bibinfo {author} {\bibfnamefont {M.}~\bibnamefont {Jarrell}},
      \bibinfo {author} {\bibfnamefont {T.~C.}\ \bibnamefont {Schulthess}},
      \bibinfo {author} {\bibfnamefont {P.~R.~C.}\ \bibnamefont {Kent}},\ and\
      \bibinfo {author} {\bibfnamefont {J.~B.}\ \bibnamefont {White}},\ }\bibfield
      {title} {\bibinfo {title} {Systematic {Study} of $d$-{Wave}
      {Superconductivity} in the {2D} {Repulsive} {Hubbard} {Model}},\ }\href
      {https://doi.org/10.1103/PhysRevLett.95.237001} {\bibfield  {journal}
      {\bibinfo  {journal} {Phys. Rev. Lett.}\ }\textbf {\bibinfo {volume} {95}},\
      \bibinfo {pages} {237001} (\bibinfo {year} {2005}{\natexlab{b}})}\BibitemShut
      {NoStop}%
    \bibitem [{Note1()}]{Note1}%
      \BibitemOpen
      \bibinfo {note} {Note that the Widom line for $N_c=1$ seems to be different
      from the ``quantum Widom line'' in Ref.~\cite
      {vucicevic:2013,Eisenlohr_Lee_Vojta_2019}.}\BibitemShut {Stop}%
    \bibitem [{\citenamefont {Rosenberg}\ \emph {et~al.}(2022)\citenamefont
      {Rosenberg}, \citenamefont {Sénéchal}, \citenamefont {Tremblay},\ and\
      \citenamefont {Charlebois}}]{rosenberg_fermi_2022}%
      \BibitemOpen
      \bibfield  {author} {\bibinfo {author} {\bibfnamefont {P.}~\bibnamefont
      {Rosenberg}}, \bibinfo {author} {\bibfnamefont {D.}~\bibnamefont
      {Sénéchal}}, \bibinfo {author} {\bibfnamefont {A.-M.~S.}\ \bibnamefont
      {Tremblay}},\ and\ \bibinfo {author} {\bibfnamefont {M.}~\bibnamefont
      {Charlebois}},\ }\bibfield  {title} {\bibinfo {title} {Fermi arcs from
      dynamical variational {Monte} {Carlo}},\ }\href
      {https://doi.org/10.1103/PhysRevB.106.245132} {\bibfield  {journal} {\bibinfo
       {journal} {Physical Review B}\ }\textbf {\bibinfo {volume} {106}},\ \bibinfo
      {pages} {245132} (\bibinfo {year} {2022})},\ \bibinfo {note} {publisher:
      American Physical Society}\BibitemShut {NoStop}%
    \bibitem [{\citenamefont {Li}\ and\ \citenamefont {Gull}(2020)}]{Li_Gull_2020}%
      \BibitemOpen
      \bibfield  {author} {\bibinfo {author} {\bibfnamefont {S.}~\bibnamefont
      {Li}}\ and\ \bibinfo {author} {\bibfnamefont {E.}~\bibnamefont {Gull}},\
      }\bibfield  {title} {\bibinfo {title} {Magnetic and charge susceptibilities
      in the half-filled triangular lattice hubbard model},\ }\href
      {https://doi.org/10.1103/PhysRevResearch.2.013295} {\bibfield  {journal}
      {\bibinfo  {journal} {Physical Review Research}\ }\textbf {\bibinfo {volume}
      {2}},\ \bibinfo {pages} {013295} (\bibinfo {year} {2020})}\BibitemShut
      {NoStop}%
    \bibitem [{\citenamefont {Vranić}\ \emph {et~al.}(2020)\citenamefont
      {Vranić}, \citenamefont {Vučičević}, \citenamefont {Kokalj},
      \citenamefont {Skolimowski}, \citenamefont {Žitko}, \citenamefont
      {Mravlje},\ and\ \citenamefont {Tanasković}}]{Vranic_Vucicevic_2020}%
      \BibitemOpen
      \bibfield  {author} {\bibinfo {author} {\bibfnamefont {A.}~\bibnamefont
      {Vranić}}, \bibinfo {author} {\bibfnamefont {J.}~\bibnamefont
      {Vučičević}}, \bibinfo {author} {\bibfnamefont {J.}~\bibnamefont
      {Kokalj}}, \bibinfo {author} {\bibfnamefont {J.}~\bibnamefont {Skolimowski}},
      \bibinfo {author} {\bibfnamefont {R.}~\bibnamefont {Žitko}}, \bibinfo
      {author} {\bibfnamefont {J.}~\bibnamefont {Mravlje}},\ and\ \bibinfo {author}
      {\bibfnamefont {D.}~\bibnamefont {Tanasković}},\ }\bibfield  {title}
      {\bibinfo {title} {Charge transport in the hubbard model at high
      temperatures: Triangular versus square lattice},\ }\href
      {https://doi.org/10.1103/PhysRevB.102.115142} {\bibfield  {journal} {\bibinfo
       {journal} {Physical Review B}\ }\textbf {\bibinfo {volume} {102}},\ \bibinfo
      {pages} {115142} (\bibinfo {year} {2020})}\BibitemShut {NoStop}%
    \bibitem [{\citenamefont {Lee}\ \emph {et~al.}(2008)\citenamefont {Lee},
      \citenamefont {Li},\ and\ \citenamefont {Monien}}]{LeeDualFermions:2008}%
      \BibitemOpen
      \bibfield  {author} {\bibinfo {author} {\bibfnamefont {H.}~\bibnamefont
      {Lee}}, \bibinfo {author} {\bibfnamefont {G.}~\bibnamefont {Li}},\ and\
      \bibinfo {author} {\bibfnamefont {H.}~\bibnamefont {Monien}},\ }\bibfield
      {title} {\bibinfo {title} {Hubbard model on the triangular lattice using
      dynamical cluster approximation and dual fermion methods},\ }\href
      {https://doi.org/10.1103/PhysRevB.78.205117} {\bibfield  {journal} {\bibinfo
      {journal} {Phys. Rev. B}\ }\textbf {\bibinfo {volume} {78}},\ \bibinfo
      {pages} {205117} (\bibinfo {year} {2008})}\BibitemShut {NoStop}%
    \bibitem [{\citenamefont {Ohashi}\ \emph {et~al.}(2008)\citenamefont {Ohashi},
      \citenamefont {Momoi}, \citenamefont {Tsunetsugu},\ and\ \citenamefont
      {Kawakami}}]{ohashi:2008}%
      \BibitemOpen
      \bibfield  {author} {\bibinfo {author} {\bibfnamefont {T.}~\bibnamefont
      {Ohashi}}, \bibinfo {author} {\bibfnamefont {T.}~\bibnamefont {Momoi}},
      \bibinfo {author} {\bibfnamefont {H.}~\bibnamefont {Tsunetsugu}},\ and\
      \bibinfo {author} {\bibfnamefont {N.}~\bibnamefont {Kawakami}},\ }\bibfield
      {title} {\bibinfo {title} {Finite temperature mott transition in hubbard
      model on anisotropic triangular lattice},\ }\href
      {https://doi.org/10.1103/PhysRevLett.100.076402} {\bibfield  {journal}
      {\bibinfo  {journal} {Physical Review Letters}\ }\textbf {\bibinfo {volume}
      {100}},\ \bibinfo {eid} {076402} (\bibinfo {year} {2008})}\BibitemShut
      {NoStop}%
    \bibitem [{\citenamefont {Kézsmárki}\ \emph {et~al.}(2006)\citenamefont
      {Kézsmárki}, \citenamefont {Shimizu}, \citenamefont {Mihály},
      \citenamefont {Tokura}, \citenamefont {Kanoda},\ and\ \citenamefont
      {Saito}}]{Shimizu_Tokura_Kanoda_Saito_2006}%
      \BibitemOpen
      \bibfield  {author} {\bibinfo {author} {\bibfnamefont {I.}~\bibnamefont
      {Kézsmárki}}, \bibinfo {author} {\bibfnamefont {Y.}~\bibnamefont
      {Shimizu}}, \bibinfo {author} {\bibfnamefont {G.}~\bibnamefont {Mihály}},
      \bibinfo {author} {\bibfnamefont {Y.}~\bibnamefont {Tokura}}, \bibinfo
      {author} {\bibfnamefont {K.}~\bibnamefont {Kanoda}},\ and\ \bibinfo {author}
      {\bibfnamefont {G.}~\bibnamefont {Saito}},\ }\bibfield  {title} {\bibinfo
      {title} {Depressed charge gap in the triangular-lattice mott insulator
      $\ensuremath{\kappa}\text{\ensuremath{-}}{(\mathrm{ET})}_{2}{\mathrm{cu}}_{2}{(\mathrm{C}\mathrm{N})}_{3}$},\
      }\href {https://doi.org/10.1103/PhysRevB.74.201101} {\bibfield  {journal}
      {\bibinfo  {journal} {Physical Review B}\ }\textbf {\bibinfo {volume} {74}},\
      \bibinfo {pages} {201101} (\bibinfo {year} {2006})}\BibitemShut {NoStop}%
    \bibitem [{\citenamefont {Pustogow}(2022)}]{Pustogow_2022}%
      \BibitemOpen
      \bibfield  {author} {\bibinfo {author} {\bibfnamefont {A.}~\bibnamefont
      {Pustogow}},\ }\bibfield  {title} {\bibinfo {title} {Thirty-year anniversary
      of $\kappa$-({BEDT-TTF})$_2${Cu}$_2$({CN})$_3$: Reconciling the spin gap in a
      spin-liquid candidate},\ }\href {https://doi.org/10.3390/solids3010007}
      {\bibfield  {journal} {\bibinfo  {journal} {Solids}\ }\textbf {\bibinfo
      {volume} {3}},\ \bibinfo {pages} {93–110} (\bibinfo {year}
      {2022})}\BibitemShut {NoStop}%
    \bibitem [{\citenamefont {Yesil}\ \emph {et~al.}(2023)\citenamefont {Yesil},
      \citenamefont {Imajo}, \citenamefont {Yamashita}, \citenamefont {Akutsu},
      \citenamefont {Saito}, \citenamefont {Pustogow}, \citenamefont {Kawamoto},\
      and\ \citenamefont
      {Nakazawa}}]{Yesil_Imajo_Yamashita_Akutsu_Saito_Pustogow_Kawamoto_Nakazawa_2023}%
      \BibitemOpen
      \bibfield  {author} {\bibinfo {author} {\bibfnamefont {E.}~\bibnamefont
      {Yesil}}, \bibinfo {author} {\bibfnamefont {S.}~\bibnamefont {Imajo}},
      \bibinfo {author} {\bibfnamefont {S.}~\bibnamefont {Yamashita}}, \bibinfo
      {author} {\bibfnamefont {H.}~\bibnamefont {Akutsu}}, \bibinfo {author}
      {\bibfnamefont {Y.}~\bibnamefont {Saito}}, \bibinfo {author} {\bibfnamefont
      {A.}~\bibnamefont {Pustogow}}, \bibinfo {author} {\bibfnamefont
      {A.}~\bibnamefont {Kawamoto}},\ and\ \bibinfo {author} {\bibfnamefont
      {Y.}~\bibnamefont {Nakazawa}},\ }\bibfield  {title} {\bibinfo {title}
      {Thermodynamic properties of the mott insulator-metal transition in a
      triangular lattice system without magnetic order},\ }\href
      {https://doi.org/10.1103/PhysRevB.107.045133} {\bibfield  {journal} {\bibinfo
       {journal} {Physical Review B}\ }\textbf {\bibinfo {volume} {107}},\ \bibinfo
      {pages} {045133} (\bibinfo {year} {2023})}\BibitemShut {NoStop}%
    \bibitem [{\citenamefont {Betts}\ \emph {et~al.}(1996)\citenamefont {Betts},
      \citenamefont {Masui}, \citenamefont {Vats},\ and\ \citenamefont
      {Stewart}}]{betts:1996}%
      \BibitemOpen
      \bibfield  {author} {\bibinfo {author} {\bibfnamefont {D.~D.}\ \bibnamefont
      {Betts}}, \bibinfo {author} {\bibfnamefont {S.}~\bibnamefont {Masui}},
      \bibinfo {author} {\bibfnamefont {N.}~\bibnamefont {Vats}},\ and\ \bibinfo
      {author} {\bibfnamefont {G.~E.}\ \bibnamefont {Stewart}},\ }\bibfield
      {title} {\bibinfo {title} {Improved finite-lattice method for estimating the
      zero-temperature properties of two-dimensional lattice models},\ }\bibfield
      {journal} {\bibinfo  {journal} {Canadian Journal of Physics}\ }\href
      {https://doi.org/10.1139/p96-010} {10.1139/p96-010} (\bibinfo {year}
      {1996})\BibitemShut {NoStop}%
    \bibitem [{\citenamefont {Eisenlohr}\ \emph {et~al.}(2019)\citenamefont
      {Eisenlohr}, \citenamefont {Lee},\ and\ \citenamefont
      {Vojta}}]{Eisenlohr_Lee_Vojta_2019}%
      \BibitemOpen
      \bibfield  {author} {\bibinfo {author} {\bibfnamefont {H.}~\bibnamefont
      {Eisenlohr}}, \bibinfo {author} {\bibfnamefont {S.-S.~B.}\ \bibnamefont
      {Lee}},\ and\ \bibinfo {author} {\bibfnamefont {M.}~\bibnamefont {Vojta}},\
      }\bibfield  {title} {\bibinfo {title} {Mott quantum criticality in the
      one-band hubbard model: Dynamical mean-field theory, power-law spectra, and
      scaling},\ }\href {https://doi.org/10.1103/PhysRevB.100.155152} {\bibfield
      {journal} {\bibinfo  {journal} {Physical Review B}\ }\textbf {\bibinfo
      {volume} {100}},\ \bibinfo {pages} {155152} (\bibinfo {year}
      {2019})}\BibitemShut {NoStop}%
    \end{thebibliography}

%

\end{document}